%% file: main.tex
\documentclass[journal,twocolumn]{IEEEtran} 

\usepackage[dvipsnames]{xcolor}
\usepackage{mathtools}
\usepackage{amsmath,amssymb,amsfonts}
\usepackage{algorithmic}
\usepackage[range-phrase=--,range-units=single,list-units=single,detect-all]{siunitx}
\usepackage{graphicx,import,svg}
\usepackage{transparent}
\usepackage{subcaption}
\usepackage{pgfplots}
\usepackage{tikz}
\usetikzlibrary{mindmap}
\usetikzlibrary{shapes,arrows}
\usetikzlibrary{arrows.meta}
\usetikzlibrary{trees}
\usepackage{cite}
\usepackage{csquotes}
\usepackage{textcomp}
\usepackage{multirow}
\usepackage{adjustbox}
\usepackage{tabularx}
\usepackage{enumitem}
\usepackage{bm}
\usepackage{pgf-pie}
\usepackage[T1]{fontenc}
\usepackage[font=footnotesize]{caption}
\usepackage{todonotes}
\usepackage{pifont}
\usepackage{wasysym}
\newcommand{\cmark}{\textcolor{green!70!black}{\ding{51}}} 
\newcommand{\xmark}{\textcolor{red}{\ding{55}}} 
\newcommand{\pmark}{\textcolor{orange}{\LEFTcircle}} 

\usepackage[hidelinks]{hyperref}
\usepackage[capitalise,noabbrev]{cleveref}
\Crefformat{figure}{#2Fig.~#1#3}
\Crefformat{equation}{#2Eq.~(#1)#3}

\pgfplotsset{compat=1.18}

\usepackage{fix-cm}

\usepackage{verbatim}

\usepackage{acro}
\usepackage{mfirstuc}
\DeclareAcronym{MC}{short=MC, long=Molecular Communication}
\DeclareAcronym{CAPD}{short=CAPD, long=Continuous Ambulatory Peritoneal Dialysis}
\DeclareAcronym{CALPUFF}{short=CALPUFF, long=California Puff}
\DeclareAcronym{VOC}{short=VOC, long=Volatile Organic Compound}
\DeclareAcronym{IoBNT}{short=IoBNT, long=Internet of Bio-Nano Things}
\DeclareAcronym{IMI}{short=IMI, long=Interferometric Mie Imaging}
\DeclareAcronym{ISI}{short=ISI, long=Inter-Symbol Interference}
\DeclareAcronym{IG}{short=IG, long=Inverse Gaussian}
\DeclareAcronym{AIGN}{short=AIGN, long=Additive Inverse Gaussian Noise}
\DeclareAcronym{CCI}{short=CCI, long=Co-Channel Interference}
\DeclareAcronym{CSK}{short=CSK, long=Concentration Shift Keying}
\DeclareAcronym{OOK}{short=OOK, long=On-Off Keying}
\DeclareAcronym{HBDB}{short=HBDB, long=Human Breathomics Data Base}
\DeclareAcronym{SAW}{short=SAW, long=Surface Acoustic Wave}
\DeclareAcronym{QCM}{short=QCM, long=Quartz Crystal Microbalance}
\DeclareAcronym{PCR}{short=PCR, long=Polymerase Chain Reaction}
\DeclareAcronym{qPCR}{short=qPCR, long=Quantitative Polymerase Chain Reaction}
\DeclareAcronym{EBT}{short=EBT, long=Exhaled Breath Temperature}
\DeclareAcronym{COPD}{short=COPD, long=Chronic Obstructive Pulmonary Disease}
\DeclareAcronym{POC}{short=POC, long=Point-Of-Care}
\DeclareAcronym{CMOS}{short=CMOS, long=Complementary Metal-Oxide Semiconductor}
\DeclareAcronym{MEMS}{short=MEMS, long=Micro-Electro-Mechanical Systems}
\DeclareAcronym{LPD}{short=LPD, long=Lagrangian Particle Dispersion}
\DeclareAcronym{LRT}{short=LRT, long=Lower Respiratory Tract}
\DeclareAcronym{PIV}{short=PIV, long=Particle Image Velocimetry}
\DeclareAcronym{URT}{short=URT, long=Upper Respiratory Tract}
\DeclareAcronym{MIMO}{short=MIMO, long=Multiple-Input Multiple-Output}
\DeclareAcronym{IoBT}{short=IoBT, long=Internet of Bio Things}
\DeclareAcronym{IoT}{short=IoT, long=Internet of Things}
\DeclareAcronym{DNA}{short=DNA, long=DeoxyriboNucleic Acid}
\DeclareAcronym{IC}{short=IC, long=Information Carrier}
\DeclareAcronym{IM}{short=IM, long=Information Molecule}
\DeclareAcronym{EBC}{short=EBC, long=Exhaled Breath Condensate}
\DeclareAcronym{RPA}{short=RPA, long=Recombinase Polymerase Amplification}
\DeclareAcronym{RNA}{short=RNA, long=RiboNucleic Acid}
\DeclareAcronym{ABMC}{short=ABMC, long=Air-Based Molecular Communication}
\DeclareAcronym{CFD}{short=CFD, long=Computational Fluid Dynamics}
\DeclareAcronym{BSA}{short=BSA, long=Body Surface Area}
\DeclareAcronym{HYSPLIT}{short=HYSPLIT, long=Hybrid$\!$ Single-Particle$\!$ Lagrangian$\!$ Integrated$\!$ Trajectory}
\DeclareAcronym{CT}{short=CT, long=Computerized Tomography}
\DeclareAcronym{MRI}{short=MRI, long=Magnetic Resonance Imaging}

\definecolor{mdblue}{HTML}{396AB1}
\definecolor{mdorange}{HTML}{DA7C30}
\definecolor{mdgreen}{HTML}{3E9651}
\definecolor{mdred}{HTML}{CC2529}
\definecolor{mdgray}{HTML}{535154}
\definecolor{mdpurple}{HTML}{6B4C9A}
\definecolor{mdbrown}{HTML}{922428}
\definecolor{mdyellow}{HTML}{948B3D}
\definecolor{darkgreen}{RGB}{1,125,32}
\pgfplotscreateplotcyclelist{colormd}{%
	mdblue,mdorange,mdgreen,mdred,mdgray,mdpurple,mdbrown,mdyellow}
\newcommand{\db}[1]{{\textcolor{olive}{#1}}}
\newcommand{\ph}[1]{{\textcolor{blue}{#1}}}

\usepackage{titlesec}
\titlespacing\section{0pt}{3pt plus 1pt minus 1pt}{2pt plus 1pt minus 1pt}
\titlespacing\subsection{0pt}{2pt plus 1pt minus 2pt}{2pt plus 0pt minus 1pt}

\begin{document}

\title{Exhaled Breath Analysis Through the Lens of Molecular Communication: A Survey}

\author{%
Sunasheer Bhattacharjee\textsuperscript{*},
Dadi Bi\textsuperscript{*},
Pit Hofmann\textsuperscript{*}, 
Alexander Wietfeld\textsuperscript{*},
\\
Sophie Becke, 
Michael Lommel,
Pengjie Zhou,
Ruifeng Zheng,
Ulrich Kertzscher,
\\
Yansha Deng,
Wolfgang Kellerer,
Frank H.P. Fitzek, and
Falko Dressler%
\thanks{\textsuperscript{*} S. Bhattacharjee, D. Bi, P. Hofmann, and A. Wietfeld contributed equally to this work and are listed as co-first authors alphabetically.}
\thanks{S. Bhattacharjee and F. Dressler are with the School of Electrical Engineering and Computer Science, TU Berlin, Germany,  e-mail: \{bhattacharjee,dressler\}@ccs-labs.org.}
\thanks{D.~Bi and Y.~Deng are with the Department of Engineering, King's College London, U.K., e-mail: \{dadi.bi,yansha.deng\}@kcl.ac.uk.}
\thanks{P. Hofmann, P. Zhou, R. Zheng, and F. Fitzek are with the Deutsche Telekom Chair of Communication Networks, Technische Universität Dresden, Germany; F. Fitzek is also with the Centre for Tactile Internet with Human-in-the-Loop (CeTI), Dresden, Germany, e-mail: \{pit.hofmann,pengjie.zhou,ruifeng.zheng,frank.fitzek\}@tu-dresden.de.}
\thanks{A. Wietfeld and W. Kellerer are with the Chair of Communication Networks, Technical University of Munich, Germany, e-mail: \{alexander.wietfeld, wolfgang.kellerer\}@tum.de.}
\thanks{S. Becke, M. Lommel, and U. Kertzscher are with the Institute of Computer-Assisted Cardiovascular Medicine, Deutsches Herzzentrum der Charité, Charité - Universitätsmedizin Berlin, Germany, e-mail: \{sophie.becke,michael.lommel\}@charite.de, ulrich.kertzscher@dhzc-charite.de.}
\thanks{Reported research was supported by the project IoBNT funded by the German Federal Ministry of Education and Research (BMBF) under grant numbers 16KIS1986K and 16KIS1994 and in the programme of ``Souverän. Digital. Vernetzt.'' Joint project 6G-life, grant numbers 16KISK001K and 16KISK002. 
This work was also supported by the German Research Foundation (DFG) as part of Germany's Excellence Strategy—EXC 2050/1—Cluster of Excellence ``Centre for Tactile Internet with Human-in-the-Loop'' (CeTI) of Technische Universität Dresden under project ID 390696704.}
}


\IEEEpubid{}

\maketitle

\begin{abstract}
\Ac{MC} has long been envisioned to enable an \ac{IoBNT} with medical applications, where nanomachines within the human body conduct monitoring, diagnosis, and therapy at micro- and nanoscale levels. 
\Ac{MC} involves information transfer via molecules and is supported by well-established theoretical models.
However, practically achieving reliable, energy-efficient, and bio-compatible communication at these scales still remains a challenge. 
\ac{ABMC} is a type of \ac{MC} that operates over larger, meter-scale distances and extends even outside the human body.
Therefore, devices and techniques to realize \ac{ABMC} are readily accessible, and associated use cases can be very promising in the near future.
Exhaled breath analysis has previously been proposed. 
It provides a non-invasive approach for health monitoring, leveraging existing commercial sensor technologies and reducing deployment barriers. 
The breath contains a diverse range of molecules and particles that serve as biomarkers linked to various physiological and pathological conditions.
The plethora of proven methods, models, and optimization approaches in \ac{MC} enable macroscale breath analysis, treating human as the transmitter, the breath as the information carrier, and macroscale sensors as the receiver.
Using \ac{ABMC} to interface with the inherent dynamic networks of cells, tissues, and organs could create a novel \ac{IoBT}, a preliminary macroscale stage of the \ac{IoBNT}.
This survey extensively reviews exhaled breath modeling and analysis through the lens of \ac{MC}, offering insights into theoretical frameworks and practical implementations from \ac{ABMC}, bringing the \ac{IoBT} a step closer to real-world use.
\end{abstract}

\begin{IEEEkeywords}
Diagnostics, Exhaled Breath, Internet of Bio-Things, Medicine, Molecular Communication.
\end{IEEEkeywords}

\IEEEpeerreviewmaketitle

\section{Introduction}
The exhaled breath is a complex mixture of various components, including gases, such as carbon dioxide ($\mathrm{C}\mathrm{O}_2$), oxygen, nitrogen, and \acp{VOC}, or particles, such as dust or pollen, cf.\ \cref{fig:breath}, that can provide information about an individual's health state.
Analyzing the exhaled breath usually takes place outside the human body, making it suitable for non-invasive analysis approaches and macroscale applications in advanced healthcare and medical environments, for example, detecting various diseases such as lung cancer, asthma, and \ac{COPD}, or monitoring the physiological and pathological states~\cite{amann2007breath}. 
This non-invasive analysis of the exhaled breath can be seen as a form of chemical communication, where the chemical signals in the breath, including sneezing and coughing, are used to transmit information about an individual's health state.
In fact, this method of interpretation is closely related to the idea of \acf{MC}.
\Ac{MC} represents an alternative communication paradigm, where chemical signals are used to transmit information via various possible \textit{\acp{IM}}. 
Inspired by our living environment, researchers try to (partially) engineer naturally occurring \ac{MC} processes for achieving synthetic \ac{MC} links and networks~\cite{Lotter2023experimental_II}. 
These engineered links and networks have been proposed to develop new diagnostic tools and medical devices that utilize the chemical signals in the exhaled breath to detect and monitor various diseases~\cite{khalidCommunicationBreathAerosol2019}.

\definecolor{nitrogenColor}{HTML}{AE1DC4}
\definecolor{oxygenColor}{HTML}{5F199E}
\definecolor{co2Color}{HTML}{0A099E}
\definecolor{vocsColor}{HTML}{FAA43F}

\begin{figure}[b]
\centering
\vspace{-0.8em}
\begin{tikzpicture}[scale=0.8, transform shape]
    \begin{axis}[
        width=6cm, height=4cm,
        xmin=0, xmax=100,
        ymin=0, ymax=1,
        axis y line=none,
        clip=false,
        legend style={at={(1.1,0.1)}, anchor=south west,
        legend image code/.code={\draw[opacity=0] (0,0) -- (0.01,0);}},
        xlabel={Cumulative Percentage [\%]}
    ]
    \addplot [fill=nitrogenColor!70] coordinates {(0,0) (0,1) (79,1) (79,0)};
    \addplot [fill=oxygenColor!70] coordinates {(79,0) (79,1) (95,1) (95,0)};
    \addplot [fill=co2Color!70] coordinates {(95,0) (95,1) (99,1) (99,0)};
    \addplot [fill=vocsColor] coordinates {(99,0) (99,1) (100,1) (100,0)};
    
    \legend{\textcolor{nitrogenColor!70}{Nitrogen (79\%)}, \textcolor{oxygenColor!70}{Oxygen (16\%)}, \textcolor{co2Color!70}{CO\textsubscript{2} (4\%)}, \textcolor{vocsColor}{VOCs (1\%)}}
    \end{axis}
\end{tikzpicture}
\caption{Cumulative percentage composition of the exhaled human breath~\cite{Paleczek2021artificial}. 
}
\label{fig:breath}
\end{figure}
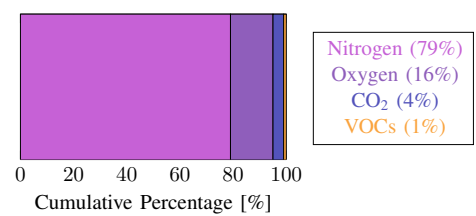

This survey provides a comprehensive analysis of existing research on human exhaled breath from the perspective of \ac{MC}, aiming to bridge the gap between related yet previously isolated research communities by establishing common terminology and promoting interdisciplinary collaboration.
An interdisciplinary overview of related research across medicine, biology, life science, chemistry, computer science, and engineering is also presented.
The detailed analysis follows the conventional, well-known information flow model for communication systems,
examining key components such as the transmitter, the channel, the propagation noise, and the receiver.

\subsection{Related Work}

This section briefly reviews existing related works and outlines the specific contribution of this survey.

A broad range of foundational surveys exists in the \ac{MC} literature. Four significant examples are highlighted, and these works provide a general view on different parts of an \ac{MC} system, such as modulation, coding, or channel modeling. However, this survey contributes as a specific \acf{ABMC} system by adopting a breath-centered perspective and expanding upon numerous crucial concepts, as outlined in the following.

In an early work, Farsad \emph{et al.}~\cite{farsad2016comprehensive} highlight the recent advancements in \ac{MC} research.
While their work generally focuses on \ac{MC}, this work discusses analyzing the human exhaled breath from an \ac{MC} point of view by connecting a natural transmitter, i.e., the human body, with an engineered receiver, for example, a sensor.

Kuran \emph{et al.}~\cite{Kuran2020survey} survey the design of modulation techniques in diffusion-based \ac{MC}. 
In their surveyed systems, \ac{IM} propagation is governed by diffusion, possibly assisted by an additional drift, mainly assuming simple end-to-end system models, i.e., a point transmitter and a spherical receiver for three-dimensional modeling. This work includes a detailed modeling of the human transmitter, the breath-based \ac{ABMC} channel, and engineered receivers, providing a specific perspective on information modulation for breath analysis.

Coding approaches for diffusion-based \ac{MC} are discussed in detail by Hofmann \emph{et al.} in~\cite{Hofmann2023coding}. 
In their work, they show that most of the coding approaches in \ac{MC} come from conventional wireless communications and are just being applied to \ac{MC} systems.
However, in the presented survey, coding approaches are also utilized, but these approaches are mapped to the naturally occurring exhalation process. 

A summary of mathematical models for diffusion-based \ac{MC} can be found in~\cite{jamaliChannelModelingDiffusive2019}. 
In similar lines, this survey also reviews basic mathematical models to provide the reader with fundamental knowledge of \ac{ABMC}. 
However, this work takes a further step by surveying advanced models for exhaled breath analysis, incorporating perspectives from disciplines such as biology and life sciences.

Several other works consider partial aspects of a breath-based \ac{ABMC} system. But, as described in Table~\ref{tab:surveys} and in the following, this survey is the first to combine the natural human transmitter with engineered receivers for breath-based disease detection.

The work in~\cite{aktas2024odor} surveys odor-based \ac{MC} in nature and examines olfactory systems, primarily based on biological systems. 
Biological systems that naturally function as \ac{MC} receivers detect and process exhaled breath for purposes like olfaction in humans and animals. 
However, these natural \ac{MC} receivers fall outside the scope of this study since natural \ac{MC} receivers are not designed to analyze the chemical or physical properties of breath for medical purposes. 
In contrast, engineered receivers are specifically designed to detect, analyze, and interpret exhaled breath, focusing on its chemical and physical characteristics. 
The magazine article in~\cite{khalidCommunicationBreathAerosol2019} initially explores the exhaled breath as a source message from the communication engineering perspective. 
The article provides a brief overview of the breath communication system, nevertheless, also by analyzing a natural receiver~\cite{khalidCommunicationBreathAerosol2019}. 
This survey provides more in-depth analysis with emphasis on engineered receivers for non-invasive disease detection via exhaled breath.

Several works identify \ac{ABMC} as a framework for pathogen or virus transmission, specifically~\cite{gulec2020molecular, barros2021molecular, schurwanz2021duality}. 
Gulec \emph{et al.}~\cite{gulec2020molecular} present an end-to-end model starting from the emitted particles by coughing or sneezing, including a channel model for air-based transmission, and a human receiver, with the goal to calculate a probability of infection. 
The survey by Barros \emph{et al.}~\cite{barros2021molecular} discusses models for virus emission from the human body, relevant experimental datasets from other disciplines, as well as in-body mechanisms for viral spread. 
Lastly, Schurwanz \emph{et al.} \cite{schurwanz2021duality} specifically target COVID-19 virus transmission as a multi-user \ac{ABMC} scenario, presenting simple channel and transmitter models and conducting coughing and sneezing experiments.
The previous three works all have in common that they are strictly limited to virus transmission and, therefore, do not consider engineered receivers for breath analysis, as is done in this survey. 
Additionally, they disregard the vast majority of other \acp{IM} related to various diseases that are emitted with the exhaled breath, a topic this survey comprehensively covers.

In terms of experimental work, Lotter \emph{et al.}~\cite{Lotter2023experimental_II,Lotter2023experimental_I} distinguish between long- and short-range experimental research, surveying experimental testbed setups including but not limited to \ac{ABMC}~\cite{Lotter2023experimental_II} and also present synthetic \ac{MC} as a framework for modeling natural \ac{MC}~\cite{Lotter2023experimental_I}.  
This survey explicitly reviews existing \ac{ABMC} testbed setups from both the \ac{MC} and the non-\ac{MC} community, providing the reader with a detailed overview of experimental work, such as specific methods, testbed design, and component analysis. By including examples from communication engineering, medicine, and life sciences, this survey attempts to provide a starting point for interdisciplinary research.
In~\cite{bhattacharjee2022digital}, Bhattacharjee \emph{et al.} consider communication techniques specifically for \ac{ABMC}, presenting an experimental testbed and analytical channel models for \ac{ABMC}, and evaluate a large number of modulation and detection techniques. 
In contrast to our work, the analysis is not specifically breath-related and lacks specific modeling of the human transmitter as well as engineered receivers for \acp{IM} such as biomarkers.

A summary of these relevant related surveys and papers and the aspects they cover is shown in Table~\ref{tab:surveys}, underlining the novel comprehensive nature of this survey compared to previous works.



\begin{table*}[ht] 
     \centering 
     \caption{Comparison of this work with existing surveys and reviews in the \ac{MC} domain, chronologically listed. Symbols indicate evaluation of the categories: \cmark~(fully addressed), \pmark~(partially addressed), and \xmark~(not addressed).} 
     \resizebox{\linewidth}{!}{%
     \begin{tabularx}{\textwidth}{p{3cm}*{13}{X}}
         \hline
         \textbf{Category / Ref.} 
         & \cite{farsad2016comprehensive}
         & \cite{jamaliChannelModelingDiffusive2019}
         & \cite{khalidCommunicationBreathAerosol2019}
         & \cite{Kuran2020survey}
         & \cite{gulec2020molecular}
         & \cite{barros2021molecular}
         & \cite{schurwanz2021duality}
         & \cite{bhattacharjee2022digital}
         & \cite{Hofmann2023coding}
         & \cite{Lotter2023experimental_I}
         & \cite{Lotter2023experimental_II}
         & \cite{aktas2024odor}
         & \textbf{This Work} \\ 
         \hline
         \textbf{Year}                     & 2016       & 2019       & 2019       & 2021       & 2021       & 2021       & 2021       & 2022       & 2023       & 2023       & 2023       & 2024       & 2025 \\[3pt]
         \textbf{Breath Analysis}          & \xmark     & \xmark     & \cmark     &   \xmark   &   \cmark   &   \cmark   &   \cmark   &   \pmark   &   \xmark   &   \xmark   &    \xmark  &   \pmark   & \cmark \\[3pt]
         \textbf{ABMC Perspective}         &   \pmark   &   \pmark   &   \cmark   &   \xmark   &   \cmark   &   \cmark   &   \cmark   &   \cmark   &  \xmark    &   \xmark   &   \cmark   &   \cmark   & \cmark \\[3pt]
         \textbf{Transmitter Models}       &   \cmark   &  \cmark    &   \pmark   &   \pmark   &   \xmark   &   \cmark   &   \pmark   &   \pmark   &  \pmark    &   \cmark   &   \cmark   &   \pmark   & \cmark \\[3pt]
         \textbf{Receiver Engineering}     &  \pmark    &   \xmark   &   \pmark   &   \cmark   &   \pmark   &   \xmark   &   \xmark   &   \pmark   &   \pmark   &   \cmark   &   \cmark   &   \cmark   & \cmark \\[3pt]
         \textbf{Channel Modeling}         &   \cmark   &   \cmark   &   \pmark   &   \pmark   &   \cmark   &   \cmark   &   \pmark   &   \cmark   &   \pmark   &   \pmark   &   \pmark   &   \cmark   & \cmark \\[3pt]
         \textbf{Biomarkers Addressed}     &   \xmark   &   \xmark   &   \xmark   &   \xmark   &   \pmark   &   \pmark   &   \xmark   &   \xmark   &   \xmark   &   \xmark   &   \xmark   &   \xmark   & \cmark \\[3pt]
         \textbf{Experimental Testbeds}    &  \pmark    &    \xmark  &   \pmark   &   \xmark   &   \xmark   &   \cmark   &   \cmark   &   \cmark   &  \xmark    &   \cmark   &   \cmark   &   \pmark   & \cmark \\[3pt]
         \textbf{Interdisciplinary}        &   \pmark   &   \pmark   &   \cmark   &   \pmark   &   \cmark   &   \cmark   &   \pmark   &   \pmark   &   \pmark   &   \cmark   &   \cmark   &   \cmark   & \cmark \\[3pt]
         \textbf{Noise and Interference}   &    \cmark  &   \cmark   &   \pmark   &   \cmark   &   \xmark   &   \xmark   &   \pmark   &   \cmark   &   \cmark   &   \pmark   &   \pmark   &   \pmark   & \cmark \\[3pt]
         \hline 
     \end{tabularx}
     }
     \label{tab:surveys}
     \vspace{-0.5cm}
 \end{table*}


\subsection{Definitions}


First of all, a common language is necessary for interdisciplinary research. The following definitions aim to remove ambiguities and clarify some frequently used terms throughout this survey.

\textbf{\textit{i)} Natural versus Synthetic \ac{ABMC}:}
Generally, \ac{ABMC} systems refer to \ac{MC} systems using air as the physical channel medium. 
Natural \ac{ABMC} describes chemical communication utilizing chemical signals for exchanging information, occurring naturally in our living environment in various scales such as gas-mediated cell signaling in the respiratory system (micro- and nanoscale) or pheromone communication between plants (macroscale)~\cite{Lotter2023experimental_I}. 
Inspired by natural \ac{MC}, the field of synthetic \ac{ABMC} evolved by ``establishing synthetic communication links''~\cite{Lotter2023experimental_I} by utilizing chemical signals~\cite{nakano2013molecular}.

\ac{ABMC} systems and their components, transmitter, channel, and receiver, can be classified as either natural or engineered. 
Thus, natural \ac{ABMC} does not include engineered components, partially engineered \ac{ABMC} includes engineering approaches for a subset of the components, and fully engineered \ac{ABMC} solely consists of engineered components.
Synthetic \ac{ABMC} includes partially engineered and fully engineered \ac{ABMC}. 
An example of fully engineered \ac{ABMC} is the world's first synthetic \ac{MC} testbed, the so-called tabletop testbed~\cite{farsad2013tabletop}.

In this work, emphasis is laid on partially engineered \ac{ABMC} by considering exhaled breath analysis. The human body is considered as a natural transmitter. 
The propagation channel can be natural, i.e., free space, or engineered, such as breathalyzer devices. 
Engineered devices in the form of sensors are specifically considered at the receiver side for exhaled breath analysis.

\textbf{\textit{ii)} Information Molecules:} 
The carriers of information in \ac{MC} encompass a diverse range of types, including magnetic nanoparticles, neurotransmitters, odor molecules, \ac{DNA} molecules, and calcium ions~\cite{xiao2023what}.
They exhibit distinct properties, such as diffusivity, biocompatibility, and magnetic susceptibility~\cite{xiao2023what}.
However, without prior knowledge of the experimental setup or the system model, it is impossible to define whether an information carrier is generally a molecule, a particle, or an ion. 
If an exact definition is not given, the term \textit{\acp{IM}} is generally used in this work to avoid ambiguity. 
The \acp{IM} are specified more precisely if a concrete model or experimental setup is available. For example, when analyzing the exhaled breath, biomarkers serve as specific \acp{IM} that provide valuable insights into physiological and pathological conditions. 
In this work, the word ``biomarker'' will be used to generalize a specific \ac{IM} used to detect specific physiological and pathological states. 


\textbf{\textit{iii)} The \textit{\acl{IoBT}}:} Originating from the well-known \ac{IoT}, an interconnected network of machines and objects with computing capabilities, the \acf{IoBNT} emerged~\cite{akyildiz2015internet,dressler2015connecting}. 
The \ac{IoBNT} describes a network of interconnected bio-nano things by using either \ac{MC} or Terahertz signaling. 
However, medical applications in biological environments on micro- and nanoscale domains require tiny machines and devices, so-called bio-nano things, and novel bio-cyber interfaces utilizing, for example, protein-based fluorescence~\cite{hofmannBiologicallyInspiredProteinBased2025} or rely on acoustic or terahertz approaches~\cite{chouhanInterfacingMolecularCommunication2023b}. 
As this work refers to the analysis of exhaled breath, the \ac{MC} system is a network of biological entities, i.e., the human body (cf.~\cref{sec:transmitter}), and engineered receivers (cf.~\cref{sec:receiver}), where the distances are on the scale of centimeters to meters.
Therefore, we use the term \acf{IoBT} as a preliminary stage of the \ac{IoBNT}. 
The \ac{IoBT} relies on existing broad knowledge of traditional receiver structures, as well as macroscale particle propagation modeling, making the initial transition more feasible compared to the \ac{IoBNT} vision.
In previous definitions, the \ac{IoBT} describes a network of ``biological computing devices''~\cite{laddomada2015crosstalk}.
This survey adopts a slightly different perspective by defining the \ac{IoBT} as a network of biological transmitters and engineered receivers (could be biological), utilizing \ac{MC} for macroscale communication.


\subsection{Structure of the Survey}

\begin{figure}[b]
    \vspace{-0.8em}
    \centering
    \begin{tikzpicture}[scale=0.63, transform shape]

    \definecolor{firstColor}{HTML}{AE1DC4}
    \definecolor{secondColor}{HTML}{0A099E}
    
    \draw[left color=firstColor!30, right color=secondColor!30] (-3,-0.5) -- (-3,-1.5) -- (3,-1.5) -- (3,-0.5) -- cycle;
    \node at (0,-1) {\shortstack{\textbf{\cref{sec:mapping}} —\\ Air-based \ac{MC} Information Flow Mapping}};

    \begin{scope}[xshift=0cm, yshift=-3cm]
    \draw[fill=firstColor!30] (-3,-0.5) -- (-3,0.5) -- (3,0.5) -- (3,-0.5) -- cycle;
    \node at (0,0) {\shortstack{\textbf{\cref{sec:transmitter}} —\\ Human Body as a Transmitter}};
    \end{scope}

    \begin{scope}[xshift=0cm, yshift=-5cm]
    \draw[fill=firstColor!30] (-3,-0.5) -- (-3,0.5) -- (3,0.5) -- (3,-0.5) -- cycle;
    \node at (0,0) {\shortstack{\textbf{\cref{sec:channel}} —\\ Communication Channel}};
    \end{scope}

    \begin{scope}[xshift=0cm, yshift=-7cm]
    \draw[fill=firstColor!30] (-3,-0.5) -- (-3,0.5) -- (3,0.5) -- (3,-0.5) -- cycle;
    \node at (0,0) {\shortstack{\textbf{\cref{sec:receiver}} —\\ Air-based \ac{MC} Receiver Engineering}};
    \end{scope}


    \draw[dashed] (-3.5,-2.25) -- (3.5,-2.25) -- (3.5,-7.75) -- (-3.5,-7.75) -- cycle;

    \begin{scope}[xshift=7.5cm, yshift=0cm]
    \draw[fill=secondColor!30] (-3,-0.5) -- (-3,-1.5) -- (3,-1.5) -- (3,-0.5) -- cycle;
    \node at (0,-1) {\shortstack{\textbf{\cref{sec:testbed}} —\\ Experimental Testbed Setups}};
    \end{scope}

    \begin{scope}[xshift=7.5cm, yshift=-5cm]
    \draw[fill=secondColor!30] (-3,-0.5) -- (-3,0.5) -- (3,0.5) -- (3,-0.5) -- cycle;
    \node at (0,0) {\shortstack{\textbf{\cref{sec:challenges}} —\\ Open Challenges and Future Research}};
    \end{scope}

    \draw[-Stealth] (0,-1.5) -- (0,-2.25);
    \draw[-Stealth] (3,-1) -- (4.5,-1);
    \draw[-Stealth] (3.5,-5) -- (4.5,-5);
    \draw[-Stealth] (0,-3.5) -- (0,-4.5);
    \draw[-Stealth] (0,-5.5) -- (0,-6.5);
    \draw[-Stealth] (8,-1.5) -- (8,-4.5);
    
\end{tikzpicture}
    \caption{Structure of the survey and recommended reading paths. 
    }
    \label{fig:overview}
\end{figure}
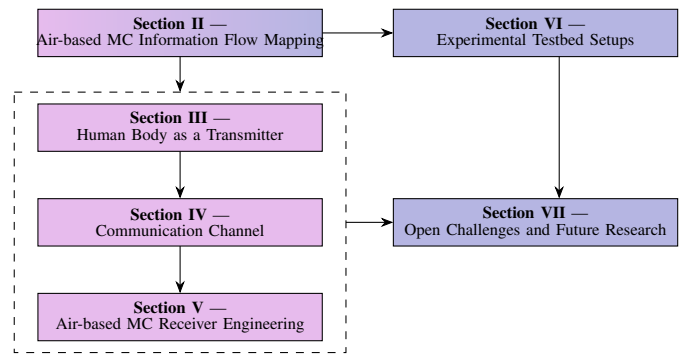 

The remainder of this survey is structured as follows.
\Cref{sec:mapping} generally maps the components of analyzing the human exhaled breath onto the \ac{MC} information flow components in a high-level manner.
These mapped components are further discussed and modeled in detail in \cref{sec:transmitter}, \cref{sec:channel}, and \cref{sec:receiver} for the human body as a transmitter, the physical environment as the propagation channel, and engineered sensors and detection technologies as receivers for disease detection, respectively. 
Furthermore, this survey provides an overview of existing testbeds in the \ac{ABMC} domain in \cref{sec:testbed}, distinguished between testbeds from the \ac{MC} field (\cref{MC:testbed}) and testbeds from other fields such as medicine or life science (\cref{nonMC:testbed}).
Finally, \cref{sec:challenges} discusses open challenges and limitations for analyzing the exhaled breath from a life science perspective in \cref{challanges:life}, and from an engineering perspective in \cref{challenges:engineering}.
\cref{sec:conclusion} concludes the survey by briefly summarizing the main impact and outlining future research.
A roadmap on how to read the survey can be found in~\cref{fig:overview}.
An overview of the frequently used abbreviations in this survey can be found in~\cref{tab:abbreviations}.


\begin{table}
    \centering
    \caption{Alphabetical list of abbreviations used throughout this survey. 
    }
    \printacronyms[template=tabular, display=used, heading=none]
    \label{tab:abbreviations}
\end{table}

\section{Air-Based Molecular Communication Information Flow Component Mapping}
\label{sec:mapping}

\begin{figure*}
    \centering
    \def\svgwidth{1\textwidth}
    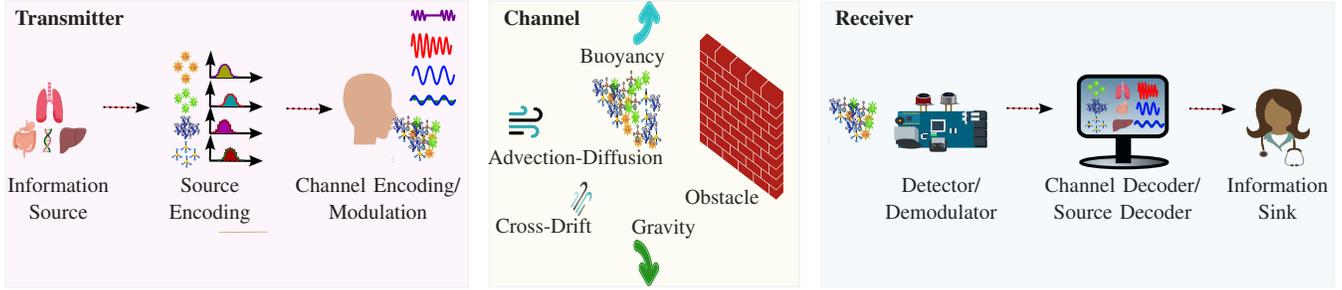
    \vspace{-1cm}
    \caption{Information flow model for analyzing the exhaled breath from a source of information via a channel to a sink of information, containing a natural transmitter, a natural channel, and an engineered receiver unit.}
    \vspace{-0.8em}
    \label{fig:respiration_new}
\end{figure*}

This section generally maps the components of the well-known information flow model, cf.~\cref{fig:respiration_new}, to the human exhaled breath analysis for medical means.
Individual components can thus be better understood, and corresponding research questions can be discussed. 


As we target a natural transmitter and engineered receiver in a natural \ac{ABMC} system, the human exhaled breath system, the transmitter in our information flow model is represented by the human body, more precisely, the human head with nose and mouth as so-called transmitter outlets, cf.~\cref{sec:transmitter}.
Thus, the transmitter is natural; the release mechanism of the \acp{IM} cannot be controlled. 
Overall, the transmitter is conceptually visualized in~\cref{fig:respiration_new}.

The \textbf{information source} represents the origin of a message located in space, intended for meaningful communication to a receiver.
From the perspective of analyzing the exhaled breath through the lens of \ac{MC}, the information source implies the underlying metabolic and health conditions, for example, diabetes, lung cancer, asthma, and viral infections.
Another valuable source of information is identifying the precise location of these conditions, which share overlapping properties and characteristics with other non-critical conditions. 
For example, acetone is a byproduct of ketogenesis during diabetic ketoacidosis, when insulin deficiency triggers fat metabolism. 
This originates from the inability of the pancreas to regulate glucose~\cite{owen1982acetone}.
At the same time, acetone is also produced during adherence to a ketogenic diet, as the liver processes fat for energy~\cite{prabhakar2015acetone}.
Moreover, the category or severity of a condition, such as the diabetes type, lung cancer stage, asthma category, or viral load, provides different information concerning the degree of progression and persistence of the condition.

Considering \textbf{source encoding} in exhaled breath analysis, the information quantification occurs with the metabolic processes converting a specific health condition and its location into a set of biomarkers in the exhaled breath.
The probability distributions of different biomarkers also indicate an individual's health condition, specific location inside the body, and severity.
In exhaled breath diagnostics, the body's metabolic activities serve as the primary information source, producing \acp{VOC} and other biomarkers that reflect specific biochemical or pathological states. 
Source encoding begins with quantifying these metabolic outputs, transforming them into discrete sets of biomarkers that encode health-related data. 
For example, the presence of acetone in breath may signify fat metabolism associated with diabetic ketoacidosis. 
At the same time, elevated nitric oxide levels can indicate inflammation in the respiratory tract, serving as a marker for asthma. 
This process effectively translates diverse health conditions into interpretable representations, capturing their spatial localization (for example, upper versus lower airway) and severity, thereby compressing complex biological data into a structured and meaningful format.

During \textbf{channel encoding}, redundant information exists in the form of recurrent occurrences of biomarkers over extended periods in quasi-reproducible concentration levels and amounts. 
The redundancy helps to mitigate the influence of noise in the transmission channel, thus minimizing the aberrations in the estimated information at the receiver side.
For example, while coughing, several coughs are exhaled, possibly representing an analogy to repetition coding.

The \textbf{modulation} process correlates to the exhalation of breath. 
The modulated physical properties of the exhaled biomarkers directly correspond to information about prevailing health conditions, their location inside the body, and their severity.
These properties of biomarkers can manifest in many forms, including concentration, type, temporal properties, the spatial regions occupied, or a hybrid combination of any of these~\cite{bhattacharjee2022digital}.
The modulation of the exhaled breath can be closely related to irregular patterns~\cite{kunczik2022breathing,kumar2022deep}. 
These can be viewed as signals that carry information about the state of health.
Analyzing their modulation characteristics can help diagnose or monitor health conditions linked to a specific breath pattern.

The \textbf{channel} for analyzing the exhaled breath is represented by the surrounding free space, containing air as the physical channel medium. 
Considering analyzing the exhaled breath, the channel is non-engineered, cf.~\cref{sec:channel}. 
The \acp{IM} propagate primarily naturally and are governed by diffusion, advection, buoyancy, gravity, and/or an initial drift velocity. 
Other effects (here called noise), such as cross-drift introduced by air conditioning systems or physical obstacles, such as walls, influence the propagation without an engineered purpose, i.e., these effects are not introduced to specifically support the propagation of the \acp{IM}. 

The only engineered component of our considered \ac{ABMC} system is the \textbf{receiver}, cf.~\cref{sec:receiver}. 
Please note that this work does not focus on natural \ac{MC} receivers.

For \textbf{demodulation} at the receiver, specialized sensors detect and demodulate information-carrying biomarkers and their physical properties in exhaled breath.
Biomarkers, which may include \acp{VOC}, nitrogen oxides, $\mathrm{C}\mathrm{O}_2$, and other gaseous metabolites, serve as critical indicators of metabolic and physiological processes in the body.
For example, metal-oxide gas sensors can detect and demodulate the received biomarkers based on their type, concentration, and temporal properties.
To ensure accurate biomarker identification and characterization, computing units on the receiver side are necessary. 
These units perform signal processing, feature extraction, and pattern recognition using machine learning algorithms~\cite{ye2021recent}. 

\textbf{Channel decoding} helps decode the information from the released biomarkers in the exhaled breath influenced by noise. 
Error detection and/or correction, correcting for noise and/or distortions, form part of the channel decoding technique.

The \textbf{source decoder} maps the detected and demodulated biomarkers back to their corresponding physiological states and locations within the human body. 
This process involves interpreting biomarker raw data as well as patterns and linking them to known health conditions, metabolic disorders, or inflammatory responses.

Finally, at the \textbf{information sink}, the information presents the individualist physiological and pathological
states, which can be further processed, for example, by healthcare providers for personalized medicine. 
Here, familiarization and training are essential, as initially, new workflows in electronic health record systems often reduce efficiency~\cite{baumann2018impact}.

\section{The Human Body as a Transmitter}
\label{sec:transmitter}

In this survey, humans are considered as the transmitter of the \ac{ABMC} system.
While \acp{IM} are emitted from various parts of the human body, this study focuses specifically on those released from the respiratory system. 
These emissions, referred to as exhaled breath, include not only normal respiration but also other mechanisms like sneezing, coughing, and talking. 

Unlike conventional \ac{MC} transmitters that intentionally encode information into a physical signal, the human transmitter shows a dual nature of signal emission, i.e., emitting signals either \textit{subconsciously and uncontrollably} (i.e., passively) or \textit{consciously and controllably} (i.e., actively). 
Examples of subconscious and uncontrollable emissions include normal breathing, which is essential for sustaining life, and involuntary actions, such as sneezing or coughing, which are triggered by reflexes or strong irritants. 
In contrast, conscious and controllable emissions can include voluntary forced breathing during medical examinations, deliberate coughing to clear the throat, or intentional talking to communicate specific information. 
However, the types of \acp{IM} emitted largely remain uncontrollable.
This section, therefore, presents descriptions and modeling approaches for them. 

In general, exhaled human breath includes both \textit{gaseous components}, such as \acp{VOC} and $\mathrm{C}\mathrm{O}_2$, as well as \textit{exhaled particles}, including aerosols and droplets, which can carry biological material. 
It is noted that the term ``particles'' is used as a general term encompassing both ``aerosols'' and ``droplets'' to avoid confusion, although they are typically differentiated by size~\cite{randall2021did}. 
The following sections identify key information sources in exhaled breath, explain their formation, and review models describing their behavior and properties.

\subsection{Information Source Identification} 
\label{sec_singlaing_molecules}


Exhaled breath analysis can reveal important health indicators. 
A detailed breakdown of the key parameters, classified in \textit{1)} biomarkers, \textit{2)} physical properties, \textit{3)} spatial origin, and \textit{4)} flow rate, is given below.

\subsubsection{Biomarkers}

Biomarkers can be distinguished in \acp{VOC}, Non-\ac{VOC} gas compositions, inflammatory markers, and microbial markers.

\textbf{\Acp{VOC}} are organic compounds in gaseous form at room temperature, emitted through various physiological processes and detectable in exhaled breath. 
They serve as useful health indicators. 
For example, specific hydrocarbons help identify asthma~\cite{dallinga2010volatile}, while oxidative stress markers and lung inflammation indicate \ac{COPD}~\cite{besa2015exhaled}. 
Alkanes and benzene derivatives are other families indicative of conditions like lung cancer~\cite{chang2018analysis}. 
Acetone, a ketone exhaled by diabetic patients, reflects fat metabolism instead of glucose breakdown~\cite{saasa2018sensing}. 
\Acp{VOC} also help diagnose liver disease~\cite{alkhouri2014analysis}, infections~\cite{zetola2017diagnosis,van2018potential}, and neurological conditions such as Alzheimer’s disease and Parkinson’s disease~\cite{bach2015measuring}.

\textbf{Non-VOC gases}, such as $\mathrm{C}\mathrm{O}_2$, offer key insights into respiratory function and gas exchange efficiency. 
Exhaled breath normally contains 4–5\% $\mathrm{C}\mathrm{O}_2$~\cite{carroll2022increased}, reflecting proper lung function. 
Elevated levels of $\mathrm{C}\mathrm{O}_2$ in blood may signal hypoventilation from conditions like \ac{COPD} or muscle weakness~\cite{simonds2013chronic}, while low levels may indicate hyperventilation due to anxiety or acidosis~\cite{brashear1983hyperventilation}. 
Additionally, abnormal $\mathrm{C}\mathrm{O}_2$ levels can also point to cardiovascular issues affecting pulmonary circulation and impairment of gas exchange~\cite{petersson2014gas}.

The presence of hydrogen sulfide ($\mathrm{H}_2\mathrm{S}$) in breath results from sulfate-reducing bacteria breaking down sulfur-containing amino acids~\cite{banik2016hydrogen}.
Elevated $\mathrm{H}_2\mathrm{S}$ levels may indicate gut infections or small intestinal bacterial overgrowth, and in the respiratory tract, it can signal sinus infections or inflammation~\cite{preti2009volatile}.

Exhaled carbon monoxide ($\mathrm{C}\mathrm{O}$) is a gaseous component of breath that reflects oxidative stress~\cite{popov2011human}.
The level of released $\mathrm{C}\mathrm{O}$ is determined through \emph{haem} metabolism catalyzed by \emph{oxygenase} enzymes, which takes place inside the human body.
Additionally, it also originates from external sources like air pollution or tobacco smoke.
A recent analysis indicates elevated exhaled $\mathrm{C}\mathrm{O}$ levels in asthmatic patients~\cite{zhang2010exhaled}. 

Elevated ammonia ($\mathrm{N}\mathrm{H}_3$) in exhaled breath can be indicative of health conditions related to metabolic and liver function~\cite{lefferts2022ammonia}.
In particular, liver dysfunction can result in increased $\mathrm{N}\mathrm{H}_3$ levels because the liver is primarily responsible for converting $\mathrm{N}\mathrm{H}_3$ into urea~\cite{hibbard2011breath}.
Conditions such as liver cirrhosis or acute liver failure can disrupt this process, leading to elevated blood $\mathrm{N}\mathrm{H}_3$ levels detectable in exhaled breath.
Similarly, renal failure can lead to high $\mathrm{N}\mathrm{H}_3$ levels as the kidneys fail to excrete urea efficiently, resulting in elevated $\mathrm{N}\mathrm{H}_3$ concentrations in the breath~\cite{essiet2013diagnosis}.

\textbf{Inflammatory markers} in exhaled breath reflect the body’s response to inflammation caused by infections, autoimmune, or chronic diseases.
For example, patients with airway inflammation in the case of asthma and \ac{COPD} release elevated levels of nitric oxide ($\mathrm{N}\mathrm{O}$) and hydrogen peroxide~\cite{boots2012versatile,brindicci2005exhaled,murata2014hydrogen}.
Inflammatory markers in the form of cytokines also help detect a wide range of health conditions such as cystic fibrosis and lung cancer~\cite{morris1981tidal}.
    
\textbf{Microbial biomarkers} include pathogen-specific compounds, microbial \ac{DNA}, or volatile metabolites produced by respiratory pathogens.
They support the diagnosis of infections like bacterial pneumonia, viral illnesses, and fungal diseases.
For example, microbial \acp{VOC} in breath aid in the early detection of bacterial infections~\cite{ahmed2023microbial}, while viral \ac{DNA} or \ac{RNA}--such as \textit{SARS-CoV-2} in exhaled aerosols--serves as a biomarker for diseases like COVID-19~\cite{sawano2021rt}.

\subsubsection{Physical Properties}

\textbf{Breath temperature and humidity} serve as valuable non-invasive indicators of health conditions, reflecting physiological and pathological processes in the respiratory and cardiovascular systems.
Elevated breath temperature can indicate airway inflammation, infections, or obstructive lung diseases like asthma and \ac{COPD}~\cite{popov2017added}.
In contrast, changes in breath humidity can signal respiratory tract hydration status, mucus production, and fluid balance issues seen in conditions such as cystic fibrosis, chronic bronchitis, and heart failure~\cite{turcios2020cystic}.
These parameters can aid in early disease detection, continuous monitoring, and management of chronic conditions.

    
\subsubsection{Spatial Origins}

\begin{figure}
    \centering
\includegraphics[width=0.4\textwidth]{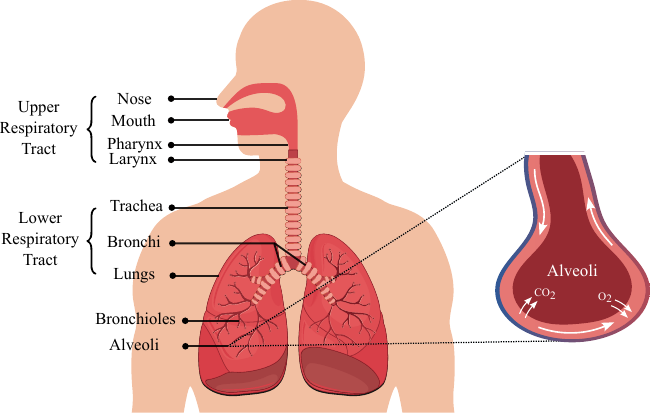}
    \caption{The respiratory system and gas exchange in the lungs.}
    \label{fig_rs}
    \vspace{-0.8em}
\end{figure}

As depicted in \cref{fig_rs}, the respiratory tract is divided into the \ac{URT} and the \ac{LRT}.
The \ac{URT} comprises the mouth (oral cavity), nose (nasal cavity), pharynx, and larynx, while the \ac{LRT} includes the trachea, right and left primary bronchi, and lungs~\cite{ahookhoshDevelopmentHumanRespiratory2020}. 

\textbf{Nasal exhaled breath}, originating from the nasal cavity and \ac{URT}, contains biomarkers influenced by the mucosal lining of the nasal passages, sinuses, and nasopharynx~\cite{barnes2019diseases}.
This breath primarily reflects \ac{URT} conditions such as allergic rhinitis, sinusitis, nasal polyps, and infections.
Elevated $\mathrm{N}\mathrm{O}$ levels in nasal breath are indicative of upper airway inflammation, aiding in the diagnosis of allergic rhinitis and sinusitis~\cite{frieri1998nitric}.
Specific VOCs and hydrogen sulfide ($\mathrm{H}_2\mathrm{S}$) can suggest bacterial infections of the sinuses~\cite{preti2009volatile}.

\textbf{Oral exhaled breath}, originating from the oral cavity, \ac{LRT}, and systemic circulation, contains biomarkers influenced by both lungs and systemic metabolic processes~\cite{wang2008analysis}.
This breath primarily reflects conditions affecting the \ac{LRT} and systemic metabolism, including lung cancer, metabolic conditions such as diabetes and liver dysfunction, and lower respiratory infections.
Elevated $\mathrm{N}\mathrm{O}$ levels in oral breath indicate lower airway inflammation, aiding in the diagnosis of asthma and \ac{COPD}~\cite{robbins1996measurement}.
Specific VOCs, such as alkanes and benzene derivatives, can indicate lung cancer~\cite{chang2018analysis}, while elevated acetone levels can signal ketosis, commonly seen in diabetes or prolonged fasting~\cite{wang2008analysis}.
VOC patterns and hydrogen cyanide help diagnose lung infections and differentiate between bacterial and viral etiologies, and elevated $\mathrm{N}\mathrm{H}_3$ levels can indicate liver dysfunction or renal failure~\cite{lefferts2022ammonia}.

\subsubsection{Flow Rate}

Finally, the flow rate may reveal significant insights into a patient's clinical status.

\textbf{Respiratory frequency} monitoring, i.e., the frequency and rhythm of breathing, can reveal many health abnormalities.
Sleep-related disorders such as central sleep apnea are examples in which sleep is disrupted by intermittent or cyclical absence of breath, affecting blood oxygen saturation levels~\cite{eckert2007central}.
Similarly, rapid and deep breathing in the case of hyperventilation is also indicative of chronic anxiety and panic attacks~\cite{brashear1983hyperventilation}.
Other health conditions that affect the respiratory rate include respiratory distress, neurological disorders, or metabolic abnormalities~\cite{nogues2008abnormalities}.

\textbf{Peak expiratory flow rate} measures the maximum speed at which an individual can exhale air from the lungs after a maximal inhalation.
This typically offers information about airflow obstruction, long-term lung complications, and asthma~\cite{morris1981tidal}.
The values vary based on age, gender, height, and smoking habits, with lower values indicating more significant airway narrowing or obstruction~\cite{boezen1994distribution}.

\textbf{Breath-holding time} is the duration a person can voluntarily hold their breath after full inhalation.
This indicates overall respiratory health and fitness, efficiency of gas exchange~\cite{hong1971alveolar}, lung capacity~\cite{whitelaw1987effect}, and the body's tolerance to elevated levels of $\mathrm{C}\mathrm{O}_2$~\cite{fowler1954breaking}.
Longer breath-holding time suggest better lung function, cardiovascular fitness, and efficient metabolic processes.
In contrast, shorter breath-holding time may indicate compromised respiratory efficiency, reduced lung capacity, or underlying health cardiovascular and/or pulmonary health issues. 

\subsection{Signaling Molecule Carrier Generation and Emission}
\label{sec:carrier}

The respiratory system plays a crucial role in the generation of exhaled particles. 

\begin{figure*}
    \centering
    \subfloat[Turbulent aerosolization in the larynx and large bronchi.\label{fig_mechanism1}]
    {\includegraphics[width=0.27\textwidth]{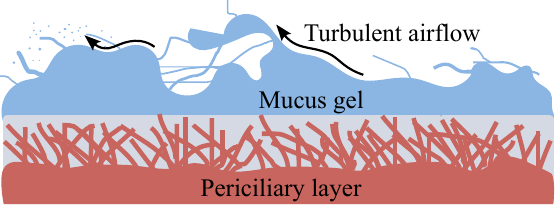}} ~ ~
    \subfloat[Rupture of fluid linings in small bronchi.\label{fig_mechanism2}]
    {\includegraphics[width=0.7\textwidth]{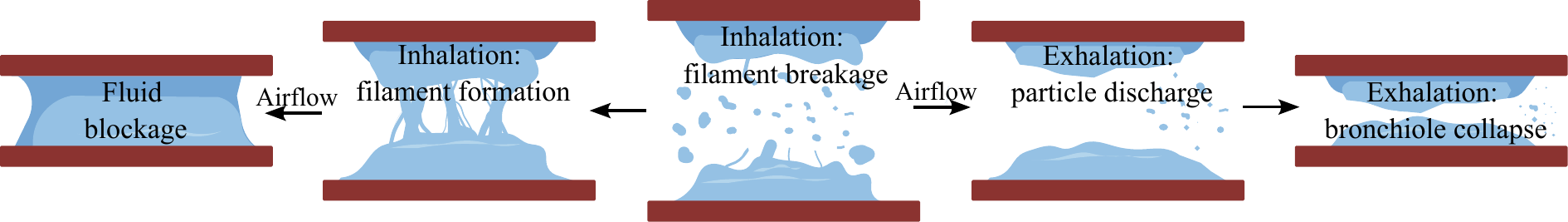}}\\ 
    \caption{Production mechanisms of exhaled particles in the respiratory tract; recreated from~\cite{morawska2022physics}.}
    \label{fig_particel_generation}
    \vspace{-0.8em}
\end{figure*}

\subsubsection{Production of Gaseous Components}
Gaseous biomarkers, produced by the body through metabolic processes or absorbed from the environment and dissolved in the bloodstream, can reach the alveoli, which are tiny, thin-walled sacs surrounded by capillaries.
These biomarkers diffuse from the capillaries into the alveolar air (see \cref{fig_rs}), driven by a concentration gradient, and are subsequently released into the airways and exhaled to the external environment~\cite{bruderer2019line}. The efficiency of this process depends on the concentration gradient, solubility, and chemical properties of the gaseous molecules.

\subsubsection{Production of Exhaled Particles}
From the nasal mucosa to the alveoli, the epithelial cells of the respiratory tract are covered by a bilayer airway surface liquid film.
This film consists of a top mucus gel layer (97$\%$ water and 3$\%$ mucins, non-mucin proteins, salts, and cellular debris) and a low-viscosity periciliary layer containing cilia beneath it~\cite{cross1994oxidants}.
It is widely recognized that respiratory particles are primarily generated from this liquid film~\cite{morawska2022physics}, and biological materials (e.g., viruses and DNA) are transported from the respiratory tract to the outside of the body by either encapsulating within exhaled particles or adsorbing onto the surface of particles.
Two primary physical mechanisms are responsible for the generation of respiratory particles.
The first is turbulent aerosolization (or shear-induced surface wave instability)~\cite{mittal2020flow}, see \cref{fig_mechanism1}.
This occurs when air flows over the liquid film at sufficient velocity, pulling portions of the film into fine ligaments that fragment into particles within the airstream~\cite{castleman1931mechanism}.
As shown in \cref{fig_mechanism2}, the second mechanism involves the rupture of the fluid lining during the reopening of closed respiratory passages.
A notable site for this process is in the terminal bronchioles, where fluid closures occur during the airway collapse following exhalation.
When reopened during inhalation, the fluid closures are ruptured, thus leading to the production of particles~\cite{bake2019exhaled}. 
A similar process likely occurs in the larynx during activities such as talking and coughing, which involve the opening and closing of the vocal folds~\cite{mittal2013fluid}. 
Additionally, this mechanism generates salivary particles through the movement and contact of the tongue and lips, especially during actions like sneezing.

\subsection{Transmitter Modeling} 
\label{sec_transmitter_mnodeling}


In the following, the transmitter models are reviewed when the human body is considered as the source of the particles presented in~\cref{sec_singlaing_molecules}.
The analysis begins with a single-transmitter scenario, covering the Wells-Riley model along with models for sneezing, coughing, and breathing.
Meanwhile, the limitations of these models are discussed and potential expansions to develop a more comprehensive model are explored, incorporating factors such as particle generation, exhaled flow dynamics, individual physiological differences, and the impact of face masks.
Finally, this section is concluded with a brief discussion on extending single-transmitter models to scenarios involving multiple transmitters. 

\subsubsection{Emission Rate Models}
The simplest transmitter model for an \ac{ABMC} is the volume transmitter assumption used in the Wells-Riley model~\cite{riley1978airborne,wells1955airborne}, which assumes that exhaled particles (e.g., virus-carrying aerosols) are instantaneously and uniformly distributed throughout an environment.
Although this is a major simplification of a human transmitter as it neglects its location and environmental geometry, it has been widely employed to predict COVID-19 transmission due to its ease of implementation and computational efficiency.

Compared to the above volume transmitter, the point transmitter assumption is a more practical model to describe the released particles from the mouth and nose of a human.
Specifically, both a sneeze and a cough are commonly modeled as an impulsive source, and the particle emission rate for a rate model~\cite{khalid2020modeling,amin2021viral,lau2022predicting} can be respectively expressed as 
\begin{equation}
    \begin{split}
        S_{\rm s}=R_{\rm s}\delta(\textbf{d}-{\textbf{d}_\text{TX}})\delta(t), \label{eq_source_sneeze}
    \end{split}
\end{equation}
\begin{equation}
    \begin{split}
        S_{\rm c}=R_{\rm c}\delta(\textbf{d}-{\textbf{d}_\text{TX}})\delta(t), \label{fl}
    \end{split}
\end{equation}
where $R_{\rm s}$ and $R_{\rm c}$ are the rates of particles emitted by sneezing and coughing, $\delta(\cdot)$ is the Kronecker delta function, $\textbf{d}$ is a vector defining a position in space, and $\textbf{d}_\text{TX}$ is the location of the release point.
Unlike sneeze and cough sources, the particle emission from a single breath is continuous.
Based on the fact that each breath takes no more than \SI{4.98}{\second} while it takes several minutes for particles to reach a receiver located a few meters away, a breath is usually approximated as a constant source with the variation of emission rate ignored~\cite{khalid2020modeling}.
Thus, a breath starting at time $t=0$ can be mathematically modeled as
\begin{align}
    S_{\rm b}=R_{\rm b}\delta(\textbf{d}-{\textbf{d}_\text{TX}})[u(t)-u(t-t_{\rm b})], \label{eq_source_breath}
\end{align}
where $R_{\rm b}$ is the averaged emission rate, $u(t)$ is the Heaviside step function, and $t_{\rm b}$ is the averaged duration of a breath.
By assuming that a sneeze, a cough, and a breath are independent of each other, the final emitted signal, including all these modes, is
\begin{align}
    S=S_{\rm s}+S_{\rm c}+S_{\rm b}. \label{eq_source}
\end{align}

Although the modeling in \eqref{eq_source_sneeze}-\eqref{eq_source_breath} can simplify propagation and receiver analyses, this remains a highly idealized assumption.
On the one hand, research has shown that a cough or sneeze cannot be accurately represented by a sudden release of particles.
Specifically, a real cough displays a complex temporal velocity profile approximated as a combination of gamma-probability-distribution functions, see~\cite[Fig. 1]{gupta2009flow}, to transport particles from the respiratory system to the ambient environment through the nasal and buccal passages.
Moreover, measurements of cough flow velocity profiles in~\cite{vansciver2011particle} indicate that the average cough width ranges from 35 mm to 45 mm, with peak velocities ranging from 1.5 m/s to 28.8 m/s (average of 10.2 m/s) and an average Reynolds number of 2.04$\times$10$^\text{4}$.
On the other hand, the breath model in \eqref{eq_source_breath} is also simplified and fails to capture many fluid dynamics characteristics of breath airflow.
First, the flow rate profile of breathing can be more accurately represented by a sinusoidal waveform rather than a rectangular waveform, i.e., $R_{\rm b}[u(t)-u(t-t_{\rm b})]$. This sinusoidal flow can be approximated as~\cite[Eq.~(1)]{guptaCharacterizingExhaledAirflow2010}
\begin{equation}
    Q_{\rm b}=\alpha \sin (\beta t). \label{eq_breath_flow_rate}
\end{equation}
Here, $\alpha$ and $\beta$ depend on the respiration frequency, defined as the number of breaths per minute, and the minute volume, which refers to the total volume of air inhaled or exhaled in one minute.
Both are influenced by individual physiological parameters.
For example, experimental studies show that increased height is correlated with a lower RF, larger body surface area is associated with a higher MV, and both RF and MV vary between male and female subjects, indicating that \eqref{eq_breath_flow_rate} is affected by body height, weight, and gender~\cite{guptaCharacterizingExhaledAirflow2010}.
Second, the flow direction of the exhaled jet during breathing is characterized by two front angles and two side angles for nose breathing, and one side angle for mouth breathing, showing minimal variation across individuals~\cite{guptaCharacterizingExhaledAirflow2010}.
The study by Xu \emph{et al.}~\cite{xuHumanExhalationCharacterization2017} also reveals interactions of the exhaled air jet with body-generated thermal plumes that influence the dispersion and mixing of particles and result in differences between exhalation standing or lying down.
These source models provide valuable insights into the transmitted signals and can serve as boundary conditions for studying subsequent propagation dynamics.

\subsubsection{Particle Concentration Models}
While \eqref{eq_source_sneeze}-\eqref{eq_source_breath} are expressed in terms of particle emission rate, they can also describe the initial quantity or concentration of released particles once these values are obtained~\cite{chen2022detection}. 
Studies measuring the concentrations of various breath metabolites, such as ammonia, acetone, methanol, ethanol, and isoprene, in healthy individuals show that their normal levels generally follow a log-normal distribution~\cite{turner2006longitudinal1,turner2006longitudinal2,turner2006longitudinal3,turner2005longitudinal}.
These baseline values can be interpreted as representing a binary bit-0 while elevated levels may correspond to a binary bit-1, potentially indicating adverse clinical conditions.
Different distributions have been observed for other types of particles.
For instance, the concentration of the COVID-19 virus in exhaled particles per breath follows a Poisson distribution, with mean values ranging from 4.9$\times$10$^{-9}$ virus copies per cm$^3$ for low emitters to 0.637 virus copies per cm$^3$ for high emitters~\cite{lotter2021statistical}. 

The size of exhaled particles is a critical parameter for estimating the biomarker load in particles, yet it has also been overlooked in current \ac{MC} modeling efforts.
Exhaled particles span a wide range of sizes, influenced by their generation mechanisms and sites of origin (see~\cite[Fig. 2]{morawska2022physics}).
The largest particles (diameter $\geq$ 100$\mu$m) are typically produced during vocalization near the front of the oral cavity, where airflow interacts with varying gaps between the lips, tongue, and teeth.
Rapid airflow, during coughing, sneezing, or sudden exhalation, generates a broad size distribution of particles via turbulent aerosolization.
For example, cough particles can range in size from less than 1$\mu$m to over 100$\mu$m~\cite{abkarian2020stretching,papineni1997size}.
Smaller particles can be generated by the vocal folds during vocalization, with sizes typically in the 1–5$\mu$m range.
Additionally, the transient closure and reopening of distal airways are thought to produce small breath particles.
These droplets, generated by the rupture of the fluid lining in the terminal bronchioles, range in size from 0.01 to 2$\mu$m in their hydrated state.
The size distribution of these particles is often well-represented by a single lognormal distribution centered around 0.7$\mu$m~\cite{papineni1997size}.

\subsubsection{Exhaled Air Trajectory Models}
When a person is considered as a point transmitter, the emitted signal can also be represented as a cloud comprising particles mixed with warm air from the mouth and nose.
This transmitter model is particularly useful for determining the cloud's trajectory by taking the effects of gravity, buoyancy, and air-droplet interactions into account~\cite{bourouibaViolentExpiratoryEvents2014,gulec2020molecular}.
To track the cloud's location in space, it is crucial to derive its time-dependent density; thus, the emitted signal is often expressed in terms of density rather than emission rate or concentration (see~\cite[eq.~(2)]{gulec2020molecular}). 
The breath cloud model is considered in detail in \cref{sec:channel}.

In addition to becoming subject to external forces such as gravity or airflow in the surroundings, as will be further described in \cref{sec:channel}, several factors are acting on the particles' trajectory after they exit the human body that are still dependent on the exhalation process. 
Furthermore, human exhalation characteristics are analyzed using Schlieren imaging~\cite{xuHumanExhalationCharacterization2017}.
The exhaled air resembles a constant jet with additional turbulent vortex rings.
The jet's width and length increase with time and distance from the body.

The centerline peak velocity $u_m$ within the jet decays with distance $x$ from the emission point, according to an exponential model $u_m \propto \left(\frac{\sqrt{A}}{x}\right)^{n_u}$, with the area of the mouth or nose $A$, and the characteristic velocity exponent $n_u$ obtained via fit to experimental data~\cite{olmedoDistributionExhaledContaminants2012, xuMeasuringExhaledBreath2015}.
The resulting initial velocity among the subjects is on the order of \SIrange{1}{2}{\meter\per\second}.
The airflow velocity $u_m$ is generally larger than the propagation velocity $u_p$ of the jet itself since the latter is subject to vortex-like patterns.
The propagation velocity correlates with physiological metrics like \ac{BSA} and RF~\cite{xuHumanExhalationCharacterization2017}. Gupta \emph{et al.}~\cite{guptaCharacterizingExhaledAirflow2010} have proposed the linear model $u_p \propto \mathrm{BSA} \cdot \mathrm{RF}$.

\subsubsection{Respiratory Tract Models}
The transmitter modeling can be further extended to include particle generation and propagation within the respiratory system. In the context of viral transmission, the concentration of particles released $C_p$ from human sneezing, coughing, and breathing is described as an advection process of virus concentration $C_v$ in the lungs~\cite{barros2021molecular}.
This model is governed by the following partial differential equations
\begin{align}
    \frac{{\rm{d}} C_p}{{\rm{d}}t}={\nabla}D_p k_v C_v
    ({\nabla}C_p),
\end{align}
and
\begin{align}
    \frac{{\rm d} C_v}{{\rm d}t}={\nabla}C_{v_0} D_p ({\nabla}C_v) + (R_v(t)*P_v(t)),
\end{align}
where $D_p$ is the diffusion coefficient of particles, $k_v$ is the rate of virus entering nasal or buccal cavity, $C_{v_0}$ is the initial concentration of virus in the nasal/oral cavity, $R_v(t)$ is the rate of virus replication in the lungs, $P_v(t)$ is the propagation of virus from the lungs to the nasal/oral cavity, and $*$ is the convolution operator.
It is noted that $k_v$ can be varied with the different respiratory events (e.g., sneezing, coughing, and breathing) and reflects the conversion speed of the virus to exhaled particles and the speed of particle release. In addition, $k_v$ also varies over time and is linked with different disease stages.
For $R_v(t)*P_v(t)$, although it currently does not exist in the literature, there is an effort from the \ac{MC} community that modeled the virus propagation within the respiratory system in a reverse direction, i.e., from the nasal cavity to alveoli~\cite{koca2021molecular}.

It is possible to describe the respiratory tract and its two segments, the \ac{LRT} and \ac{URT} in more detail by looking at it from a geometric modeling perspective.
Each segment exhibits distinct anatomical properties, flow dynamics, and modeling challenges, contributing to the complexity of simulating particle emission processes.

\begin{figure}
    \centering
    \includegraphics[width=0.5\linewidth]{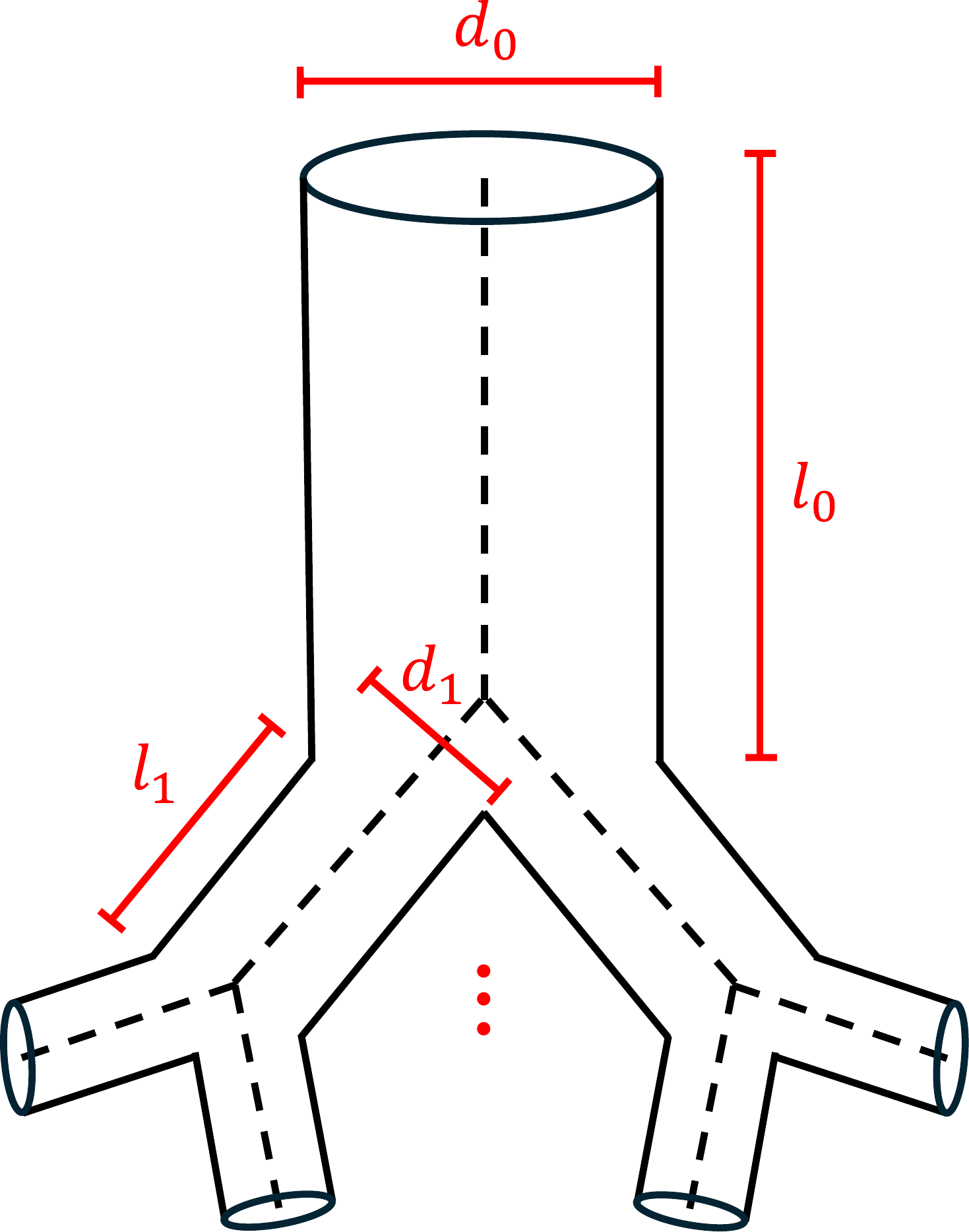}
    \caption{\textit{Weibel A} symmetric bifurcation model for the \ac{LRT}~\cite{weibelMorphometryHumanLung1963}.}
    \label{fig:weibel}
    \vspace{-0.8em}
\end{figure}

The \ac{LRT} forms a highly branched system designed for efficient air distribution.
Its bifurcation dynamics create complex secondary flows and inertial effects, which are further influenced by patient-specific factors such as asymmetrical lung volumes.
Investigations such as those in~\cite {debackerFlowAnalysesLower2008} and the review in~\cite{ahookhoshDevelopmentHumanRespiratory2020} reveal critical insights.
In terms of branching and flow dynamics, the \ac{LRT} demonstrates intricate bifurcation behavior, with airflow distribution heavily dependent on individual anatomical variations.
\Ac{CFD} simulations have highlighted significant disparities in flow rates between the left and right lungs, particularly in individuals with pathological conditions.
Models that consider pressure-driven flow conditions more accurately represent real-world scenarios compared to uniform flow assumptions~\cite{debackerFlowAnalysesLower2008}.
There are several modeling approaches.
A widely used framework for studying the lower airways is the symmetric branching model proposed by Weibel (Model A)~\cite{weibelMorphometryHumanLung1963}, where airway diameters and lengths decrease geometrically with each generation
\begin{equation}
    d_n = d_0 \cdot 2^{-\frac{n}{3}}, \quad l_n = l_0 \cdot 2^{-\frac{n}{3}}
\end{equation}
where $d_n$ and $l_n$ are the diameter and length of the $n$-th generation of bifurcation, with $d_0$ and $l_0$ as the initial diameter and length.
The principle of the model is also depicted in \cref{fig:weibel}.
However, this model assumes regular, symmetric branching patterns, which rarely occur in nature.
The less common Weibel B model incorporates asymmetries to better reflect real anatomical structures~\cite{weibelMorphometryHumanLung1963}.
Advancements in medical imaging techniques, such as computed tomography, now allow for creating highly detailed, patient-specific airway geometries~\cite{ahookhoshDevelopmentHumanRespiratory2020}.
These models enable simulations that incorporate individualized boundary conditions, leveraging turbulence models such as the kappa-epsilon ($k$-$\varepsilon$) model, which considers differential equations for the turbulent kinetic energy $\kappa$ and the dissipation $\varepsilon$ for accurate flow dynamics~\cite{launderNumericalComputationTurbulent1974, xuInvestigationInhalationExhalation2020}.

The \ac{URT}, encompassing several bifurcations and asymmetries, plays a vital role in humidifying and filtering inhaled air.
It also serves as an initial region where turbulence significantly influences the behavior of exhaled particles.
The \ac{URT} exhibits distinct flow characteristics during inhalation and exhalation.
Inhalation is associated with intense turbulence caused by the narrowing of the glottis, whereas exhalation tends to exhibit more uniform flow patterns~\cite{xuInvestigationInhalationExhalation2020}.
Despite this distinction, turbulence remains a key factor in accurately modeling particle transmission, as demonstrated in computational studies~\cite{xuInvestigationInhalationExhalation2020}.
In terms of modeling approaches, advances in medical imaging, such as \ac{CT} and \ac{MRI}, now enable the development of realistic three-dimensional reconstructions of \ac{URT} geometries~\cite{ahookhoshDevelopmentHumanRespiratory2020}.
These models provide a powerful tool for predicting deposition patterns in aerosol therapy and exploring molecular signaling pathways.
Realistic morphometric and image-based methods have made it possible to perform experimental and computational studies with high anatomical fidelity~\cite{ahookhoshDevelopmentHumanRespiratory2020}.

The effects of human physiology factors in terms of \ac{URT} geometry and saliva property on the particle emission during a sneeze are introduced and investigated numerically in~\cite{fontes2020study}, since these factors are associated with the illness, anatomy, stress condition, and sex of an individual.
This study proposes a \ac{URT} model that includes simplified representations of the pharynx, nasal cavity, and buccal cavity, connected to form an integrated flow network.
To assess the impact of the geometric features within the nasal or buccal passages of \ac{URT}, four conditions (i.e., open or blocked nasal passage with or without teeth) are considered.
Results indicate that the presence of teeth can increase exit velocities from the lower mouth regions, potentially leading to a higher release of particles.
Additionally, variations in saliva properties—such as fluid density, viscosity, and surface tension—are found to influence both the number and size of emitted particles.

\subsubsection{Further Model Refinements}
    \textbf{Face Masks:} After particles are released from the mouth or nose, some can be blocked and deposited by a face mask, preventing them from contributing to the final emitted signal.
    When a mask fits perfectly to the face, the emitted signal depends primarily on the mask's filtration efficiency, with surgical masks and N95 respirators providing filtration rates of 95$\%$ and 97$\%$, respectively~\cite{li2006vivo}.
    However, a mask is not always perfectly fitted, and particles that escape through gaps between the mask's edge and the skin also contribute to the emitted signal.
    To model this scenario, the area around the mouth or nose can be represented as a planar surface, with the gaps modeled as a cuboid extending from the mask surface to the environment (see~\cite[Fig. 1]{peric2020analytical,lotter2021statistical}).
    The probabilities of particles traveling through either the gap or the mask are calculated as functions of various mask parameters and the exhaled airflow rate.
    Consequently, the emission of a single particle into the ambient environment (i.e., the propagation channel) is modeled as a Bernoulli random variable.
    
    \textbf{Mobile Transmitter:} An underlying assumption in all the transmitter models discussed so far is that the human transmitter remains static.
    However, this assumption may not hold in real-life scenarios where individuals can move within an environment.
    Consequently, incorporating human mobility into the model is necessary, making the emitted signal a function of both location and time.
    While human mobility in \ac{ABMC} systems has yet to be investigated, introducing this factor would significantly increase the complexity of transmitter modeling.
    However, the mobility issue has so far received attention in liquid-based \ac{MC} systems, such as the movement of nanobots within blood vessels~\cite{mosayebi2018early}. We believe that the methodologies developed for liquid-based \ac{MC} systems could offer valuable insights for modeling the movement of human transmitters in air-based scenarios.
    For more details on the analysis of mobile liquid-based \ac{MC} systems, we refer the readers to~\cite{nakano2016performance,nakano2019methods,cao2019diffusive,huang2021membrane,ahmadzadeh2017diffusive}. 
    
    \textbf{Multi-Transmitter Scenarios:} The transmitter models discussed above focused on a single transmitter.
    In scenarios where multiple individuals (i.e., multiple transmitters) are present at different locations, the final emitted signal to a receiver can often be approximated as the sum of signals from each individual, assuming that their emissions are independent.
    However, this assumption may not hold in certain scenarios, where interactions between emitters or environmental factors could influence the emitted signals.


%
%
%

\section{Communication Channel}
\label{sec:channel}

Building upon the generic channel modeling principles introduced in the following subsection, the concepts that apply to breath-related \ac{ABMC} are examined in this section.
Research from diverse fields, such as aerosol science, fluid mechanics, and indoor air quality studies, is incorporated.
The focus is on the effects of ventilation, obstacles, and human interactions on biomarker transport in ambient and controlled setups.
These factors impact the behavior of the channel, directly or indirectly, in the context of exhaled breath analysis for health diagnostics from an \ac{MC} perspective.
Therefore, this section provides an overview of the key physical channel characteristics, presents noise and interference sources and mechanisms, and considers more detailed aspects of breath cloud modeling and mathematical channel models.

\subsection{Propagation Mechanisms and Interference}
\label{subsec:propa}

For \ac{ABMC} systems, diffusion, advection, gravitational forces, initial drift velocity, and buoyancy are well-known propagation mechanisms~\cite{gulec2020molecular}, cf.~\cref{fig:respiration_new}.


Diffusion describes the random movement of the \acp{IM} due to thermal energy, also termed Brownian motion, and occurs in fluid-based media. 
Diffusion is stated as a potential method for transporting \acp{IM} in \ac{MC} systems due to its energy efficiency, as it relies solely on the environment's thermal energy~\cite{nakano2013molecular}. 
The rate of diffusion depends on the physical state of the propagation medium. 
In addition, a flow component is often present in the fluid-based propagation medium, accelerating the diffusion process in a directed way. 
This is especially helpful in transporting \acp{IM} over larger distances with a more channelized approach to reach a particular target. 
Together with diffusion, this process is termed the advection-diffusion process.
Based on the diffusion process, the advection-diffusion process can be modeled by introducing flow velocity components in each spatial dimension. 
An initial drift velocity also impacts the propagation of the \acp{IM}, changing their position before the effect of diffusion or advection comes into play.
Finally, a gravitational force and a buoyant force may affect the propagation of the \acp{IM}~\cite{gulec2020molecular}.
As these two forces act in opposite directions, the resulting force on the \acp{IM} is the difference between the two.

Besides propagation of \acp{IM}, interference significantly impacts the received signal in \ac{MC} channels. 
These channels exhibit memory due to delayed \acp{IM} from previous emissions, causing \ac{ISI}. 
\Ac{ISI} occurs when residual signals from past transmissions overlap with current ones, altering \ac{IM} concentration from a single transmitter. In multi-transmitter systems, \ac{CCI} also arises from signal overlap between different transmitters, further complicating \ac{MC} system performance.


\subsection{Physical Channel Characteristics}

\begin{figure}
    \centering
    \includegraphics[width=0.85\linewidth]{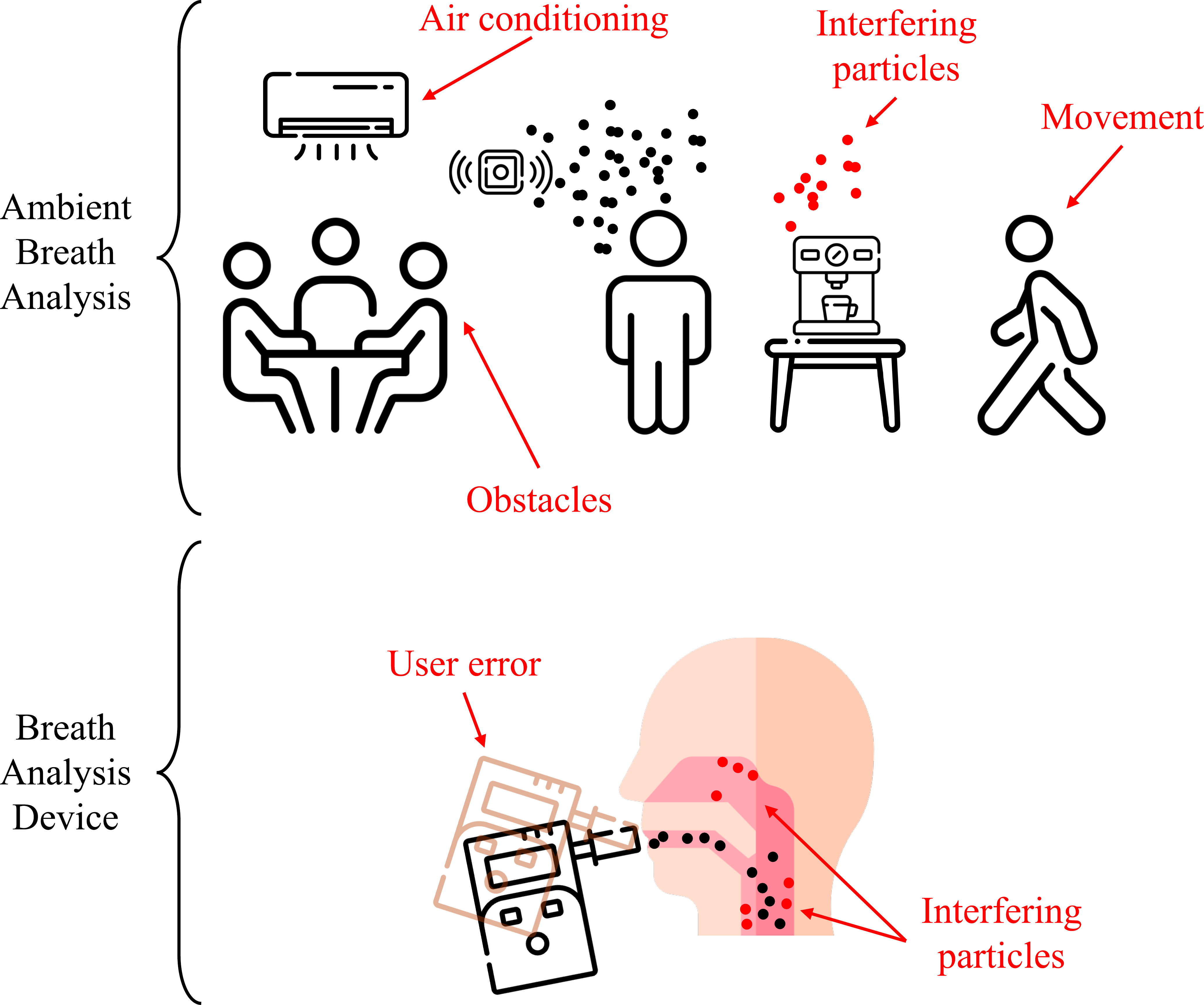}
    \caption{Visualization\protect\footnotemark{} of various channel characteristics and noise sources in different breath analysis scenarios.}
    \label{fig:channels}
    \vspace{-0.8em}
\end{figure}

\ac{ABMC} in human-centric settings is highly affected by the channel's physical characteristics.
These vary widely depending on the application, ranging from large, open indoor spaces for ambient breath analysis to narrow, controlled pathways in direct breath analysis devices. 
An overview is shown in \cref{fig:channels}.

\footnotetext{Icons by Freepik, Mayor Icons, DinosoftLabs, and Muh Zakaria.}

\subsubsection{Ambient Breath Analysis} 

In indoor settings, \acp{IM} disperse through both bounded and unbounded spaces, with ventilation systems and dynamic emissions influencing VOC distribution. 
As noted in \cite{liangModelingVolatileOrganic2012}, modeling ventilation strategies helps predict \ac{VOC} concentration changes and their impact on indoor air quality.
Particle loss due to diffusion, surface absorption, and settling is significant, but certain ventilation methods can reduce particle flow to critical zones, lowering concentrations in specific areas \cite{wongEffectsMedicalStaffs2022}.

Obstacles like humans, furniture, and structures cause turbulence and alter particle trajectories, contributing to channel variability. 
Indoor sources, such as coffee machines or garbage cans, release additional \acp{VOC}, changing \ac{IM} composition \cite{liangModelingVolatileOrganic2012}. 
Noise sources, including heating, ventilation, and human movements, further disrupt airflow patterns, adding variability \cite{wongEffectsMedicalStaffs2022}. 
Therefore, modeling both steady and transient air circulation changes is essential for accurately capturing the distribution of target molecules.

\subsubsection{Breath Analysis Devices}\label{sec:breathalyzer}

Unlike ambient breath analysis, breath analyzers (e.g., breathalyzers) use controlled channels, like short tubes, to minimize particle loss and enhance detection precision.
While these environments limit interference, non-target \acp{VOC} from the respiratory tract can still impact performance \cite{khalidModelingViralAerosol2020}, requiring specialized filters to isolate target compounds.

Noise sources in controlled breath analysis systems often stem from user behavior and handling errors. 
Variations in exhalation speed, volume, or technique affect airflow and measurement accuracy, while transient events like coughing or sneezing introduce aerosol surges \cite{khalidModelingViralAerosol2020}.
Hence, effective device design must account for both physical channel constraints and human factors to ensure robust molecular detection in real-world scenarios.

\subsection{Noise Analysis and Interference}

Noise and interference in the transmission channel arise from background biomarkers, cross-drift, turbulence, temperature changes, and obstacles, hindering accurate detection of target biomarkers and their properties in exhaled breath.

\subsubsection{Molecular Noise}

Molecular noise in \ac{MC} arises from collisions, reactions, and thermal fluctuations, posing key challenges in diffusion-based systems. 
Pierobon \emph{et al.}~\cite{pierobon2011diffusion} identify two primary noise sources: particle sampling noise (from transmitter emission) and particle counting noise (from signal propagation), modeled using physical and stochastic approaches.
Physical models mathematically analyze noise processes, while stochastic models describe them via random processes. 
Later, Pierobon \emph{et al.}~\cite{pierobon2011noise} introduce reception noise from ligand-receptor binding, modeled through ligand-receptor and stochastic chemical kinetics. 
These approaches provide mathematical formulations for noise source simulation and closed-form solutions for stochastic noise modeling. 
Singhal \emph{et al.}~\cite{singhal2014effect} explore noise at a microscopic level using Langevin force analysis. 
For diffusion-based \ac{MC} using molecule release time, an inverse Gaussian distribution is used to model system noise, leading to the development of the additive inverse Gaussian noise channel model~\cite{srinivas2012molecular}.

\subsubsection{Environmental Interference}

Environmental interference, such as \acp{VOC} from building materials, furniture, and activities, can overlap with target biomarkers, increasing false positives and requiring advanced filtering~\cite{bhardwaj2021recent}. 
Their emissions vary with environmental conditions and material properties, making them a challenge to model~\cite{liangModelingVolatileOrganic2012}.
They closely resemble \ac{CCI}, as described in \cref{subsec:propa}, since non-target molecules disrupt reception.

As discussed in~\cref{sec:breathalyzer}, large-scale ambient factors like ventilation and human activity~\cite{wongEffectsMedicalStaffs2022} make the propagation channel highly dynamic. 
This necessitates adaptive real-time or long-term averaging models that ignore short-term fluctuations.
These effects resemble \ac{ISI} in \ac{MC}~(\cref{subsec:propa}), where lingering \acp{IM} impact current and future readings.

Finally, the detection systems introduce inherent noise due to sensor variability affected by flow rate, binding efficiency, and air velocity~\cite{bhardwaj2021recent}. 
Sudden events such as coughing or sneezing add further disruptions through burst particle emission into the channel~\cite{khalidModelingViralAerosol2020}.

\subsection{Breath Cloud}

\begin{figure}
    \centering
    \includegraphics[width=\linewidth]{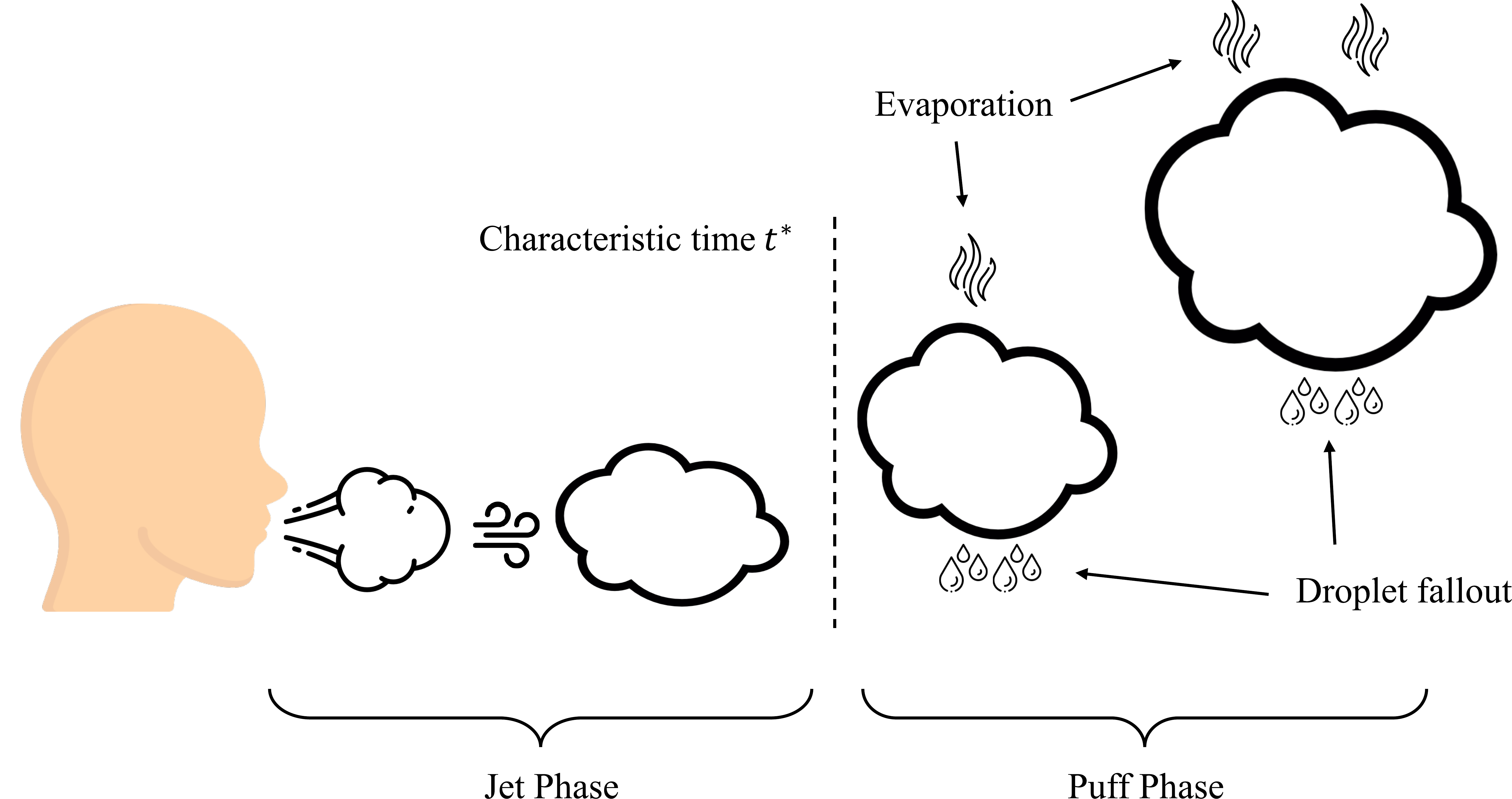}
    \caption{Visualization\protect\footnotemark{} of the breath cloud model including the different phases after violent expiratory events as well as fallout and evaporation effects.}
    \label{fig:breath_cloud}
    \vspace{-0.8em}
\end{figure}
\footnotetext{Icons by Freepik.}

The dynamics of the breath cloud is key to \ac{ABMC}, affecting the spread of exhaled particles. 
\Cref{fig:breath_cloud} depicts a summary of the most important effects influencing the breath cloud.
The droplet behavior varies by size: large droplets (>\SI{100}{\micro\meter}) quickly settle due to gravity, while smaller ones (<\SI{10}{\micro\meter}) evaporate into airborne droplet nuclei that remain suspended in air for long periods~\cite{cavazzutiStatisticalAnalysisExhaled2023}.

Buoyancy shapes the path of the exhaled breath cloud, lifting it upward and helping smaller droplets travel farther. 
The warm, humid microenvironment created by the breath cloud also slows droplet evaporation by reducing moisture loss. 
These thermo-hygrometric effects are essential for understanding droplet behavior and the propagation of the resultant signal~\cite{bourouibaViolentExpiratoryEvents2014, cavazzutiStatisticalAnalysisExhaled2023}.

The breath cloud expands self-similarly, keeping its approximate shape as it grows~\cite{cavazzutiStatisticalAnalysisExhaled2023}.
It is well established that the cloud's speed decreases with distance, while its radius increases proportionally, making this pattern key to predicting particle dispersion and spatial distribution.

Different respiratory and expiratory events uniquely shape the dynamics of the breath cloud.
Breathing produces slow, stable clouds, while speaking increases droplet production and spread.
In contrast, violent expiratory events, such as coughing and sneezing, create turbulent multiphase clouds that contain hot, moist air and droplets of various sizes. 
These clouds initially behave as momentum-driven jets before transitioning into buoyancy-dominated puffs as they decelerate~\cite{bourouibaViolentExpiratoryEvents2014}.

During breathing, exhaled droplets are affected by gravity, drag, evaporation, and turbulent dispersion.
In abstract terms, the weight force scales cubically with the droplet diameter and depends on the density difference between the droplet and the surrounding air, causing larger droplets to settle more rapidly. 
The drag force, which scales with the square of the diameter and the relative airflow velocity, exerts a stronger influence on smaller droplets. 
Droplet evaporation is governed by a convection-diffusion process.
The droplet radius $D_p$ over time $t$ is
\begin{equation}
    D_p(t) = \sqrt{D_{p,0}^2 - Kt},
\end{equation}
where $D_{p,0}$ is the initial droplet diameter and $K$ is the evaporation rate~\cite{cavazzutiStatisticalAnalysisExhaled2023}. 
Turbulent dispersion introduces additional velocity fluctuations that further affect the trajectories of the smallest droplets. 

Violent expiratory events like coughing and sneezing produce exhaled clouds with two phases~\cite{bourouibaViolentExpiratoryEvents2014}. 
In the initial \textit{jet phase}, high momentum of the exhaled breath propels the cloud forward, with the cloud radius increasing roughly linearly with distance. 
As the cloud slows, buoyancy takes over, transitioning to the \textit{puff phase} at a characteristic time
\begin{equation}
    t^* = \frac{I_0}{B_0},
\end{equation}
where $I_0$ represents the initial momentum and $B_0$ the buoyant force.
Droplets eventually settle when their terminal settling speed, determined by gravitational acceleration, droplet diameter, dynamic viscosity, and density differences, is reached.

Bourouiba \emph{et al.}~\cite{bourouibaViolentExpiratoryEvents2014} present the Stokes settling speed, showing that larger droplets descend faster due to their size and density difference from air. 

Furthermore, Bourouiba \emph{et al.}~\cite{bourouibaViolentExpiratoryEvents2014} present a formula for the Stokes settling speed, at which droplets start descending, depending most importantly on the droplet diameter and the difference between droplet and surrounding air density.
They also propose a continuous fallout model to describe how suspended particle numbers decrease over time, highlighting the interplay between droplet settling and cloud expansion.

The formation of a turbulent multiphase flow in the breath cloud is a hallmark of violent respiratory events. 
This complex flow contains droplets of various sizes entrained within hot, moist air, creating challenges for accurately predicting droplet trajectories and evaporation rates. 
Detailed models~\cite{bourouibaViolentExpiratoryEvents2014, cavazzutiStatisticalAnalysisExhaled2023} help understand these dynamics, facilitating better predictions of droplet behavior and channel characteristics.

\subsection{Channel Modeling}

This section reviews channel modeling in \ac{ABMC}, focusing on aerosol transmission systems. 
Here, the human respiratory tract acts as the transmitter, releasing \acp{IM} via breathing, coughing, or sneezing (\cref{sec:transmitter}), with breathing being a key source of continuous dispersion.
To model this scenario, several atmospheric dispersion models can be adapted for \ac{MC} applications, offering effective tools to describe molecular propagation in such channels.

The \textbf{Gaussian plume model} describes the steady-state distribution of continuously emitted particles, assuming Gaussian spread in horizontal and vertical directions. 
It factors in wind speed, atmospheric stability, and emission rate.
Thus, the concentration of \acp{IM} at a point $(x,y,z)$ downwind from a source with height $H$ can be given as
\begin{multline}
C(x, y, z) = \frac{Q}{2 \pi \sigma_y \sigma_z u} \exp\left(-\frac{y^2}{2\sigma_y^2}\right) \\ \cdot \left[ \exp\left(-\frac{(z - H)^2}{2\sigma_z^2}\right) + \exp\left(-\frac{(z + H)^2}{2\sigma_z^2}\right) \right],
\end{multline}
where $Q$ is the emission rate, $u$ the wind speed, and $\sigma_y$ and $\sigma_z$ represent horizontal and vertical dispersion parameters, respectively~\cite{khalid2018system, seinfeldAtmosphericChemistryPhysics2016}.

\textbf{\ac{LPD} models} simulate trajectories of individual \acp{IM}, accounting for turbulent diffusion and advection processes based on local wind fields and turbulence parameters.
\Ac{LPD} models help predict stochastic molecular behaviors such as propagation delay and signal attenuation~\cite{linNearfieldToolSimulating2003}.

\textbf{\ac{CFD} models} numerically solve fluid flow equations, capturing complex effects such as vortices and boundary interactions.
These simulations are especially useful in environments with obstacles, incorporating molecular dispersion under complex airflow conditions~\cite{dboukCoughingAirborneDroplet2020}.

The \textbf{\ac{HYSPLIT} model} combines Eulerian and Lagrangian methodologies to simulate long-range transport, dispersion, and deposition of airborne \acp{IM}, if adapted for~\ac{MC}, taking into account changing atmospheric conditions~\cite{steinNOAAsHYSPLITAtmospheric2015}.

The \textbf{\ac{CALPUFF} modeling system}, like the \ac{HYSPLIT} model, is a non-steady-state dispersion model used by the United States Environmental Protection Agency to capture time- and space-varying meteorological effects influencing \ac{IM} transport and transformation.
Within \ac{MC} systems, \ac{CALPUFF} can help characterize dynamic channels affected by available weather conditions~\cite{agencyCalpuffDispersionModel2012}.

\textbf{Artificial neural network-based models} use trained neural networks to capture complex dispersion patterns from turbulent flows.
Leveraging experimental or simulation data, these models predict \ac{IM} behavior, offering a data-driven addition to traditional physical models in \ac{ABMC}~\cite{crawfordUseGaussianMixture2020}.

\section{Air-Based Molecular Communication Receiver Engineering}
\label{sec:receiver}
In a communication-theoretic model, the interpretation of exhaled breath depends on its application. 
Hence, a distinction is made between natural and engineered \ac{MC} receivers.
Natural receivers, such as the olfactory systems in humans, animals, and microbial interactions, process breath for biological or environmental purposes. 
As they are not designed for medical analysis, they are beyond the scope of this work. 
The interested reader can refer to Aktas \emph{et al.}~\cite{aktas2024odor}on odor-based communication in nature.

In comparison, engineered receivers are built systems for a purpose, in this case, to detect, analyze, and interpret the physical and chemical properties of exhaled breath.
This work explicitly focuses on engineered receivers for non-invasive breath analysis for medical applications.
The reception process on the receiver side can be distinguished into three parts: \textit{A)} breath sampling, \textit{B)} disease detection through sensors, and \textit{C)} decision making~\cite{khalidCommunicationBreathAerosol2019,khalid2018system}. 
While the breath sampling is discussed in~\cref{subsec:sampling}, \cref{subsec:detection_technology} presents various biomarker detection techniques and sensing technologies, \cref{subsec:disease} focuses on disease detection (utilizing the surveyed sensors in \cref{subsec:detection_technology}) and decision making. 
Data set availability for analyzing the exhaled breath is reported in~\cref{subsec:disease} and different approaches for modeling the receiver are discussed in~\cref{subsec:model_receiver}.

\subsection{Exhaled Breath Sampling}
\label{subsec:sampling}

Controlling the exhaled breath sampling rate is a key aspect of the receiver's front end~\cite{khalid2018system}.
Several methods, categorized as \textit{i)} filters, \textit{ii)} condensation, \textit{iii)} electrostatic precipitation, and \textit{iv)} solid and liquid impactors, are commonly used~\cite{Ladhani2020-ha, Verreault2008-ko}.
The suitability of a breath sampling method depends on the selected biomarkers and the chosen analysis technique~\cite{Wang2024-wn}. Sensors often exhibit noisy behavior and systematic drift~\cite{padilla2010drift}, which needs to be addressed through sophisticated processing.

\subsubsection{Filters}
This method captures biomarkers 
from exhaled breath by passing it through a filter material~\cite{Malik2021-pf}. 
Filter efficacy depends on mechanisms such as interception, inertial impaction, diffusion, gravitational settling, and electrostatic attraction. 
Filters vary in pore size, material, and thickness, enabling long-term sampling of particles ranging from \SIrange{10}{900}{\nano\meter}~\cite{Verreault2008-ko}. 
Different approaches using filters for pathogen detection are described in the literature. 
For instance, Malik \emph{et al.}~\cite{Malik2021-pf} use an electret air filter-based device to determine differences in the viral load of SARS-CoV-2 in exhaled breath compared to pharyngeal swabs. 

\subsubsection{Sampling via Condensation}  
By cooling exhaled air onto inert surfaces, water vapor condenses, allowing the collection of biomarkers 
from the exhaled breath. 
Ten minutes of tidal breathing produces \SIrange{1}{3}{\milli\liter} of breath condensate~\cite{nwanochie2022}. 
\Ac{EBC} collection can be integrated into face masks~\cite{Daniels2021-bq} or \ac{POC} devices with detection sensors~\cite{Ghumra2023-nk}. 
\Ac{EBC} can contain respiratory pathogens like SARS-CoV-2, rhinovirus, and mycobacterium tuberculosis~\cite{nwanochie2022} and proteins/enzymes reflecting lung health~\cite{kubavn2013exhaled}. 
It aids in diagnosing, monitoring disease, and evaluating treatment efficacy.

\subsubsection{Electrostatic Precipitation}  
Electrostatic precipitation involves passing a particle-laden gas through electrodes with a high-voltage potential, causing the particles to migrate to one electrode based on their precharge~\cite{calvert1990glossary}. 
This method samples particles in the \SIrange{2}{100}{\nano\meter} range~\cite{tsi3089}. 
In a pilot study, Ladhani \emph{et al.}~\cite{Ladhani2020-ha} use an electrostatic sampling of exhaled breath to detect pathogens \textit{Staphylococcus aureus} in a primary care setting.

\subsubsection{Solid and Liquid Impactors}
Unlike filters, which capture particles through multiple deposition mechanisms, impactors use inertial impaction to separate particles by size.
Filters collect a broad range of particle sizes, while impactors enable size fractionation, making them useful for distinguishing between different respiratory particle sizes.
Solid impactors, such as slit, cyclone, or Andersen samplers, accelerate particles through narrow openings, forcing them to collide with a surface for separation. 
Filters are then washed or used for plaque assays~\cite{Verreault2008-ko}.
Starkenborg \emph{et al.}~\cite{Stakenborg2022-jw} combine a portable silicon sieve-based breath sampler with molecular analysis to detect SARS-CoV-2 at the point of need.
Liquid impactors, like all-glass impingers, direct particles through a narrow orifice into a liquid medium, capturing larger ones by impact and smaller ones via diffusion~\cite{Verreault2008-ko}.


\subsection{Detection Technologies}
\label{subsec:detection_technology}

The research community has explored popular detection technologies that efficiently help monitor biomarkers in exhaled breath. 
In the following, we present an overview of common detection technologies and explain the functionality of the sensors and technologies.


\textbf{Metal-oxide sensors} are widely used for gas detection due to their high sensitivity, broad gas range, and cost-effectiveness~\cite{mahajan2020metal}. 
They detect gases by measuring resistance changes in metal-oxide layers like tin dioxide or zinc oxide.
Operating at temperatures up to \SI{450}{\celsius} enhances their sensitivity~\cite{wang2010metal}.
Their applications span environmental monitoring, industrial safety, healthcare, and automotive industries~\cite{saxena2023review}. 
While they offer high sensitivity and robustness, challenges include temperature and humidity sensitivity, long-term drift, and high power
consumption~\cite{singh2021recent}. 
Advances focus on nanostructured metal-oxides, noble metal catalysts incorporation, and improved portability~\cite{zhang2020diversiform}.

\textbf{Conducting polymers}, such as polyaniline, polypyrrole, and polythiophene, conduct electricity through conjugated $\pi$-electron systems, making them ideal for sensor applications~\cite{namsheer2021conducting}. 
Their conductivity changes upon interaction with gases or \acp{VOC}~\cite{macdiarmid1997polyaniline}.
Conducting polymers are synthesized via chemical or electrochemical polymerization, and can be processed into various forms like thin films and nanostructures~\cite{cho2008fast}. 
Their applications can be found in environmental monitoring, medical diagnostics, and industrial safety, offering high sensitivity, selectivity, and room-temperature operation~\cite{dincer2019disposable}. 
Despite advantages such as mechanical flexibility and ease of fabrication, challenges like environmental sensitivity and long-term stability persist~\cite{alam2022recent}. 
Recent advances in nanocomposites and molecularly imprinted polymers improve stability, sensitivity, and integration into portable and wearable devices~\cite{ahmad2015nanocomposites}.

\textbf{\Ac{QCM} sensors} detect tiny mass changes by measuring shifts in resonant frequency~\cite{bragazzi2015quartz}. 
They rely on the piezoelectric effect, where quartz crystals oscillate at a specific frequency that varies with surface mass changes~\cite{speight2012survey}. 
\Ac{QCM} sensors are widely used in chemistry, biology, and environmental monitoring, however, they face challenges like environmental sensitivity~\cite{wessels2013practical} and surface fouling~\cite{kraus2001use}.
Advances in nanoparticles and molecularly imprinted polymers enhance their sensitivity, selectivity, and versatility~\cite{akgonullu2022recent,fu2003quartz}.

\textbf{\Ac{SAW} sensors} detect target biomarkers by measuring changes in acoustic waves traveling along a piezoelectric substrate~\cite{mandal2022surface}.
These sensors use the piezoelectric effect to generate and detect mechanical vibrations, with interactions altering wave properties like frequency, phase, and amplitude~\cite{nair2021acoustic}.
This enables high sensitivity to mass loading, viscosity, and conductivity changes~\cite{zhou2019structural}.
\Ac{SAW} sensors are widely used in chemical and biological sensing, environmental monitoring, and medical diagnostics~\cite{mandal2022surface}.
Challenges like environmental sensitivity and surface fouling persist~\cite{sankaranarayanan2008flow}, but advancements in materials~\cite{damasceno2023recent}, surface functionalization~\cite{mcgill1998performance}, and microfluidic integration aim to improve performance~\cite{lange2006integration}.

\textbf{Mass spectrometers} identify and quantify chemical compounds by analyzing their mass-to-charge ratio~\cite{griffiths2008brief}.
They operate through ionization, ion separation, and detection, using techniques like electron and electrospray ionization~\cite{mirsaleh2008electron}.
Used across diverse scientific disciplines, mass spectrometers find applications in the areas of chemical analysis, environmental monitoring, clinical diagnostics, and forensics~\cite{mbughuni2016mass}.
They offer high sensitivity and specificity, but face challenges like cost, complexity, and sample preparation requirements~\cite{de2007mass,hiraoka2013fundamentals}. 
Recent advancements focus on enhancing their sensitivity, resolution, automation, and miniaturization for portable applications, thereby expanding their utility in multiple fields~\cite{elpa2020automation,maciel2020miniaturized}.

\textbf{Optical sensors} convert light into electrical signals by detecting changes in light properties such as intensity, wavelength, phase, or polarization~\cite{santos2014handbook}.
They consist of a light source, sensing element, and detector, with interactions leading to absorption, reflection, fluorescence, or scattering. 
Types include absorption-based, fluorescence, surface plasmon resonance, fiber optic, and photonic crystal sensors~\cite{khonina2023optical,nair2010photonic,homola2008surface}. 
They are fabricated using materials like semiconductors, metals, and polymers~\cite{dey2018semiconductor}, employing techniques such as photolithography and chemical vapor deposition~\cite{du2022advances}. 
Applications span environmental monitoring, biomedical diagnostics, industrial process control, food safety, and security~\cite{gupta2017optical,narsaiah2012optical}. 
Advantages include high sensitivity, fast response time, non-destructive analysis, and versatility, while challenges involve environmental interference, fabrication complexity, cost, and the need for regular calibration~\cite{khonina2023optical}. 
Recent advancements focus on enhancing sensitivity and miniaturization, leveraging nanophotonics, plasmonics, and advanced materials for next-generation optical sensors with improved performance.

\textbf{Gas chromatography} is a widely employed technique for separating and analyzing volatile compounds, and it can be integrated with the aforementioned sensors to create a versatile analytical platform~\cite{olle2020advancements,akbar2015chip}. 
In gas chromatography, a continuous flow of an inert or non-reactive gas, known as the mobile phase, transports the vaporized sample mixture through a narrow tube called the column. 
The column's inner surface is coated with a stationary phase, and the separation occurs based on the chemical and physical properties of the sample components and their interactions with the stationary phase. 
As a result, individual components traverse the column at different rates and reach the coupled sensor at distinct times. 
For further details on gas chromatography principles and operation, readers are referred to~\cite{mcnair2019basic}.

\subsection{Detecting Diseases}
\label{subsec:disease}

Detecting diseases in the exhaled breath includes detecting specific \acp{IM}, such as biomarkers or microbial DNA sequences, and physical breath parameters, such as breathing patterns, temperature, or humidity. 
Breath analysis may even reveal genetic material~\cite{koblentz2003pathogens}, which is beyond the scope of this survey, however can be highly relevant in terms of security considerations. This survey highlights selected relevant studies focusing on biomarkers and microbial DNA to illustrate detection possibilities, given the vast research in this field~\cite[Fig.~1]{rydosz2018sensors}.

\subsubsection{Detection through Specific Information Molecules}
One major avenue of research for disease detection in exhaled breath is to utilize targeted molecular analysis, and identify \acp{IM} that correspond to specific diseases. The detection of specific \acp{IM} can be split into the fields of biomarkers and microbial DNA, which will both be highlighted in this section.
Closely related to the detection itself is the difficulty of obtaining data sets. 
Therefore, in this section, existing databases related to analyzing the exhaled breath are also highlighted based on the \ac{IM}.

\paragraph{Biomarkers} 

\textbf{Diabetes:} Early studies use iodine titration to detect breath acetone, linking it to diabetes~\cite{hubbard1920determination}.
Reviews confirm higher acetone levels in Type I diabetes but not in Type II~\cite{wang2013breath, ghimenti2013monitoring, sun2015determination}.
However, classification based on acetone alone is unreliable~\cite{ghimenti2013monitoring}.
Additional compounds like 3-hydroxy-butan-2-one and butane-2,3-dione have been identified~\cite{ghimenti2013monitoring}.
Advances in sensor-based detection further refine breath analysis~\cite{rydosz2018sensors}.

\textbf{Asthma:} Harkins \emph{et al.}~\cite{harkins2004exhaled} compare nitric oxide concentration in the exhaled breath of asthmatic patients to control subjects. 
The clinical trial in~\cite{harkins2004exhaled} demonstrates that asthmatic patients with an exacerbation have a higher mean concentration of nitric oxide in the exhaled breath than patients without an exacerbation.

\textbf{Cystic Fibrosis:} 
Hydrogen cyanide in exhaled breath serves as a marker for \textit{Pseudomonas aeruginosa} infection in cystic fibrosis patients~\cite{gilchrist2012investigation}. 
Smith \emph{et al.}~\cite{smith2013hydrogen} confirm its presence in both mouth and nasal exhalations, correlating its levels with the bacterium, which also produces hydrogen cyanide \textit{in vitro}.

\textbf{Lung Cancer:} Dent \emph{et al.}~\cite{dent2013exhaled} identify distinct \ac{VOC} patterns and genetic changes in lung cancer patients' exhaled breath condensate.
A study by Sakumura \emph{et al.}~\cite{sakumura2017diagnosis} shows 89.0\% accuracy in lung cancer screening using gas chromatography and mass spectrometry to detect five \acp{VOC}.
A comprehensive review by Buszewski \emph{et al.}~\cite{buszewski2007human} categorizes lung cancer biomarkers into six classes: alcohols, aldehydes, ketones, hydrocarbons, esters, and heterocycles. 
Moreover, elevated interleukin-6 levels in exhaled breath may also serve as an early detection marker~\cite{carpagnano2002interleukin}.

\textbf{Colorectal Cancer:} A study on methane concentration in the exhaled breath of colorectal cancer patients reveals that 80\% of patients have higher methane levels than control subjects, suggesting a difference in anaerobic intestinal flora~\cite{Haines1977breath}. 
However, subsequent research does not confirm a correlation between elevated methane levels and colorectal cancer, with 63\% of patients and 56\% of controls showing detectable methane concentrations in their breath~\cite{Haines1977breath,Sivertsen1992breath}.

\textbf{Kidney Disease:} The study in~\cite{Grabowska2013detection} on biomarkers in the exhaled breath of chronic kidney disease patients identifies triethylamine, aliphatic hydrocarbons, and sulfur components in their breath, which are absent or at lower levels in control subjects. 
These findings~\cite{Grabowska2013detection} suggest that these compounds may indicate chronic kidney disease or coexisting conditions.

\textbf{Myocardial Infarction:} Weitz \emph{et al.}~\cite{Weitz1991high} identify pentane in the exhaled breath of myocardial infarction patients, although also present in the exhaled breath of control subjects. 
Further research finds no significant difference in pentane concentrations between patients and controls~\cite{Mendis1995expired}, but a higher concentration of isoprene is observed in the patients' exhaled breath~\cite{Mendis1995expired}.

\textbf{Obstructive Sleep Apnea:} 
Olopade \emph{et al.}~\cite{Olopade1997exhalded} discuss increased pentane and nitric oxide levels in the breath of obstructive sleep apnea patients post-sleep, unlike controls.
Additionally, controls show elevated oral nitric oxide pre- and post-sleep. Carpagnano \emph{et al.}~\cite{carpagnano2002increased} report higher interleukin-6 and 8-isoprostane levels in sleep apnea patients, correlating with disease severity

\textbf{Renal Failure:} Davies \emph{et al.}~\cite{Davies1997quantitative} show that patients with chronic renal failure, especially those on \ac{CAPD}, display elevated levels of ammonia in exhaled breath.
Similarly, hemodialysis patients also show increased ammonia and ethylene levels\cite{Popa2011ethylene}. 
Kearney \emph{et al.}~\cite{kearney2002breath} report lower breath ammonia in \textit{H. pylori}-positive patients compared to uninfected controls.

\textbf{Liver Disease:} Isoprene is identified as a biomarker for liver disease, detectable in the exhaled breath~\cite{kapishon2013atmospheric}. 
Its formation occurs ``along the mevalonic pathway in cholesterol synthesis''~\cite{kapishon2013atmospheric, selvaraj2020advances}.

\textbf{Schizophrenia:} Phillips, Sabas \& Greenberg present in~\cite{phillips1993increased} a study showing that the mean alveolar gradient of pentane and carbon disulfide in the exhaled breath of patients suffering from schizophrenia is higher compared to the exhaled breath control subjects.

\textbf{Ulcerative Colitis:} Sedghi\emph{et al.}~\cite{sedghi1994elevated} investigate ethane levels in the exhaled breath of ulcerative colitis patients. 
Reactive oxygen species can damage tissues during colitis, triggering \enquote{lipid peroxidation of omega-3 and omega-6 fatty acids}~\cite{sedghi1994elevated}.
Ethane, a byproduct of this process, is determined to measure this peroxidation. 
The results of the study reveal elevated levels in colitis patients compared to control subjects.

Although smoking is not a disease, it increases the risk of stroke and cancer, making it relevant to this work.
Vasthare \emph{et al.} review carbon monoxide sensors for detecting smokers via exhaled breath analysis~\cite{vasthare2018carbon}. 
These sensors effectively distinguish smokers from non-smokers and support smoking cessation programs~\cite{vasthare2018carbon}.
Furthermore, Ryter \emph{et al.}~\cite{ryter2013carbon} identify carbon monoxide as a breath marker for oxidative stress or \enquote{stimulation by pro-inflammatory cytokines}~\cite{selvaraj2020advances}, for example, in smoking cessation.

\paragraph{Microbial DNA}

    \textbf{Bacterial Infections:} During breathing, different bacterial pathogens can be emitted, including, among others, \textit{Hemophilus influenzae}, \textit{Escherichia coli}, and \textit{Legionella pneumophila}. 
    \textit{Hemophilus influenzae}, for example, can cause pneumonia and meningitis.
    The study in~\cite{zheng2018bacterial} demonstrates their detection in the exhaled breath using a protocol that integrates the collection of \ac{EBC} and loop-mediated isothermal amplification.
    In~\cite{cheng2025visual}, a respirator is proposed to detect mycobacterial DNA in exhaled breath samples. 
    The proposed respirator includes an exhalation reservoir, a microfluidic chip, recombinase polymerase amplification assays, and lateral flow assays, enabling the detection of mycobacterium tuberculosis DNA directly from exhaled breath samples.
    
    \textbf{Viral Infections:} Besides bacterial infections, viral infections can be detected by analyzing the exhaled breath. 
    Studies, primarily published during or after the COVID-19 pandemic, demonstrate that SARS-CoV-2 \ac{RNA} can be detected in exhaled breath samples using \acp{PCR}~\cite{Ghumra2023-nk,Stakenborg2022-jw,bellarmino2024covid}. 
    Therefore, \ac{EBC} has also been explored as a diagnostic medium, as it contains lower respiratory droplets, potentially improving detection sensitivity for SARS-CoV-2~\cite{sawano2021rt,ryan2020use}.
    Apart from SARS-CoV-2, for example, the \textit{Influenza} virus \ac{RNA} is detected in exhaled breath using \ac{qPCR}~\cite{fabian2008influenza}. 
    Additionally, exhaled breath is used to detect \textit{Human papillomavirus}, known as HPV, in the \ac{EBC} of lung cancer patients~\cite{carpagnano2011hpv}.
    
    \textbf{Fungal Infections:} Fungal infections are mainly detected by analyzing the exhaled breath in terms of \acp{VOC}~\cite{diefenderfer2024new}.
    Besides, for detecting fungi, the \ac{EBC} is used~\cite{carpagnano2016analysis,bhimji2018asper,bitterman2024exhaled}, whereby most of the studies focus on \textit{Aspergillus}~\cite{bhimji2018asper,bitterman2024exhaled}, a genus of mold (fungus) that includes a wide variety of species. 
    One of the most well-known and clinically significant species is \textit{Aspergillus fumigatus}. 
    It can cause \textit{aspergillosis}, which affects the lungs and other organs. 
    The most common form of aspergillosis is invasive pulmonary aspergillosis, which primarily affects immunocompromised individuals, such as those with cancer or organ transplants.

\paragraph{Data Set Availability}

For a diagnostic validation, structured and well-documented data sets are crucial.
BreathBase\textregistered Data~\cite{breathbase_data} and the \ac{HBDB}~\cite{hbdb_data} are key databases in this field. 
BreathBase\textregistered Data holds over 160,000 breath profiles with clinical data, covering conditions like cancer, \ac{COPD}, asthma, and interstitial lung diseases.
The \ac{HBDB}, on the other hand, catalogs 1,143 VOCs linked to 60 diseases, sourced from 2,766 studies, using text mining to map breath compounds to diseases and biochemical pathways~\cite{hbdb_data}.

\subsubsection{Detection through Physical Breath Parameters}

Besides detecting specific \acp{IM}, various diseases can be detected through more physical and superficial parameters, whose detection techniques are more easily available without extensive molecular analysis. 

\paragraph{Physical Properties}

Besides detecting disease by analyzing \acp{IM} in the exhaled breath, the physical properties of the exhaled breath, such as temperature and humidity, can also be used for disease detection.

\textbf{Temperature:}
Popov~\cite{popov2011human} identifies \ac{EBT} as a potential disease marker. 
Studies show elevated \ac{EBT} in asthmatic patients, decreasing with asthma control~\cite{popov2017added,melo2010exhaled,garcia2013increased,wojsyk2017evaluation}, and lower \ac{EBT} in \ac{COPD}, rising during exacerbations~\cite{paredi2003exhaled,klokstad2010effect,lazar2014exhaled}.
Higher \ac{EBT} also correlates with non-small-cell lung cancer severity~\cite{carpagnano2014exhaled,carpagnano2015exhaled}. 
The study in~\cite{popov2016relationship} reveals that \ac{EBT} rises before ear temperature in infections, making it a potential early diagnostic marker.

\textbf{Humidity:} Breath humidity reflects respiratory tract hydration, mucus production, and fluid balance~\cite{turcios2020cystic}, aiding in detecting conditions like cystic fibrosis, chronic bronchitis, and heart failure. 
However, it plays a minor role in disease diagnosis. Instead, humidity can affect detection reliability by acting as a cross-reactive parameter~\cite{dixit2021exhaled,lagopati2023role}, potentially interfering with \acp{VOC} analysis in diabetes monitoring~\cite{dixit2021exhaled}.

\paragraph{Spatial Origins}
The spatial origin of exhaled breath (nasal or oral) impacts both oral and systemic health. 
Differentiating between the spatial origin, i.e., nasal or oral breathing, involves analyzing \acp{VOC}, temperature, breathing patterns, and acoustic signals. 
\Ac{VOC} concentration and composition varies based on the exhalation route~\cite{sukul2017oral}.
Utilizing thermography can differentiate between nasal and oral breathing by monitoring the temperature changes in the two facial regions~\cite{huang2021nose}. 
Apart from the surveyed sensors in~\cref{subsec:detection_technology}, acoustic sensors can also help distinguish between nasal and mouth breathing by analyzing breath sounds.~\cite{curran2012using} demonstrated that the spatial origins can be distinguished using acoustic sensors placed on specific body locations, followed by signal processing.

\paragraph{Flow Rate}

A wide range of sensors can detect disease based on the dynamics of breath flow, specifically breath rate, such as temperature, pressure, strain, acoustic, or optical sensors~\cite{yu2017wrinkled}.
Distinct patterns in the respiratory rate and rhythm in the exhaled breath can indicate specific respiratory conditions, for example, associated with disorders such as Biot's respiration, tachypnea, bradypnea, and Cheyne-Stokes respiration~\cite{whited2023abnormal,buaruk2023respiratory}. 
Analyzing breath patterns may also help diagnose neurological diseases, such as Parkinson's disease through nocturnal patterns~\cite{yang2022artificial} or Alzheimer's disease through changes in breath rate~\cite{bjerkan2024neurovascular}.

\subsection{Receiver Modeling}
\label{subsec:model_receiver}

For the various sensor types listed in \cref{subsec:detection_technology}, the actual form between the concentration of target molecules $C$ and the measured response can vary depending on the underlying physical or chemical transduction mechanism, but a generic approximation that has been widely adopted in the literature can be expressed as 
\begin{align}
    \Delta \mathcal{M} \approx \mathcal{K}C^\alpha,
    \label{eq:receiver}
\end{align}
where $\mathcal{K}$ and $\alpha$ are sensor specific parameters. For example, in metal-oxide sensors, $\Delta \mathcal{M}$ corresponds to the change in sensor resistance, with $\mathcal{K}$ as the gas-specific response coefficient and $\alpha$ as the sensitivity parameter~\cite{purnamadjaja2007guiding}.
In \ac{QCM} and \ac{SAW} sensors, $\Delta \mathcal{M}$ represents the frequency shift due to deposited gas mass, with $\mathcal{K}$ dependent on the crystal electrode area, the shear modulus of quartz, and quartz density~\cite{na2021overview}.
For mass spectrometers, 
$\mathcal{K}$ is an instrument-specific sensitivity factor influenced by ionization efficiency, transmission efficiency, and detector response. When $\Delta \mathcal{M}$ represents the change in absorbance for optical sensors, $\mathcal{K}$ is determined by the molar absorptivity and light path length as described by the Beer-Lambert law \cite{swinehart1962beer}.

\section{Experimental Testbed Setups}
\label{sec:testbed}

This section discusses experimental platforms for air-based communication from the \ac{MC} community and other fields (e.g., medicine and epidemiology) in \cref{MC:testbed} and \cref{nonMC:testbed}, respectively. 
Here, the \ac{MC} testbeds refer to those clearly defining a communication system.
While previous surveys, for example,~\cite{Lotter2023experimental_II}, emphasize environmental constraints, this review focuses on equipment rather than communication performance.
Testbeds from other fields are incorporated for their relevance, though this review is not exhaustive.
Instead, representative platforms are highlighted to compare testbeds and explore cross-domain insights in \cref{testbed:discussion}.

\subsection{Testbeds from the Molecular Communication Field}
\label{MC:testbed}

The \ac{ABMC} testbed by Farsad \emph{et al.}~\cite{farsad2013tabletop} is the first of its kind, using an electronic sprayer to release alcohol molecules detected by metal oxide sensors.
This platform is helpful for initially studying engineered \ac{ABMC} systems. 
A key finding from the experiments is the system's nonlinear behavior, challenging traditional linear \ac{MC} models and highlighting the need for new analytical approaches.
Building on this work, Koo \emph{et al.}~\cite{koo2016molecular} enhance the platform with additional spray nozzles and metal oxide sensors, creating a \ac{MIMO} \ac{MC} system. 
This setup enables the study of multiple individual cases and the determination of interlink interference. 
In addition, by introducing distinct biomarkers per transmitter, this setup could investigate interference in cases where sensors lack specificity and support the design of algorithms to accurately decode signals from multiple sources.

Unlike the static transmitters in~\cite{farsad2013tabletop, koo2016molecular}, Qui \emph{et al.}~\cite{qiu2016mobile} explore mobility by mounting an electronic sprayer on a robot.
Similar to the work in~\cite{farsad2013tabletop}, alcohol molecules are released and detected by a metal oxide sensor.
This testbed reveals mobility-induced transmission errors, necessitating the development of forward error correction codes for reliable communication in mobile \ac{MC} systems.
By emulating human movement, this platform provides a framework for studying dynamic scenarios and bridging static-mobile transmitter modeling.

Purnamadjaja \emph{et al.}~\cite{purnamadjaja2007guiding} present a mobile \ac{ABMC} system where a robot leader releases pheromones to influence other mobile robots equipped with metal oxide sensors. 
Inspired by queen bee pheromones, this system demonstrates how air-based signaling can guide coordinated actions. 
The concept of leader-follower dynamics offers valuable inspiration for mobile distributed sensing systems aimed at detecting exhaled breath biomarkers.
In hospitals or crowded spaces, a \enquote{central sensor} could detect critical biomarkers and relay data to mobile sensors, enabling spatial mapping of targeted biomarkers for disease monitoring and outbreak detection.

Shakya \emph{et al.}~\cite{shakya2018correlated} investigate a vapor propagation system using temporally modulated chemical plumes to correlate signals with transmitted patterns.
This approach enhances detection at meter-scale distances, even with non-selective sensors.
Their testbed features an electronically controlled transmitter that releases isopropanol vapor in binary patterns via a solenoid valve into a \SI{2.5}{\centi\meter} acrylic tube with adjustable wind speeds.
High-speed photoionization detectors measure vapor concentrations, enabling the detection of \SI{1}{\text{ppb}} isopropanol against a \SI{1}{\text{ppm}} ethanol background in turbulence, demonstrating potential for low-cost chemical sensing and source localization.
Ozmen \emph{et al.}~\cite{ozmen2018high} extend this to \ac{MC}, achieving low bit error rates at data rates up to \SI{40}{\text{bit}\per\second} at a distance of up to \SI{1.4}{\meter}.
With high temporal sensitivity, the system enables selective identification of disease-related biomarkers, supporting non-invasive diagnostics.

Cole \emph{et al.}~\cite{cole2009biomimetic} investigate \ac{MC} as an infochemical communication system inspired by insect pheromone biosynthesis and detection.
They develop chemoemitters that generate precise pheromone signals using \acp{VOC} and chemoreceivers modeled after insect olfactory systems with \ac{SAW} devices.
Preliminary results demonstrate the integration of these synthetic modules into a \enquote{moth-on-a-chip} system, highlighting potential applications in pest management and swarm robotics.
Similar chemoreceiver principles can be applied to detect \acp{VOC} in exhaled breath, where \ac{SAW} devices selectively capture biomarkers linked to metabolic processes or disease states.
These interactions produce measurable acoustic wave changes, enabling non-invasive health monitoring.

Giannoukos \emph{et al.}~\cite{giannoukos2017molecular} explore scent-based \ac{MC}, developing an advanced odor emitter for controlled olfactory transmission. 
Using a portable mass spectrometer, they generate, encode, and spatially encrypt \acp{VOC} like acetone and toluene in dynamic patterns over tubular chambers up to \SI{3}{\meter}.
They also examine signal modulation via flow rate and temperature, with applications in entertainment, security, and medicine. 
This work is extended in~\cite{giannoukos2018chemical}, where chemical signals encoded with the American Standard Code for Information Interchange transmit digital information via odors over \SI{4}{\meter}, using \ac{OOK} and \ac{CSK} for secure communication. 
Similarly, in~\cite{zhou2024pathogen}, a controlled tubular transmission safely transports biomarkers like bacteria and viruses to a mass spectrometer for analysis, reducing environmental exposure.

Bhattacharjee \emph{et al.}~\cite{bhattacharjee2020testbed} develop a macroscopic \ac{ABMC} testbed using fluorescein for data transmission in industrial networks. 
A spray nozzle serves as the transmitter, with high-speed cameras enabling detection over centimeter-to-meter distances. 
Inspired by this, Schurwanz \emph{et al.}~\cite{schurwanz2021duality} apply \ac{MC} concepts to respiratory particle transmission from coughing, speaking, and breathing. 
Fluorescent dye simulates aerosol dispersion, with the \emph{Pogona} simulator~\cite{drees2020efficient} assessing transmission scenarios.
Results confirm the effectiveness of masks and ventilation in reducing aerosol spread. 
Schurwanz \emph{et al.}~\cite{schurwanz2021infectious} further model viral transmission as a multiuser \ac{MC} system, where infected individuals broadcast pathogen-laden particles. 
Infection follows a threshold detection model, with viral load affecting disease severity, guiding strategies to limit information transfer between infected and healthy individuals.

\subsection{Testbeds from Other Fields}
\label{nonMC:testbed}

Xie \emph{et al.}~\cite{xie2009exhaled} investigate the emission of respiratory particles by healthy individuals during talking and coughing using a sealed box.
In the experiments, large particles are captured on microscope slides and water-sensitive paper, which change color upon contact, while smaller particles are analyzed in real-time with a dust monitor to determine their size distribution.

Gupta \emph{et al.}~\cite{gupta2009flow} measure exhalation airflow using a spiro\-meter with capillary tubes, where a pressure drop across the tubes indicates flow rate according to Poiseuille's law.
The direction of airflow is visualized using \SI{120}{\hertz} photography with cigarette smoke as a tracer.

\begin{figure}
    \centering
    \includegraphics[width=0.9\linewidth]{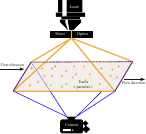}
    \caption{Illustration of a particle image velocimetry (PIV) setup.}
    \label{fig_piv}
    \vspace{-0.8em}
\end{figure}

\Ac{PIV} is widely used to visualize exhaled airflow, employing a double-pulsed laser, optics, particle seeding, and a camera, as shown in \cref{fig_piv}.
Introduced into airflow, tiny seeding particles serve as tracers that follow the flow without altering its dynamics. 
Successive laser-illuminated images are captured, and flow velocity vectors are computed based on particle displacement patterns between frames.
In the past, \ac{PIV} has also been used to measure cough airflow velocity~\cite{zhu2006study}, cough jet characteristics~\cite{vansciver2011particle}, and exhaled air speed during coughing and speaking~\cite{chao2009characterization}.
Furthermore, Chao \emph{et al.}~\cite{chao2009characterization} use \ac{IMI} with \ac{PIV} to measure particle size, revealing mean diameter of cough droplets to be \SI{13.5}{\micro\meter} with a concentration of \SIrange{2.4}{5.2}{\per\cubic\centi\meter}, and the mean diameter of speech droplets to be \SI{16}{\micro\meter} with a concentration range of \SIrange{0.004}{0.223}{\per\cubic\centi\meter}.

\begin{figure}
    \centering
\includegraphics[width=0.8\linewidth]{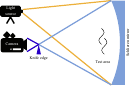}
    \caption{Illustration of a Schlieren imaging setup.}
    \label{fig_si}
    \vspace{-0.8em}
\end{figure}

Schlieren imaging is another effective way of visualizing exhaled airflow by detecting light refraction due to density differences between exhaled and ambient air, as demonstrated by Tang \emph{et al.}~\cite{tang2011qualitative} (cf.\ \cref{fig_si}).
This method evaluates the efficacy of interventions like tissue barriers in reducing the transmission of air-based infections.
For example, layering four tissue sheets effectively contains sneeze puffs, demonstrating the benefits of simple protective measures~\cite{tang2011qualitative}.

Zhang \emph{et al.}~\cite{zhang2019distribution} investigate ventilation effects on respiratory particle dispersion using a thermal manikin (\SI{30}{\celsius}) and a commercial particle generator.
A test chamber with a ventilation system serves as the propagation channel, with particle concentrations measured at various locations.
Findings highlight ventilation's role in reducing particle spread from coughing and breathing through the nose.

Perelló-Roig \emph{et al.}~\cite{perello2021cmos} develop a portable \ac{CMOS}–\ac{MEMS} gas sensor for detecting \acp{VOC}, targeting acetone for non-invasive diabetes diagnosis.
Fabricated with \SI{0.35}{\micro\meter} \ac{CMOS} technology, the system uses two 4-anchored \ac{MEMS} plate resonators with poly-4-vinylheduorocumyl alcohol coating to enhance acetone selectivity while reducing butane interference.
Experiments show a manifold increase in acetone response with a detection limit as low as \SI{20}{\text{ppb}}, demonstrating its potential for cost-effective breath analysis and \ac{POC} diagnostics.

Jaeschke \emph{et al.}~\cite{jaeschke2018novel} present a compact electronic nose for breath analysis with a modular design that supports up to thirty metal oxide sensors.
Its temperature-controlled chamber detects acetone, isoprene, pentane, and isopropanol, with integrated pressure, humidity, and temperature sensors.
Pattern recognition enables \ac{VOC} discrimination in varying humidity, highlighting its potential as a \ac{POC} diagnostic tool.

Additionally, Shang \emph{et al.}~\cite{shang2023chemiresistive} introduce a portable, wireless breath sensor array system for non-invasive lung cancer screening. 
It integrates chemiresistive thin films of gold nanoparticles, wireless transduction, and flow control for breath sampling. 
The system detects cancer-specific \ac{VOC} with a \SI{6}{\text{ppb}} limit, showing high sensitivity and selectivity. 
Simulations confirm tunable chemiresistive responses, while experiments with spiked human breath samples validate disease discrimination.
Its stable performance, portability, and cost-effectiveness highlight its potential as a \ac{POC} tool for early lung cancer detection.

Lommel \emph{et al.}\cite{lommel2021novel} introduce a method to study virus spread via aerosols in indoor settings using an emitter manikin releasing $\mathrm{NaCl}$-water tracer particles. 
Recipients measure inhaled tracer levels at different distances, enabling real-time evaluation of protection measures like ventilation. 
This is supported by an analytical model that predicts time-dependent aerosol concentrations based on distance. 
Schulz \emph{et al.}~\cite{schulz2024} extend this system to study aerosol dispersion in theatres, opera houses, arenas, and lecture halls.

Noti \emph{et al.}~\cite{noti2013high} present an investigation on the effects of humidity on influenza-laden aerosols in a simulated exam room with coughing and breathing manikins. 
For examination, a bio-aerosol sampler is used to fractionate the aerosol. 
Infectious and noninfectious virus amounts are measured via real-time quantitative
polymerase chain reaction and viral plaque assay.

To investigate the interpersonal particle transport in an aircraft cabin mockup, Zhang \emph{et al.} in~\cite{zhang2024experimental} use a particle generator and nanofiber air filters. 
Arizona test dust is used in a fan-driven particle generator to measure the size-dependent particle removal efficiency of the filtration units.

\subsection{Similarities and Differences of the Testbeds}
\label{testbed:discussion}

\Ac{ABMC} and non-\ac{MC} platforms share a focus on controlled signal detection but differ in objectives. 
For instance, \ac{MC} systems, like those by Farsad \textit{et al.}~\cite{farsad2013tabletop} and Koo \textit{et al.}~\cite{koo2016molecular}, use electronic sprays and sensors for molecular diffusion studies, while non-\ac{MC} platforms, such as Gupta's spirometer~\cite{gupta2009flow} and Adrian's \ac{PIV} setup~\cite{adrian2011particle}, analyze airflow and particle dynamics. 
\Ac{MC} testbeds prioritize encoding and transmission, as shown by Shakya \textit{et al.}~\cite{shakya2018correlated} and Ozmen \textit{et al.}~\cite{ozmen2018high}, while non-\ac{MC} systems focus on real-time monitoring, exemplified by Schlieren imaging~\cite{tang2011qualitative} and \ac{CMOS}–\ac{MEMS} gas sensors~\cite{perello2021cmos}.

Advancements in sensor technology and airflow modeling can enhance \ac{MC} platforms. 
High-sensitivity VOC detection in breath analysis~\cite{jaeschke2018novel,shang2023chemiresistive} could improve biomarker detection, while \ac{PIV} and Schlieren imaging offer insights into molecular signal dynamics. 
Likewise, \ac{MC} techniques in signal modulation and interference management~\cite{koo2016molecular,qiu2016mobile} could benefit targeted drug delivery and swarm robotics. 
Cross-disciplinary collaboration can drive innovations in communication, diagnostics, and environmental monitoring.

\section{Open Challenges and Future Research Roadmap}
\label{sec:challenges}

\begin{figure}
\begin{center}
\resizebox{0.9\columnwidth}{!}
{
\begin{tikzpicture}
[mindmap,
grow cyclic,
every node/.style=concept,
concept color=white!40,
level 1/.append style={sibling angle=360/5},
]
\node [root concept] {Challenges for Life Sciences}
child [concept color=pink!40]{
node {Disease Remediation Feedback}}
child [concept color=pink!40]{
node {Biomarker Concentration and Quantification}}
child [concept color=pink!40]{
node {Binders}}
child [concept color=pink!40]{
node {Disease Specificity}}
child [concept color=pink!40]{
node {Target Group Diversity}
};
\end{tikzpicture}
}
\end{center}
\caption{Life science challenges pertaining to non-invasive diagnosis of human health anomalies by detecting exhaled breath biomarkers. 
}
\label{fig:life_science_challenges}
\vspace{-0.8em}
\end{figure}
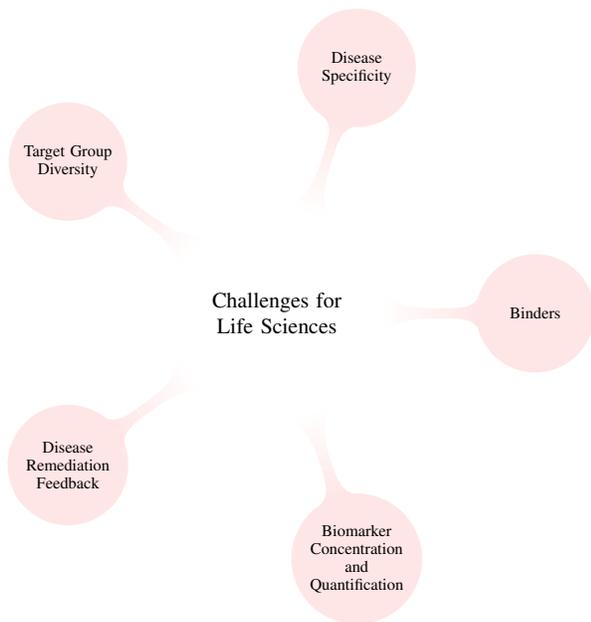
\begin{figure}
\begin{center}
\resizebox{\columnwidth}{!}
{
\begin{tikzpicture}
[mindmap,
grow cyclic,
every node/.style=concept,
concept color=white!40,
level 1/.append style={sibling angle=360/9},
]
\node [root concept] {Challenges for Engineering Sciences}
child [concept color=cyan!40]{
node {Portability of Devices}}
child [concept color=cyan!40]{
node {Sample Collection}}
child [concept color=cyan!40]{
node {Data Privacy Concerns}
}
child [concept color=cyan!40]{
node {Calibration Drift}
}
child [concept color=cyan!40]{
node {Inbound Communication}
}
child [concept color=cyan!40]{
node {Network Architecture and Protocols}
}
child [concept color=cyan!40]{
node {Real-Time Monitoring}
}
child [concept color=cyan!40]{
node {System Reliability and Resilience}
}
child [concept color=cyan!40]{
node {Stochastic Channel Modeling}
};
\end{tikzpicture}
}
\end{center}
\caption{Engineering challenges pertaining to non-invasive diagnosis of human health anomalies by detecting exhaled breath biomarkers. 
}
\label{fig:engineering_challenges}
\vspace{-0.8em}
\end{figure}
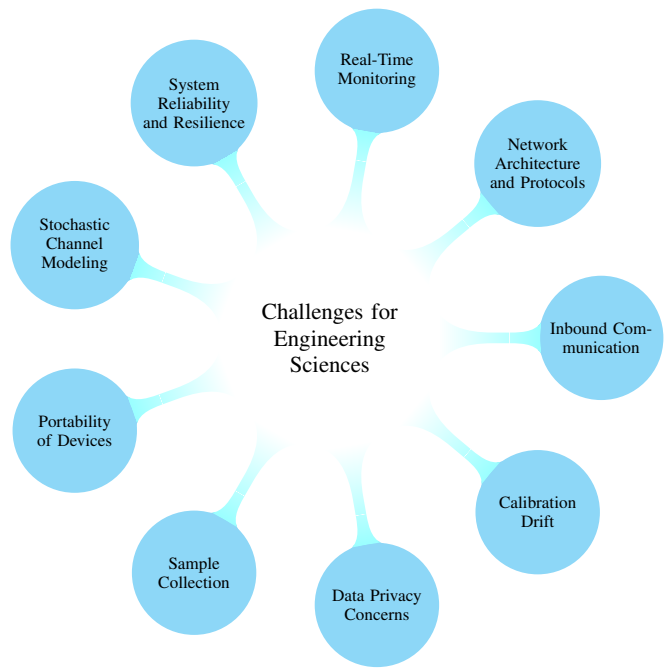

Exhaled breath contains biomarkers that provide insights into respiratory diseases, infections, and metabolic processes. 
Within the \ac{IoBT}, this data enables non-invasive detection and real-time monitoring, supporting precise diagnosis and treatment. 
However, as the field is still emerging, challenges remain for life sciences and engineering communities, from understanding biomarker complexities to overcoming hurdles in sensing, communication, and data analysis. 
This section discusses these challenges in general with illustrations in \cref{fig:life_science_challenges,fig:engineering_challenges}.

\subsection{Challenges for Life Sciences}
\label{challanges:life}

\subsubsection{Target Group Diversity}
Detecting health conditions through exhaled breath is challenging due to metabolic, physiological and environmental variability across ages, sex, and regions~\cite{fens2013exhaled,das2020non}.
Diet, lifestyle, and local disease prevalence further complicate biomarker interpretation~\cite{queiroz2021time}.
Baseline differences and co-existing conditions can obscure disease markers, while sensor calibration and machine learning models variability add complexity.
Therefore, standardizing breath collection and ensuring adaptable diagnostic models against various diversities are essential for accuracy.
    
\subsubsection{Disease Specificity}
Disease specificity in breath-based diagnosis is challenging as many inflammatory disorders, infections, and metabolic issues produce similar \acp{VOC} and byproducts.
Although the concentration levels vary, their differentiation remains complex.
For instance, $\mathrm{C}\mathrm{O}$ release makes asthma and \ac{COPD} hard to differentiate~\cite{boots2012versatile}.
Common \acp{VOC} like acetone arise from standard biological processes, complicating diagnosis~\cite{samudrala2014breath}.
Some diseases even lack distinct breath signatures, requiring additional diagnostic tools for accuracy.

\subsubsection{Binders}
Many breath diagnostic devices rely on implementing binding agents such as antibodies or aptamers to recognize pathogens specifically. 
These binders act as one of the bottlenecks in developing breath diagnostic devices. 
The emergence of new diseases and variants leads to the need to develop new reliable and exceptionally stable binders~\cite{zhang2022advanced}.

\subsubsection{Biomarker Concentration and Quantification}
Low biomarker concentrations in exhaled breath make detection challenging, requiring highly sensitive sensors.
Quantitative measurement is crucial for monitoring disease treatment and severity~\cite{ates2021integrated}.
Hence, breath sensors must detect and differentiate various pathogens (e.g., \emph{respiratory syncytial virus}, influenza) for broader health applications.

\subsubsection{Disease Remediation Feedback}
The \Ac{IoBNT} envisions swarming nanodevices for health monitoring and treatment~\cite{akyildiz2015internet}.
Challenges include detecting anomalies in exhaled breath and coordinating responses like targeted drug delivery~\cite{mills1999targeted}, toxin removal, and tissue repair.
A robust feedback system is needed to relay biomarker data, ensuring real-time adaptation while addressing bio-compatibility and control issues.

\subsection{Challenges for Engineering Sciences}
\label{challenges:engineering}

\subsubsection{Stochastic Channel Modeling} 
This survey revealed the inherent complexity and dynamic nature of the breath communication channel, driven by environmental~\cite{liangModelingVolatileOrganic2012, wongEffectsMedicalStaffs2022} and physiological~\cite{khalidModelingViralAerosol2020} variability. A primary lesson learned is that accurate modeling of turbulence, airflow dynamics~\cite{cavazzutiStatisticalAnalysisExhaled2023, bourouibaViolentExpiratoryEvents2014}, and environmental disturbances can be extremely challenging.
Hence, broad stochastic models~\cite{khalid2018system} and detailed simulations~\cite{dboukCoughingAirborneDroplet2020} are crucial for accurate biomarker detection.
Additionally, research into dynamic models~\cite{linNearfieldToolSimulating2003, steinNOAAsHYSPLITAtmospheric2015} is needed to ensure consistent and reliable diagnoses despite these varying factors.

\subsubsection{System Reliability and Resilience} 
In parallel to the channel models, we have identified several factors of instability for breath-based \ac{ABMC}, including dynamic environments~\cite{wongEffectsMedicalStaffs2022, cavazzutiStatisticalAnalysisExhaled2023}, molecular noise~\cite{pierobon2011noise, srinivas2012molecular}, sensor sensitivity~\cite{purnamadjaja2007guiding}, and differences in channel geometry~\cite{khalidModelingViralAerosol2020}.
Ensuring reliable and resilient exhaled breath monitoring and analysis requires addressing signal weakening, interference, and network disruptions, particularly at the engineered receiver side. 
Reliability translates to data integrity, fault tolerance, and consistent performance under varying conditions, while resilience allows recovery from failures and ensures long-term adaptability and operability. 
Key strategies include error detection and correction, mitigating false diagnoses, redundant data transmission, and adaptive techniques to maintain communication despite environmental challenges.

\subsubsection{Real-Time Monitoring}

Breath-based disease detection faces the challenge of real-time monitoring~\cite{das2020non,skarysz2022fast}.
Real-time health monitoring ensures communication within a pre-defined time frame, even when processing or transmission delays influence it. 
Real-time systems are based on determinism, i.e., designed to ensure a response within predefined time limits.
For example, in analyzing exhaled breath, real-time monitoring ensures the reception of specific required health information within a predefined time frame. 
Addressing these challenges requires advancements in low-latency communication protocols or edge computing to minimize delays~\cite{intharawijitr2017simulation}. 
Overall, low latency may not be sufficient for real-time operation if delays are unpredictable, sometimes exceeding the threshold, making the system unreliable for medical applications.
In fact, in many cases, detecting a disease while it is still clinically silent would be desirable. Concepts such as \textit{negative latency}~\cite{schulzNegativeLatencyTactile2024} describe the prediction of a communication system event before it occurs. This could be mapped to breath analysis to provide guarantees on high-confidence disease detection.


\subsubsection{Network Architecture and Protocols} 
We propose a macro-scale communication approach for \ac{ABMC} breath analysis. 
Integrating this system into an \ac{IoBT} requires only a few adaptations to network architectures and protocols due to the use of macroscale communication channels and commercially available sensing technology.
We believe a first practical step can be taken towards a future \ac{IoBNT}, which could still take years of research, regulation, and standardization to be commercialized.
Instead of requiring novel bio-cyber interfaces~\cite{hofmannBiologicallyInspiredProteinBased2025, chouhanInterfacingMolecularCommunication2023b}, and significant progress in creating networks from bio-nano machines~\cite{akyildiz2015internet}, the \ac{IoBT} is more easily connected to existing network infrastructure.
However, some challenges remain. Exhaled breath analysis in the \ac{IoBT} will require integrating molecular sensors, RF-based communication, and various sensors within wireless and cloud networks to improve real-time diagnostics. 
Challenges include defining architectures, standardizing protocols, ensuring interoperability, and managing data transmission, including compatibility with electronic health records~\cite{baumann2018impact} and clinical decision support systems. 
Overcoming this problem demands the establishment of universal data formats and communication protocols.
Additionally, scaling this system to millions of users requires efficient big data management, secure communication, and advanced analytics for accurate diagnostics.


\subsubsection{Inbound Communication}
This survey examines health diagnostics by detecting specific \acp{IM} and breath parameters in an outbound communication system.
We have identified the human as a transmitter of \acp{IM}, the respiratory system and surrounding space as a channel, and engineered sensors as the receivers, making diseases and conditions within the human body the communicable information.
However, a different perspective could also consider the inbound communication system, where the lungs act as a natural channel and receiver for \acp{IM} like pollen or pollutants from the outside environment, acting as the transmitter. 
The lung, specifically the alveoli, facilitates gas exchange, inhaled external \acp{IM} need to be processed by the human respiratory system, and health issues can arise during this process. 
However, inbound communication analysis could involve the detection of biomarkers inside the body to assess external influences, such as harmful exogenous gases or allergens such as pollen. 
The engineered system now resides inside the human body, requiring miniaturized bio-compatible sensors.
This would bridge the gap between an \ac{IoBT} and \ac{IoBNT} adding significantly to the system complexity.
Reliable molecular data transmission, and accurate correlations between biomarkers and external gases within this type of system add another significant dimension that should be the focus of a future survey.


\subsubsection{Calibration Drift}
Breath biomarker sensors face calibration drift due to aging, environment, and contamination~\cite{padilla2010drift}.
Advancements in sensor materials, design, and self-calibration algorithms are essential to maintaining accuracy.


\subsubsection{Data Privacy Concerns} 
The vast selection of identifiable diseases and conditions we discussed in this survey underscores significant privacy risks associated with the inherently sensitive nature of breath biomarker data. 
Intercepting exhaled breath can expose sensitive health data, posing privacy risks.
Patient confidentiality and data security must be integrated at all stages, from sensor design to data management and transmission protocols, to prevent tampering that could lead to data leaks or false diagnoses.
Breath analysis may also reveal genetic material from pathogens~\cite{koblentz2003pathogens}, raising security concerns.
Malicious actors could exploit stolen data for targeted bio-threats.
Strong encryption, restricted access, data anonymization, and strict biosecurity protocols are essential to mitigate these risks.

\subsubsection{Sample Collection}
To detect biomarkers from exhaled breath, air samples are collected using filters or condensers.
A challenge is the presence of interfering particles such as dust or pathogens~\cite{bhardwaj2021recent}, which can cause filter clogging, false readings, or contamination, compromising accuracy.
The low concentration of biomarkers requires large sample volumes, leading to long sampling times, reducing practicality for widespread use.
Improved sampling methods are needed to address these issues.

\subsubsection{Portability of Devices}
Sensors and devices for exhaled breath analysis must be efficient, robust, cost-effective, and portable for \ac{POC} use.
Future efforts should focus on applying novel technologies for \emph{in situ} detection in clinical settings~\cite{bhardwaj2021recent}.
To ensure practicality, sensors and setups should be reusable, user-friendly, and operate without laboratory equipment.

\section{Conclusion}
\label{sec:conclusion}

This work provides an extensive survey of exhaled breath through the lens of \ac{MC}, highlighting key biomarkers, physical properties, and spatial origins within the human respiratory system.
Detailed models of particle generation from coughing, sneezing, and breathing are discussed, alongside key propagation mechanisms such as diffusion, advection, gravitational forces, and buoyancy within \ac{ABMC} channels.
Additionally, the survey reviews engineered receiver techniques, including breath sampling methods, sensor technologies, and decision-making processes for accurate disease detection.

The integration of \ac{ABMC} with exhaled breath is a major research topic linking engineered communication systems and biological entities to create interactive health-monitoring networks. 
Leveraging natural physiological processes and advanced sensor technologies could enhance personalized healthcare, improve patient outcomes, and facilitate more accessible and efficient diagnostics.
Despite highlighting significant advancements, the survey also outlines ongoing challenges and open research directions in modeling accuracy, system robustness, environmental interference management, and practical testbed implementations.
Standardization efforts and interdisciplinary collaborations between \ac{MC} researchers, medical experts, and engineers will be crucial for translating findings into clinical practice.
By systematically addressing these challenges, the application of breath analysis within an \ac{IoBT} framework will advance non-invasive diagnostic methodologies and personalized healthcare solutions.


\bibliographystyle{IEEEtran}
\bibliography{output.bbl}


\end{document}

%% file: respiration_new.pdf_tex
\begingroup%
  \makeatletter%
  \providecommand\color[2][]{%
    \errmessage{(Inkscape) Color is used for the text in Inkscape, but the package 'color.sty' is not loaded}%
    \renewcommand\color[2][]{}%
  }%
  \providecommand\transparent[1]{%
    \errmessage{(Inkscape) Transparency is used (non-zero) for the text in Inkscape, but the package 'transparent.sty' is not loaded}%
    \renewcommand\transparent[1]{}%
  }%
  \providecommand\rotatebox[2]{#2}%
  \newcommand*\fsize{\dimexpr\f@size pt\relax}%
  \newcommand*\lineheight[1]{\footnotesize\fontsize{\fsize}{#1\fsize}\selectfont}%
  \ifx\svgwidth\undefined%
    \setlength{\unitlength}{992.12598425bp}%
    \ifx\svgscale\undefined%
      \relax%
    \else%
      \setlength{\unitlength}{\unitlength * \real{\svgscale}}%
    \fi%
  \else%
    \setlength{\unitlength}{\svgwidth}%
  \fi%
  \global\let\svgwidth\undefined%
  \global\let\svgscale\undefined%
  \makeatother%
  \begin{picture}(1,0.25714286)%
    \lineheight{1}%
    \setlength\tabcolsep{0pt}%
    \put(0,0){\includegraphics[width=\unitlength,page=1]{respiration_new.pdf}}%
    \put(0.46516744,0.20056199){\color[rgb]{0,0,0}\transparent{0.871369}\makebox(0,0)[t]{\lineheight{1.29999995}\smash{\begin{tabular}[t]{c}Buoyancy\end{tabular}}}}%
    \put(0.49533218,0.07483893){\color[rgb]{0,0,0}\transparent{0.871369}\makebox(0,0)[t]{\lineheight{1.29999995}\smash{\begin{tabular}[t]{c}Gravity\end{tabular}}}}%
    \put(0,0){\includegraphics[width=\unitlength,page=2]{respiration_new.pdf}}%
    \put(0.53812527,0.09700778){\color[rgb]{0,0,0}\transparent{0.871369}\makebox(0,0)[t]{\lineheight{1.29999995}\smash{\begin{tabular}[t]{c}Obstacle\end{tabular}}}}%
    \put(0,0){\includegraphics[width=\unitlength,page=3]{respiration_new.pdf}}%
    \put(0.40742218,0.07606554){\color[rgb]{0,0,0}\transparent{0.871369}\makebox(0,0)[t]{\lineheight{1.29999995}\smash{\begin{tabular}[t]{c}Cross-Drift\end{tabular}}}}%
    \put(0.43078287,0.12609511){\color[rgb]{0,0,0}\transparent{0.871369}\makebox(0,0)[t]{\lineheight{1.29999995}\smash{\begin{tabular}[t]{c}Advection-Diffusion\end{tabular}}}}%
    \put(0.69876311,0.1060944){\color[rgb]{0,0,0}\transparent{0.871369}\makebox(0,0)[t]{\lineheight{1.29999995}\smash{\begin{tabular}[t]{c}Detector/\\Demodulator\end{tabular}}}}%
    \put(0.83180963,0.1060944){\color[rgb]{0,0,0}\transparent{0.871369}\makebox(0,0)[t]{\lineheight{1.29999995}\smash{\begin{tabular}[t]{c}Channel Decoder/\\Source Decoder\end{tabular}}}}%
    \put(0,0){\includegraphics[width=\unitlength,page=4]{respiration_new.pdf}}%
    \put(0.94549665,0.10604639){\color[rgb]{0,0,0}\transparent{0.871369}\makebox(0,0)[t]{\lineheight{1.29999995}\smash{\begin{tabular}[t]{c}Information\\Sink\end{tabular}}}}%
    \put(0,0){\includegraphics[width=\unitlength,page=5]{respiration_new.pdf}}%
    \put(0.05158294,0.1060944){\color[rgb]{0,0,0}\transparent{0.871369}\makebox(0,0)[t]{\lineheight{1.29999995}\smash{\begin{tabular}[t]{c}Information\\Source\end{tabular}}}}%
    \put(0.0599034,0.22836542){\color[rgb]{0,0,0}\transparent{0.871369}\makebox(0,0)[t]{\lineheight{1.29999995}\smash{\begin{tabular}[t]{c}\textbf{Transmitter}\end{tabular}}}}%
    \put(0.40613003,0.22836542){\color[rgb]{0,0,0}\transparent{0.871369}\makebox(0,0)[t]{\lineheight{1.29999995}\smash{\begin{tabular}[t]{c}\textbf{Channel}\end{tabular}}}}%
    \put(0.64974577,0.22836051){\color[rgb]{0,0,0}\transparent{0.871369}\makebox(0,0)[t]{\lineheight{1.29999995}\smash{\begin{tabular}[t]{c}\textbf{Receiver}\end{tabular}}}}%
    \put(0,0){\includegraphics[width=\unitlength,page=6]{respiration_new.pdf}}%
    \put(0.16271903,0.1060944){\color[rgb]{0,0,0}\transparent{0.871369}\makebox(0,0)[t]{\lineheight{1.29999995}\smash{\begin{tabular}[t]{c}Source \\Encoding\end{tabular}}}}%
    \put(0,0){\includegraphics[width=\unitlength,page=7]{respiration_new.pdf}}%
    \put(0.2860156,0.1060944){\color[rgb]{0,0,0}\transparent{0.871369}\makebox(0,0)[t]{\lineheight{1.29999995}\smash{\begin{tabular}[t]{c}Channel Encoding/\\Modulation\end{tabular}}}}%
    \put(0,0){\includegraphics[width=\unitlength,page=8]{respiration_new.pdf}}%
  \end{picture}%
\endgroup%

%% file: main.bbl
\begin{thebibliography}{100}
\providecommand{\url}[1]{#1}
\csname url@samestyle\endcsname
\providecommand{\newblock}{\relax}
\providecommand{\bibinfo}[2]{#2}
\providecommand{\BIBentrySTDinterwordspacing}{\spaceskip=0pt\relax}
\providecommand{\BIBentryALTinterwordstretchfactor}{4}
\providecommand{\BIBentryALTinterwordspacing}{\spaceskip=\fontdimen2\font plus
\BIBentryALTinterwordstretchfactor\fontdimen3\font minus \fontdimen4\font\relax}
\providecommand{\BIBforeignlanguage}[2]{{%
\expandafter\ifx\csname l@#1\endcsname\relax
\typeout{** WARNING: IEEEtran.bst: No hyphenation pattern has been}%
\typeout{** loaded for the language `#1'. Using the pattern for}%
\typeout{** the default language instead.}%
\else
\language=\csname l@#1\endcsname
\fi
#2}}
\providecommand{\BIBdecl}{\relax}
\BIBdecl

\bibitem{amann2007breath}
A.~Amann, P.~Spanel, and D.~Smith, ``Breath analysis: {The} approach towards clinical applications,'' \emph{Mini-Rev. in Med. Chem.}, vol.~7, no.~2, pp. 115--129, Feb. 2007.

\bibitem{Lotter2023experimental_II}
S.~Lotter \emph{et~al.}, ``Experimental research in synthetic molecular communications – {Part II},'' \emph{IEEE Nanotechnol. Mag.}, vol.~17, no.~3, pp. 54--65, Apr. 2023.

\bibitem{khalidCommunicationBreathAerosol2019}
M.~Khalid, O.~Amin, S.~Ahmed, B.~Shihada, and M.-S. Alouini, ``Communication through breath: {Aerosol} transmission,'' \emph{IEEE Commun. Mag.}, vol.~57, no.~2, pp. 33--39, Feb. 2019.

\bibitem{Paleczek2021artificial}
A.~Paleczek, D.~Grochala, and A.~Rydosz, ``Artificial breath classification using {XGBoost} algorithm for diabetes detection,'' \emph{Sensors}, vol.~21, no.~12, p. {Art. no. 4187}, Jun. 2021.

\bibitem{farsad2016comprehensive}
N.~Farsad, H.~B. Yilmaz, A.~Eckford, C.-B. Chae, and W.~Guo, ``A comprehensive survey of recent advancements in molecular communication,'' \emph{IEEE Commun. Surv. Tuts.}, vol.~18, no.~3, pp. 1887--1919, Feb. 2016.

\bibitem{Kuran2020survey}
M.~S. Kuran, H.~B. Yilmaz, I.~Demirkol, N.~Farsad, and A.~Goldsmith, ``A survey on modulation techniques in molecular communication via diffusion,'' \emph{IEEE Commun. Surv. Tuts.}, vol.~23, no.~1, pp. 7--28, Dec. 2020.

\bibitem{Hofmann2023coding}
P.~Hofmann, J.~A. Cabrera, R.~Bassoli, M.~Reisslein, and F.~H.~P. Fitzek, ``Coding in diffusion-based molecular nanonetworks: {A} comprehensive survey,'' \emph{IEEE Access}, vol.~11, pp. 16\,411--16\,465, Feb. 2023.

\bibitem{jamaliChannelModelingDiffusive2019}
V.~Jamali, A.~Ahmadzadeh, W.~Wicke, A.~Noel, and R.~Schober, ``Channel modeling for diffusive molecular communication---{A} tutorial review,'' \emph{Proc. IEEE}, vol. 107, no.~7, pp. 1256--1301, Jul. 2019.

\bibitem{aktas2024odor}
D.~Aktas \emph{et~al.}, ``Odor-based molecular communications: {State}-of-the-art, vision, challenges, and frontier directions,'' \emph{IEEE Commun. Surv. Tuts.}, pp. 1--1, Oct. 2024, early access.

\bibitem{gulec2020molecular}
F.~Gulec and B.~Atakan, ``A molecular communication perspective on airborne pathogen transmission and reception via droplets generated by coughing and sneezing,'' \emph{IEEE Trans. Mol. Biol. Multi-Scale Commun.}, vol.~7, no.~3, pp. 175--184, May 2021.

\bibitem{barros2021molecular}
M.~T. Barros, M.~Veleti{\'c}, M.~Kanada, M.~Pierobon, S.~Vainio, I.~Balasingham, and S.~Balasubramaniam, ``Molecular communications in viral infections research: {Modeling}, experimental data, and future directions,'' \emph{IEEE Trans. Mol. Biol. Multi-Scale Commun.}, vol.~7, no.~3, pp. 121--141, Apr. 2021.

\bibitem{schurwanz2021duality}
M.~Schurwanz, P.~A. Hoeher, S.~Bhattacharjee, M.~Damrath, L.~Stratmann, and F.~Dressler, ``Duality between coronavirus transmission and air-based macroscopic molecular communication,'' \emph{IEEE Trans. Mol. Biol. Multi-Scale Commun.}, vol.~7, no.~3, pp. 200--208, Apr. 2021.

\bibitem{Lotter2023experimental_I}
S.~Lotter \emph{et~al.}, ``Experimental research in synthetic molecular communications – {Part I},'' \emph{IEEE Nanotechnol. Mag.}, vol.~17, no.~3, pp. 42--53, Apr. 2023.

\bibitem{bhattacharjee2022digital}
S.~Bhattacharjee, M.~Damrath, L.~Stratmann, P.~A. Hoeher, and F.~Dressler, ``Digital communication techniques in macroscopic air-based molecular communication,'' \emph{IEEE Trans. Mol. Biol. Multi-Scale Commun.}, vol.~8, no.~4, pp. 276--291, Nov. 2022.

\bibitem{nakano2013molecular}
T.~Nakano, A.~W. Eckford, and T.~Haraguchi, \emph{Molecular communication}.\hskip 1em plus 0.5em minus 0.4em\relax Cambridge University Press, 2013.

\bibitem{farsad2013tabletop}
N.~Farsad, W.~Guo, and A.~W. Eckford, ``Tabletop molecular communication: {Text} messages through chemical signals,'' \emph{PloS One}, vol.~8, no.~12, Dec. 2013, {Art. no. e82935}.

\bibitem{xiao2023what}
H.~Xiao, K.~Dokaj, and O.~B. Akan, ``What really is “molecule” in molecular communications? {The} quest for physics of particle-based information carriers,'' \emph{IEEE Trans. Mol. Biol. Multi-Scale Commun.}, vol.~10, no.~1, pp. 43--74, Dec. 2023.

\bibitem{akyildiz2015internet}
I.~F. Akyildiz, M.~Pierobon, S.~Balasubramaniam, and Y.~Koucheryavy, ``{The Internet of Bio-Nano Things},'' \emph{IEEE Commun. Mag.}, vol.~53, no.~3, pp. 32--40, Mar. 2015.

\bibitem{dressler2015connecting}
F.~Dressler and S.~Fischer, ``{Connecting In-Body Nano Communication with Body Area Networks: Challenges and Opportunities of the Internet of Nano Things},'' \emph{Nano Communic. Netwo.}, vol.~6, pp. 29--38, 6 2015.

\bibitem{hofmannBiologicallyInspiredProteinBased2025}
P.~Hofmann, J.~A. Cabrera, G.~Schulte, and F.~H.~P. Fitzek, ``A {{Biologically Inspired}} and {{Protein-Based Bio-Cyber Interface}} for the {{Internet}} of {{Bio-Nano Things}},'' \emph{Biosens. and Bioelectr.: X}, vol.~22, p. 100572, Mar. 2025.

\bibitem{chouhanInterfacingMolecularCommunication2023b}
L.~Chouhan and M.-S. Alouini, ``Interfacing of {{Molecular Communication System With Various Communication Systems Over Internet}} of {{Every Nano Things}},'' \emph{IEEE Internet of Things Journal}, vol.~10, no.~16, pp. 14\,552--14\,568, Aug. 2023.

\bibitem{laddomada2015crosstalk}
M.~Laddomada and M.~Pierobon, ``A crosstalk-based linear filter in biochemical signal transduction pathways for the {Internet of Bio-Things},'' in \emph{Proc. IEEE Int. Conf. on Acoust., Speech and Signal Process. (ICASSP)}, Aug. 2015, pp. 5520--5524.

\bibitem{owen1982acetone}
O.~Owen \emph{et~al.}, ``Acetone metabolism during diabetic ketoacidosis,'' \emph{Diabetes}, vol.~31, no.~3, pp. 242--248, Mar. 1982.

\bibitem{prabhakar2015acetone}
A.~Prabhakar \emph{et~al.}, ``Acetone as biomarker for ketosis buildup capability-{A} study in healthy individuals under combined high fat and starvation diets,'' \emph{Nutr. J.}, vol.~14, pp. 1--11, Apr. 2015.

\bibitem{kunczik2022breathing}
J.~Kunczik, K.~Hubbermann, L.~M{\"o}sch, A.~Follmann, M.~Czaplik, and C.~Barbosa~Pereira, ``Breathing pattern monitoring by using remote sensors,'' \emph{Sensors}, vol.~22, no.~22, Nov. 2022, {Art. no. 8854}.

\bibitem{kumar2022deep}
A.~K. Kumar, M.~Ritam, L.~Han, S.~Guo, and R.~Chandra, ``Deep learning for predicting respiratory rate from biosignals,'' \emph{Comput. in Biol. and Med.}, vol. 144, May 2022, {Art. no. 105338}.

\bibitem{ye2021recent}
Z.~Ye, Y.~Liu, and Q.~Li, ``Recent progress in smart electronic nose technologies enabled with machine learning methods,'' \emph{Sensors}, vol.~21, no.~22, Nov. 2021, {Art. no. 7620}.

\bibitem{baumann2018impact}
L.~A. Baumann, J.~Baker, and A.~G. Elshaug, ``The impact of electronic health record systems on clinical documentation times: {A} systematic review,'' \emph{Heal. Polic.}, vol. 122, no.~8, pp. 827--836, Aug. 2018.

\bibitem{randall2021did}
K.~Randall, E.~T. Ewing, L.~C. Marr, J.~L. Jimenez, and L.~Bourouiba, ``How did we get here: {What} are droplets and aerosols and how far do they go? {A} historical perspective on the transmission of respiratory infectious diseases,'' \emph{Interface Focus.}, vol.~11, no.~6, Oct. 2021, {Art. no. 20210049}.

\bibitem{dallinga2010volatile}
J.~Dallinga, C.~Robroeks, J.~Van~Berkel, E.~Moonen, R.~Godschalk, Q.~J{\"o}bsis, E.~Dompeling, E.~Wouters, and F.~Van~Schooten, ``Volatile organic compounds in exhaled breath as a diagnostic tool for asthma in children,'' \emph{Clin. \& Exp. Allergy}, vol.~40, no.~1, pp. 68--76, Jan. 2010.

\bibitem{besa2015exhaled}
V.~Besa \emph{et~al.}, ``Exhaled volatile organic compounds discriminate patients with chronic obstructive pulmonary disease from healthy subjects,'' \emph{Int. J. of Chronic Obstr. Pulm. Dis.}, pp. 399--406, Feb. 2015.

\bibitem{chang2018analysis}
J.-E. Chang \emph{et~al.}, ``Analysis of volatile organic compounds in exhaled breath for lung cancer diagnosis using a sensor system,'' \emph{Sensors and Actuators B: Chem.}, vol. 255, pp. 800--807, Feb. 2018.

\bibitem{saasa2018sensing}
V.~Saasa, T.~Malwela, M.~Beukes, M.~Mokgotho, C.-P. Liu, and B.~Mwakikunga, ``Sensing technologies for detection of acetone in human breath for diabetes diagnosis and monitoring,'' \emph{Diagn.}, vol.~8, no.~1, Jan. 2018, {Art. no. 12}.

\bibitem{alkhouri2014analysis}
N.~Alkhouri \emph{et~al.}, ``Analysis of breath volatile organic compounds as a noninvasive tool to diagnose nonalcoholic fatty liver disease in children,'' \emph{Eur. j. of Gastroenterol. \& Hepatol.}, vol.~26, no.~1, pp. 82--87, Jan. 2014.

\bibitem{zetola2017diagnosis}
N.~M. Zetola \emph{et~al.}, ``Diagnosis of pulmonary tuberculosis and assessment of treatment response through analyses of volatile compound patterns in exhaled breath samples,'' \emph{J. of Infect.}, vol.~74, no.~4, pp. 367--376, Apr. 2017.

\bibitem{van2018potential}
P.~M. van Oort \emph{et~al.}, ``The potential role of exhaled breath analysis in the diagnostic process of pneumonia—{A} systematic review,'' \emph{J. Breath Res.}, vol.~12, no.~2, Feb. 2018, {Art. no. 024001}.

\bibitem{bach2015measuring}
J.-P. Bach \emph{et~al.}, ``Measuring compounds in exhaled air to detect {Alzheimer's disease and Parkinson’s} disease,'' \emph{PloS One}, vol.~10, no.~7, Jul. 2015, {Art. no. e0132227}.

\bibitem{carroll2022increased}
G.~T. Carroll, D.~L. Kirschman, and A.~Mammana, ``Increased {CO\textsubscript{2}} levels in the operating room correlate with the number of healthcare workers present: {An} imperative for intentional crowd control,'' \emph{Patient Saf. in Surg.}, vol.~16, no.~1, Nov. 2022, {Art. no. 35}.

\bibitem{simonds2013chronic}
A.~K. Simonds, ``Chronic hypoventilation and its management,'' \emph{Eur. Respir. Rev.}, vol.~22, no. 129, pp. 325--332, Sep. 2013.

\bibitem{brashear1983hyperventilation}
R.~E. Brashear, ``Hyperventilation syndrome,'' \emph{Lung}, vol. 161, pp. 257--273, Dec. 1983.

\bibitem{petersson2014gas}
J.~Petersson and R.~W. Glenny, ``Gas exchange and ventilation--{Perfusion} relationships in the lung,'' \emph{Eur. Respir. J.}, vol.~44, no.~4, pp. 1023--1041, Oct. 2014.

\bibitem{banik2016hydrogen}
G.~D. Banik \emph{et~al.}, ``Hydrogen sulphide in exhaled breath: {A} potential biomarker for small intestinal bacterial overgrowth in {IBS},'' \emph{J. Breath Res.}, vol.~10, no.~2, May 2016, {Art. no. 026010}.

\bibitem{preti2009volatile}
G.~Preti, E.~Thaler, C.~W. Hanson, M.~Troy, J.~Eades, and A.~Gelperin, ``Volatile compounds characteristic of sinus-related bacteria and infected sinus mucus: {Analysis} by solid-phase microextraction and gas chromatography--mass spectrometry,'' \emph{J. of Chromatogr. B}, vol. 877, no.~22, pp. 2011--2018, Jul. 2009.

\bibitem{popov2011human}
T.~A. Popov, ``Human exhaled breath analysis,'' \emph{Ann. of Allergy, Asthma \& Immunol.}, vol. 106, no.~6, pp. 451--456, Jun. 2011.

\bibitem{zhang2010exhaled}
J.~Zhang \emph{et~al.}, ``Exhaled carbon monoxide in asthmatics: {A} meta-analysis,'' \emph{Respir. Res.}, vol.~11, pp. 1--10, Apr. 2010.

\bibitem{lefferts2022ammonia}
M.~J. Lefferts and M.~R. Castell, ``Ammonia breath analysis,'' \emph{Sensors \& Diagn.}, vol.~1, no.~5, pp. 955--967, Aug. 2022.

\bibitem{hibbard2011breath}
T.~Hibbard and A.~J. Killard, ``Breath ammonia analysis: {Clinical} application and measurement,'' \emph{Crit. Rev. in Anal. Chem.}, vol.~41, no.~1, pp. 21--35, Feb. 2011.

\bibitem{essiet2013diagnosis}
I.~O. Essiet, ``Diagnosis of kidney failure by analysis of the concentration of ammonia in exhaled human breath,'' \emph{J. of Emerg. Trends in Eng. and Appl. Sci.}, vol.~4, no.~6, pp. 859--862, Dec. 2013.

\bibitem{boots2012versatile}
A.~W. Boots, J.~J. van Berkel, J.~W. Dallinga, A.~Smolinska, E.~F. Wouters, and F.~J. van Schooten, ``The versatile use of exhaled volatile organic compounds in human health and disease,'' \emph{J. Breath Res.}, vol.~6, no.~2, Jun. 2012, {Art. no. 027108}.

\bibitem{brindicci2005exhaled}
C.~Brindicci, K.~Ito, O.~Resta, N.~Pride, P.~Barnes, and S.~Kharitonov, ``Exhaled nitric oxide from lung periphery is increased in {COPD},'' \emph{Eur. Respir. J.}, vol.~26, no.~1, pp. 52--59, Jul. 2005.

\bibitem{murata2014hydrogen}
K.~Murata, K.~Fujimoto, Y.~Kitaguchi, T.~Horiuchi, K.~Kubo, and T.~Honda, ``Hydrogen peroxide content and {pH} of expired breath condensate from patients with asthma and {COPD},'' \emph{COPD: J. of Chronic Obstr. Pulm. Dis.}, vol.~11, no.~1, pp. 81--87, Feb. 2014.

\bibitem{morris1981tidal}
M.~Morris and D.~Lane, ``Tidal expiratory flow patterns in airflow obstruction,'' \emph{Thorax}, vol.~36, no.~2, pp. 135--142, Feb. 1981.

\bibitem{ahmed2023microbial}
W.~M. Ahmed \emph{et~al.}, ``Microbial volatiles as diagnostic biomarkers of bacterial lung infection in mechanically ventilated patients,'' \emph{Clin. Infect. Dis.}, vol.~76, no.~6, pp. 1059--1066, Mar. 2023.

\bibitem{sawano2021rt}
M.~Sawano, K.~Takeshita, H.~Ohno, and H.~Oka, ``{RT-PCR diagnosis of COVID-19 from exhaled breath condensate: A clinical study},'' \emph{J. Breath Res.}, vol.~15, no.~3, Jun. 2021, {Art. no. 037103}.

\bibitem{popov2017added}
T.~A. Popov, T.~Z. Kralimarkova, M.~Labor, and D.~Plavec, ``The added value of exhaled breath temperature in respiratory medicine,'' \emph{J. Breath Res.}, vol.~11, no.~3, Aug. 2017, {Art. no. 034001}.

\bibitem{turcios2020cystic}
N.~L. Turcios, ``Cystic fibrosis lung disease: {An} overview,'' \emph{Respir. care}, vol.~65, no.~2, pp. 233--251, Feb. 2020.

\bibitem{ahookhoshDevelopmentHumanRespiratory2020}
K.~Ahookhosh, O.~Pourmehran, H.~Aminfar, M.~Mohammadpourfard, M.~M. Sarafraz, and H.~Hamishehkar, ``Development of human respiratory airway models: {{A}} review,'' \emph{Eur. J. of Pharm. Sci.}, vol. 145, Mar. 2020, {Art. no. 105233}.

\bibitem{barnes2019diseases}
L.~Barnes, ``Diseases of the nasal cavity, paranasal sinuses, and nasopharynx,'' in \emph{Surg. Pathol. of the Head and Neck}, 2019, pp. 353--432.

\bibitem{frieri1998nitric}
M.~Frieri, ``Nitric oxide in allergic rhinitis and asthma,'' in \emph{Allergy and Asthma Proc.}, vol.~19, no.~6, Dec. 1998, {Art. no. 349}.

\bibitem{wang2008analysis}
T.~Wang, A.~Pysanenko, K.~Dryahina, P.~{\v{S}}pan{\v{e}}l, and D.~Smith, ``Analysis of breath, exhaled via the mouth and nose, and the air in the oral cavity,'' \emph{J. Breath Res.}, vol.~2, no.~3, Sep. 2008, {Art. no. 037013}.

\bibitem{robbins1996measurement}
R.~A. Robbins \emph{et~al.}, ``Measurement of exhaled nitric oxide by three different techniques,'' \emph{Am. J. Respir. Crit. Care Med.}, vol. 153, no.~5, pp. 1631--1635, May 1996.

\bibitem{eckert2007central}
D.~J. Eckert, A.~S. Jordan, P.~Merchia, and A.~Malhotra, ``Central sleep apnea: {Pathophysiology} and treatment,'' \emph{Chest}, vol. 131, no.~2, pp. 595--607, Feb. 2007.

\bibitem{nogues2008abnormalities}
M.~A. Nogu{\'e}s and E.~Benarroch, ``Abnormalities of respiratory control and the respiratory motor unit,'' \emph{The Neurol.}, vol.~14, no.~5, pp. 273--288, Sep. 2008.

\bibitem{boezen1994distribution}
H.~Boezen, J.~Schouten, D.~Postma, and B.~Rijcken, ``Distribution of peak expiratory flow variability by age, gender and smoking habits in a random population sample aged 20-70 yrs,'' \emph{Eur. Respir. J.}, vol.~7, no.~10, pp. 1814--1820, Oct. 1994.

\bibitem{hong1971alveolar}
S.~Hong \emph{et~al.}, ``Alveolar gas exchanges and cardiovascular functions during breath holding with air,'' \emph{J. of Appl. Physiol.}, vol.~30, no.~4, pp. 540--547, Apr. 1971.

\bibitem{whitelaw1987effect}
W.~Whitelaw, B.~McBride, and G.~Ford, ``Effect of lung volume on breath holding,'' \emph{J. of Appl. Physiol.}, vol.~62, no.~5, pp. 1962--1969, May 1987.

\bibitem{fowler1954breaking}
W.~S. Fowler, ``Breaking point of breath-holding,'' \emph{J. of Appl. Physiol.}, vol.~6, no.~9, pp. 539--545, Mar. 1954.

\bibitem{morawska2022physics}
L.~Morawska, G.~Buonanno, A.~Mikszewski, and L.~Stabile, ``The physics of respiratory particle generation, fate in the air, and inhalation,'' \emph{Nat. Rev. Phys.}, vol.~4, no.~11, pp. 723--734, Aug. 2022.

\bibitem{bruderer2019line}
T.~Bruderer, T.~Gaisl, M.~T. Gaugg, N.~Nowak, B.~Streckenbach, S.~M{\"u}ller, A.~Moeller, M.~Kohler, and R.~Zenobi, ``On-line analysis of exhaled breath: {Focus} review,'' \emph{Chem. Rev.}, vol. 119, no.~19, pp. 10\,803--10\,828, Oct. 2019.

\bibitem{cross1994oxidants}
C.~E. Cross, A.~van~der Vliet, C.~A. O'Neill, S.~Louie, and B.~Halliwell, ``Oxidants, antioxidants, and respiratory tract lining fluids,'' \emph{Environ. Heal. Perspect.}, vol. 102, no. Art. no. Suppl. 10, pp. 185--191, Dec. 1994.

\bibitem{mittal2020flow}
R.~Mittal, R.~Ni, and J.-H. Seo, ``The flow physics of {COVID-19},'' \emph{J. of Fluid Mech.}, vol. 894, May 2020, {Art. no. F2}.

\bibitem{castleman1931mechanism}
R.~Castleman, ``The mechanism of the atomization of liquids,'' \emph{Bur. of Stand. J. of Res.}, vol.~6, no.~3, pp. 369--376, Mar. 1931.

\bibitem{bake2019exhaled}
B.~Bake, P.~Larsson, G.~Ljungkvist, E.~Ljungstr{\"o}m, and A.-C. Olin, ``Exhaled particles and small airways,'' \emph{Respir. Res.}, vol.~20, pp. 1--14, Jan. 2019.

\bibitem{mittal2013fluid}
R.~Mittal, B.~D. Erath, and M.~W. Plesniak, ``Fluid dynamics of human phonation and speech,'' \emph{Annu. Rev. of Fluid Mech.}, vol.~45, no.~1, pp. 437--467, Oct. 2012.

\bibitem{riley1978airborne}
E.~Riley, G.~Murphy, and R.~Riley, ``Airborne spread of measles in a suburban elementary school,'' \emph{Am. J. Epidemiol.}, vol. 107, no.~5, pp. 421--432, May 1978.

\bibitem{wells1955airborne}
W.~F. Wells, ``Airborne contagion and air hygiene: {An} ecological study of droplet infections,'' \emph{J. of the Am. Med. Assoc.}, vol. 159, no.~1, Sep. 1955, {Art. no. 90}.

\bibitem{khalid2020modeling}
M.~Khalid, O.~Amin, S.~Ahmed, B.~Shihada, and M.-S. Alouini, ``Modeling of viral aerosol transmission and detection,'' \emph{IEEE Trans. Commun.}, vol.~68, no.~8, pp. 4859--4873, May 2020.

\bibitem{amin2021viral}
O.~Amin, H.~Dahrouj, N.~Almayouf, T.~Y. Al-Naffouri, B.~Shihada, and M.-S. Alouini, ``Viral aerosol concentration characterization and detection in bounded environments,'' \emph{IEEE Trans. Mol. Biol. Multi-Scale Commun.}, vol.~7, no.~3, pp. 185--199, May 2021.

\bibitem{lau2022predicting}
Z.~Lau, I.~M. Griffiths, A.~English, and K.~Kaouri, ``Predicting the spatio-temporal infection risk in indoor spaces using an efficient airborne transmission model,'' \emph{Proc. R. Soc. A}, vol. 478, no. 2259, Mar. 2022, {Art. no. 20210383}.

\bibitem{gupta2009flow}
J.~K. Gupta, C.-H. Lin, and Q.~Chen, ``Flow dynamics and characterization of a cough,'' \emph{Indoor air}, vol.~19, no.~6, pp. 517--525, Dec. 2009.

\bibitem{vansciver2011particle}
M.~VanSciver, S.~Miller, and J.~Hertzberg, ``Particle image velocimetry of human cough,'' \emph{Aerosol Sci. and Technol.}, vol.~45, no.~3, pp. 415--422, Jan. 2011.

\bibitem{guptaCharacterizingExhaledAirflow2010}
J.~K. Gupta, C.-H. Lin, and Q.~Chen, ``Characterizing exhaled airflow from breathing and talking,'' \emph{Indoor Air}, vol.~20, no.~1, pp. 31--39, 2010.

\bibitem{xuHumanExhalationCharacterization2017}
C.~Xu, P.~V. Nielsen, L.~Liu, R.~L. Jensen, and G.~Gong, ``Human exhalation characterization with the aid of schlieren imaging technique,'' \emph{Build. and Environ.}, vol. 112, pp. 190--199, Feb. 2017.

\bibitem{chen2022detection}
X.~Chen, M.~Wen, F.~Ji, Y.~Huang, Y.~Tang, and A.~W. Eckford, ``Detection interval of aerosol propagation from the perspective of molecular communication: {How} long is enough?'' \emph{IEEE J. Sel. Areas Commun.}, vol.~40, no.~11, pp. 3255--3270, Oct. 2022.

\bibitem{turner2006longitudinal1}
C.~Turner, P.~{\v{S}}pan{\v{e}}l, and D.~Smith, ``A longitudinal study of ammonia, acetone and propanol in the exhaled breath of 30 subjects using selected ion flow tube mass spectrometry, {SIFT-MS},'' \emph{Physiol. Meas.}, vol.~27, no.~4, Apr. 2006, {Art. no. 321}.

\bibitem{turner2006longitudinal2}
------, ``A longitudinal study of methanol in the exhaled breath of 30 healthy volunteers using selected ion flow tube mass spectrometry, {SIFT-MS},'' \emph{Physiol. Meas.}, vol.~27, no.~7, Jul. 2006, {Art. no. 637}.

\bibitem{turner2006longitudinal3}
------, ``A longitudinal study of ethanol and acetaldehyde in the exhaled breath of healthy volunteers using selected-ion flow-tube mass spectrometry,'' \emph{Rapid Commun. in Mass Spectrom.}, vol.~20, no.~1, pp. 61--68, Nov. 2005.

\bibitem{turner2005longitudinal}
------, ``A longitudinal study of breath isoprene in healthy volunteers using selected ion flow tube mass spectrometry {(SIFT-MS)},'' \emph{Physiol. Meas.}, vol.~27, no.~1, Jan. 2006, {Art. no. 13}.

\bibitem{lotter2021statistical}
S.~Lotter, L.~Brand, M.~Sch{\"a}fer, and R.~Schober, ``Statistical modeling of airborne virus transmission through imperfectly fitted face masks,'' in \emph{Proc. ACM Int. Conf. on Nanoscale Comput. and Commun. (NanoCom)}, Sep. 2021, pp. 1--7.

\bibitem{abkarian2020stretching}
M.~Abkarian and H.~A. Stone, ``Stretching and break-up of saliva filaments during speech: {A} route for pathogen aerosolization and its potential mitigation,'' \emph{Phys. Rev. Fluids}, vol.~5, no.~10, Oct. 2020, {Art. no. 102301}.

\bibitem{papineni1997size}
R.~S. Papineni and F.~S. Rosenthal, ``The size distribution of droplets in the exhaled breath of healthy human subjects,'' \emph{J. of Aerosol Med.}, vol.~10, no.~2, pp. 105--116, Jan. 1997.

\bibitem{bourouibaViolentExpiratoryEvents2014}
L.~Bourouiba, E.~Dehandschoewercker, and J.~W.~M. Bush, ``Violent expiratory events: {On} coughing and sneezing,'' \emph{J. of Fluid Mech.}, vol. 745, pp. 537--563, Apr. 2014.

\bibitem{olmedoDistributionExhaledContaminants2012}
I.~Olmedo, P.~V. Nielsen, M.~{Ruiz de Adana}, R.~L. Jensen, and P.~Grzelecki, ``Distribution of exhaled contaminants and personal exposure in a room using three different air distribution strategies,'' \emph{Indoor Air}, vol.~22, no.~1, pp. 64--76, Feb. 2012.

\bibitem{xuMeasuringExhaledBreath2015}
C.~Xu, P.~V. Nielsen, G.~Gong, L.~Liu, and R.~L. Jensen, ``Measuring the exhaled breath of a manikin and human subjects,'' \emph{Indoor Air}, vol.~25, no.~2, pp. 188--197, Apr. 2015.

\bibitem{koca2021molecular}
C.~Koca, M.~Civas, S.~M. Sahin, O.~Ergonul, and O.~B. Akan, ``Molecular communication theoretical modeling and analysis of {SARS-CoV2} transmission in human respiratory system,'' \emph{IEEE Trans. Mol. Biol. Multi-Scale Commun.}, vol.~7, no.~3, pp. 153--164, Apr. 2021.

\bibitem{debackerFlowAnalysesLower2008}
J.~W. De~Backer \emph{et~al.}, ``Flow analyses in the lower airways: {{Patient-specific}} model and boundary conditions,'' \emph{Med. Eng. \& Phys.}, vol.~30, no.~7, pp. 872--879, Sep. 2008.

\bibitem{weibelMorphometryHumanLung1963}
E.~R. Weibel, \emph{Morphometry of the human lung}.\hskip 1em plus 0.5em minus 0.4em\relax Springer, 1963.

\bibitem{launderNumericalComputationTurbulent1974}
B.~E. Launder and D.~B. Spalding, ``The numerical computation of turbulent flows,'' \emph{Comput. Methods in Appl. Mech. and Eng.}, vol.~3, no.~2, pp. 269--289, Mar. 1974.

\bibitem{xuInvestigationInhalationExhalation2020}
X.~Xu, J.~Wu, W.~Weng, and M.~Fu, ``Investigation of inhalation and exhalation flow pattern in a realistic human upper airway model by {{PIV}} experiments and {{CFD}} simulations,'' \emph{Biomech. and Model. in Mechanobiol.}, vol.~19, no.~5, pp. 1679--1695, Oct. 2020.

\bibitem{fontes2020study}
D.~Fontes, J.~Reyes, K.~Ahmed, and M.~Kinzel, ``A study of fluid dynamics and human physiology factors driving droplet dispersion from a human sneeze,'' \emph{Phys. of Fluids}, vol.~32, no.~11, Nov. 2020.

\bibitem{li2006vivo}
Y.~Li \emph{et~al.}, ``In vivo protective performance of {N95} respirator and surgical facemask,'' \emph{Am. J. Ind. Med.}, vol.~49, no.~12, pp. 1056--1065, Dec. 2006.

\bibitem{peric2020analytical}
R.~Peri{\'c} and M.~Peri{\'c}, ``Analytical and {{Numerical Investigation}} of the {{Airflow}} in {{Face Masks}} used for {{Protection}} against {{COVID-19 Virus}} --{{Implications}} for {{Mask Design}} and {{Usage}},'' \emph{J. Appl. Fluid Mech.}, vol.~13, no.~06, Nov. 2020.

\bibitem{mosayebi2018early}
R.~Mosayebi, A.~Ahmadzadeh, W.~Wicke, V.~Jamali, R.~Schober, and M.~Nasiri-Kenari, ``Early cancer detection in blood vessels using mobile nanosensors,'' \emph{IEEE Trans. Nanobiosci.}, vol.~18, no.~2, pp. 103--116, Dec. 2018.

\bibitem{nakano2016performance}
T.~Nakano \emph{et~al.}, ``Performance evaluation of leader--follower-based mobile molecular communication networks for target detection applications,'' \emph{IEEE Trans. Commun.}, vol.~65, no.~2, pp. 663--676, Nov. 2016.

\bibitem{nakano2019methods}
T.~Nakano, Y.~Okaie, S.~Kobayashi, T.~Hara, Y.~Hiraoka, and T.~Haraguchi, ``Methods and applications of mobile molecular communication,'' \emph{Proc. IEEE}, vol. 107, no.~7, pp. 1442--1456, Jun. 2019.

\bibitem{cao2019diffusive}
T.~N. Cao \emph{et~al.}, ``Diffusive mobile {MC} with absorbing receivers: {Stochastic} analysis and applications,'' \emph{IEEE Trans. Mol. Biol. Multi-Scale Commun.}, vol.~5, no.~2, pp. 84--99, Mar. 2020.

\bibitem{huang2021membrane}
X.~Huang, Y.~Fang, A.~Noel, and N.~Yang, ``Membrane fusion-based transmitter design for static and diffusive mobile molecular communication systems,'' \emph{IEEE Trans. Commun.}, vol.~70, no.~1, pp. 132--148, Oct. 2021.

\bibitem{ahmadzadeh2017diffusive}
A.~Ahmadzadeh, V.~Jamali, A.~Noel, and R.~Schober, ``Diffusive mobile molecular communications over time-variant channels,'' \emph{IEEE Commun. Lett.}, vol.~21, no.~6, pp. 1265--1268, Mar. 2017.

\bibitem{liangModelingVolatileOrganic2012}
W.~Liang, P.~Gao, J.~Guan, and X.~Yang, ``Modeling volatile organic compound ({{VOC}}) concentrations due to material emissions in a real residential unit. {{Part I}}: {{Methodology}} and a preliminary case study,'' \emph{Build. Simul.}, vol.~5, no.~4, pp. 351--357, Dec. 2012.

\bibitem{wongEffectsMedicalStaffs2022}
K.~Y. Wong \emph{et~al.}, ``Effects of medical staff's turning movement on dispersion of airborne particles under large air supply diffuser during operative surgeries,'' \emph{Environ. Sci. and Pollut. Res.}, vol.~29, no.~54, pp. 82\,492--82\,511, Nov. 2022.

\bibitem{khalidModelingViralAerosol2020}
M.~Khalid, O.~Amin, S.~Ahmed, B.~Shihada, and M.-S. Alouini, ``Modeling of viral aerosol transmission and detection,'' \emph{IEEE Trans. Commun.}, vol.~68, no.~8, pp. 4859--4873, Aug. 2020.

\bibitem{pierobon2011diffusion}
M.~Pierobon and I.~F. Akyildiz, ``Diffusion-based noise analysis for molecular communication in nanonetworks,'' \emph{IEEE Trans. Signal Process.}, vol.~59, no.~6, pp. 2532--2547, Feb. 2011.

\bibitem{pierobon2011noise}
------, ``Noise analysis in ligand-binding reception for molecular communication in nanonetworks,'' \emph{IEEE Trans. Signal Process.}, vol.~59, no.~9, pp. 4168--4182, Jun. 2011.

\bibitem{singhal2014effect}
A.~Singhal, R.~K. Mallik, and B.~Lall, ``Effect of molecular noise in diffusion-based molecular communication,'' \emph{IEEE Wirel. Commun. Lett.}, vol.~3, no.~5, pp. 489--492, Aug. 2014.

\bibitem{srinivas2012molecular}
K.~V. Srinivas, A.~W. Eckford, and R.~S. Adve, ``Molecular communication in fluid media: {The} additive inverse gaussian noise channel,'' \emph{IEEE Trans. Inf. Theory}, vol.~58, no.~7, pp. 4678--4692, Apr. 2012.

\bibitem{bhardwaj2021recent}
S.~K. Bhardwaj \emph{et~al.}, ``Recent progress in nanomaterial-based sensing of airborne viral and bacterial pathogens,'' \emph{Environ. Int.}, vol. 146, Jan. 2021, {Art. no. 106183}.

\bibitem{cavazzutiStatisticalAnalysisExhaled2023}
M.~Cavazzuti, L.~Campanelli, and P.~Tartarini, ``A statistical analysis of exhaled respiratory droplet trajectory distribution in view of assessing the risk of infectious disease transmission,'' \emph{J. of Phys.: Conf. Ser.}, vol. 2648, no.~1, Dec. 2023, {Art. no. 012031}.

\bibitem{khalid2018system}
M.~Khalid, O.~Amin, S.~Ahmed, and M.-S. Alouini, ``System modeling of virus transmission and detection in molecular communication channels,'' in \emph{Proc. IEEE Int. Conf. on Commun. (ICC)}, May 2018, pp. 1--6.

\bibitem{seinfeldAtmosphericChemistryPhysics2016}
J.~H. Seinfeld and S.~N. Pandis, \emph{Atmospheric Chemistry and Physics: {From} Air Pollution to Climate Change}, 3rd~ed.\hskip 1em plus 0.5em minus 0.4em\relax Newark: Wiley, Apr. 2016.

\bibitem{linNearfieldToolSimulating2003}
J.~C. Lin \emph{et~al.}, ``A near-field tool for simulating the upstream influence of atmospheric observations: {{The}} stochastic time-inverted lagrangian transport ({{STILT}}) model,'' \emph{J. of Geophys. Res.: Atmospheres}, vol. 108, no. D16, Aug. 2003.

\bibitem{dboukCoughingAirborneDroplet2020}
T.~Dbouk and D.~Drikakis, ``On coughing and airborne droplet transmission to humans,'' \emph{Phys. of Fluids}, vol.~32, no.~5, p. 053310, May 2020.

\bibitem{steinNOAAsHYSPLITAtmospheric2015}
A.~F. Stein, R.~R. Draxler, G.~D. Rolph, B.~J.~B. Stunder, M.~D. Cohen, and F.~Ngan, ``{{NOAA}}'s {{HYSPLIT}} atmospheric transport and dispersion modeling system,'' \emph{Bulletin of the American Meteorological Society}, Dec. 2015.

\bibitem{agencyCalpuffDispersionModel2012}
U.~S. E.~P. Agency, \emph{Calpuff Dispersion Model, {User's} Guide}.\hskip 1em plus 0.5em minus 0.4em\relax Bibliogov, Sep. 2012.

\bibitem{crawfordUseGaussianMixture2020}
A.~Crawford, ``The use of {{Gaussian}} mixture models with atmospheric lagrangian particle dispersion models for density estimation and feature identification,'' \emph{Atmosphere}, vol.~11, no.~12, p. 1369, Dec. 2020.

\bibitem{Ladhani2020-ha}
L.~Ladhani, G.~Pardon, P.~Moons, H.~Goossens, and W.~van~der Wijngaart, ``Electrostatic sampling of patient breath for pathogen detection: {A} pilot study,'' \emph{Front. Mech. Eng.}, vol.~6, Jun. 2020.

\bibitem{Verreault2008-ko}
D.~Verreault, S.~Moineau, and C.~Duchaine, ``Methods for sampling of airborne viruses,'' \emph{Microbiol. Mol. Biol. Rev.}, vol.~72, no.~3, pp. 413--444, Sep. 2008.

\bibitem{Wang2024-wn}
R.~Wang and M.~D. Davis, ``A concise review of exhaled breath testing for respiratory clinicians and researchers,'' \emph{Respir. Care}, vol.~69, no.~5, pp. 613--620, Apr. 2024.

\bibitem{padilla2010drift}
M.~Padilla, A.~Perera, I.~Montoliu, A.~Chaudry, K.~Persaud, and S.~Marco, ``Drift compensation of gas sensor array data by orthogonal signal correction,'' \emph{Chemom. and Intell. Lab. Syst.}, vol. 100, no.~1, pp. 28--35, Jan. 2010.

\bibitem{Malik2021-pf}
M.~Malik, A.-C. Kunze, T.~Bahmer, S.~Herget-Rosenthal, and T.~Kunze, ``{SARS-CoV-2}: {Viral} loads of exhaled breath and oronasopharyngeal specimens in hospitalized patients with {COVID-19},'' \emph{Int. J. Infect. Dis.}, vol. 110, pp. 105--110, Sep. 2021.

\bibitem{nwanochie2022}
E.~Nwanochie and J.~C. Linnes, ``Review of non-invasive detection of {SARS-CoV-2} and other respiratory pathogens in exhaled breath condensate,'' \emph{J. Breath Res.}, vol.~16, no.~2, Mar. 2022.

\bibitem{Daniels2021-bq}
J.~Daniels \emph{et~al.}, ``A mask-based diagnostic platform for point-of-care screening of {Covid-19},'' \emph{Biosens. Bioelectron.}, vol. 192, p. Art. no. 113486, Nov. 2021.

\bibitem{Ghumra2023-nk}
D.~P. Ghumra \emph{et~al.}, ``Rapid direct detection of {SARS-CoV-2} aerosols in exhaled breath at the point of care,'' \emph{ACS Sensors}, vol.~8, no.~8, pp. 3023--3031, Aug. 2023.

\bibitem{kubavn2013exhaled}
P.~Kub{\'a}{\v{n}} and F.~Foret, ``Exhaled breath condensate: {Determination} of non-volatile compounds and their potential for clinical diagnosis and monitoring. {A} review,'' \emph{Anal. Chim. Acta}, vol. 805, pp. 1--18, Dec. 2013.

\bibitem{calvert1990glossary}
J.~G. Calvert, ``Glossary of atmospheric chemistry terms ({Recommendations 1990}),'' \emph{Pure and Appl. Chem.}, vol.~62, no.~11, pp. 2167--2219, Oct. 1990.

\bibitem{tsi3089}
\BIBentryALTinterwordspacing
``{TSI 3089 Nanometer Aerosol Sampler},'' 2024, accessed: 2024-12-17. [Online]. Available: \url{https://www.kenelec.com.au/products/tsi-3089-nanometer-aerosol-sampler/}
\BIBentrySTDinterwordspacing

\bibitem{Stakenborg2022-jw}
T.~Stakenborg \emph{et~al.}, ``Molecular detection of {SARS-COV-2} in exhaled breath at the point-of-need,'' \emph{Biosens. Bioelectron.}, vol. 217, p. Art. no. 114663, Dec. 2022.

\bibitem{mahajan2020metal}
S.~Mahajan and S.~Jagtap, ``Metal-oxide semiconductors for carbon monoxide {(CO)} gas sensing: {A} review,'' \emph{Appl. Mater. Today}, vol.~18, Mar. 2020, {Art. no. 100483}.

\bibitem{wang2010metal}
C.~Wang, L.~Yin, L.~Zhang, D.~Xiang, and R.~Gao, ``Metal oxide gas sensors: {Sensitivity} and influencing factors,'' \emph{Sensors}, vol.~10, no.~3, pp. 2088--2106, Mar. 2010.

\bibitem{saxena2023review}
P.~Saxena and P.~Shukla, ``A review on recent developments and advances in environmental gas sensors to monitor toxic gas pollutants,'' \emph{Environ. Prog. \& Sustain. Energy}, vol.~42, no.~5, Apr. 2023, {Art. no. e14126}.

\bibitem{singh2021recent}
A.~Singh, S.~Sikarwar, A.~Verma, and B.~C. Yadav, ``The recent development of metal oxide heterostructures based gas sensor, their future opportunities and challenges: {A} review,'' \emph{Sensors and Actuators A: Phys.}, vol. 332, Dec. 2021, {Art. no. 113127}.

\bibitem{zhang2020diversiform}
D.~Zhang, Z.~Yang, S.~Yu, Q.~Mi, and Q.~Pan, ``Diversiform metal oxide-based hybrid nanostructures for gas sensing with versatile prospects,'' \emph{Coord. Chem. Rev.}, vol. 413, Jun. 2020, {Art. no. 213272}.

\bibitem{namsheer2021conducting}
K.~Namsheer and C.~S. Rout, ``Conducting polymers: {A} comprehensive review on recent advances in synthesis, properties and applications,'' \emph{RSC Adv.}, vol.~11, no.~10, pp. 5659--5697, Feb. 2021.

\bibitem{macdiarmid1997polyaniline}
A.~G. MacDiarmid, ``Polyaniline and polypyrrole: {Where} are we headed?'' \emph{Synth. Met.}, vol.~84, no. 1-3, pp. 27--34, Jan. 1997.

\bibitem{cho2008fast}
S.~I. Cho and S.~B. Lee, ``Fast electrochemistry of conductive polymer nanotubes: {Synthesis}, mechanism, and application,'' \emph{Accounts of Chem. Res.}, vol.~41, no.~6, pp. 699--707, May 2008.

\bibitem{dincer2019disposable}
C.~Dincer, R.~Bruch, E.~Costa-Rama, M.~T. Fern{\'a}ndez-Abedul, A.~Merko{\c{c}}i, A.~Manz, G.~A. Urban, and F.~G{\"u}der, ``Disposable sensors in diagnostics, food, and environmental monitoring,'' \emph{Adv. Mater.}, vol.~31, no.~30, Jul. 2019, {Art. no. 1806739}.

\bibitem{alam2022recent}
M.~W. Alam \emph{et~al.}, ``Recent progress, challenges, and trends in polymer-based sensors: {A} review,'' \emph{Polym.}, vol.~14, no.~11, May 2022, {Art. no. 2164}.

\bibitem{ahmad2015nanocomposites}
R.~Ahmad, N.~Griffete, A.~Lamouri, N.~Felidj, M.~M. Chehimi, and C.~Mangeney, ``Nanocomposites of gold nanoparticles @ molecularly imprinted polymers: {Chemistry}, processing, and applications in sensors,'' \emph{Chem. of Mater.}, vol.~27, no.~16, pp. 5464--5478, May 2015.

\bibitem{bragazzi2015quartz}
N.~L. Bragazzi, D.~Amicizia, D.~Panatto, D.~Tramalloni, I.~Valle, and R.~Gasparini, ``Quartz-crystal microbalance {(QCM)} for public health: {An} overview of its applications,'' \emph{Adv. in Protein Chem. and Struct. Biol.}, vol. 101, pp. 149--211, Sep. 2015.

\bibitem{speight2012survey}
R.~E. Speight and M.~A. Cooper, ``A survey of the 2010 quartz crystal microbalance literature,'' \emph{J. of Mol. Recognit.}, vol.~25, no.~9, pp. 451--473, Aug. 2012.

\bibitem{wessels2013practical}
A.~Wessels, B.~Kl{\"o}ckner, C.~Siering, and S.~R. Waldvogel, ``Practical strategies for stable operation of {HFF-QCM} in continuous air flow,'' \emph{Sensors}, vol.~13, no.~9, pp. 12\,012--12\,029, Sep. 2013.

\bibitem{kraus2001use}
P.~Kraus, F.~Cooper, D.~Emmons, S.~Ferguson, R.~McClain, and J.~Spates, ``Use of quartz crystal microbalance sensors for monitoring fouling and viscoelastic phenomena in industrial process applications,'' in \emph{Proc. ISA/IEEE. Sensors for Ind. Conf.}, Nov. 2001, pp. 241--245.

\bibitem{akgonullu2022recent}
S.~Akg{\"o}n{\"u}ll{\"u}, E.~{\"O}zg{\"u}r, and A.~Denizli, ``Recent advances in quartz crystal microbalance biosensors based on the molecular imprinting technique for disease-related biomarkers,'' \emph{Chemosens.}, vol.~10, no.~3, Mar. 2022, {Art. no. 106}.

\bibitem{fu2003quartz}
Y.~Fu and H.~O. Finklea, ``Quartz crystal microbalance sensor for organic vapor detection based on molecularly imprinted polymers,'' \emph{Anal. Chem.}, vol.~75, no.~20, pp. 5387--5393, Sep. 2003.

\bibitem{mandal2022surface}
D.~Mandal and S.~Banerjee, ``Surface acoustic wave {(SAW)} sensors: {Physics}, materials, and applications,'' \emph{Sensors}, vol.~22, no.~3, Jan. 2022, {Art. no. 820}.

\bibitem{nair2021acoustic}
M.~P. Nair, A.~J. Teo, and K.~H.~H. Li, ``Acoustic biosensors and microfluidic devices in the decennium: {Principles} and applications,'' \emph{Micromachines}, vol.~13, no.~1, Dec. 2021, {Art. no. 24}.

\bibitem{zhou2019structural}
Z.~Zhou, J.~Tan, J.~Zhang, and M.~Qin, ``Structural optimization and analysis of surface acoustic wave biosensor based on numerical method,'' \emph{Int. J. of Distrib. Sens. Networks}, vol.~15, no.~9, Sep. 2019, {Art. no. 1550147719875648}.

\bibitem{sankaranarayanan2008flow}
S.~K. Sankaranarayanan, S.~Cular, V.~R. Bhethanabotla, and B.~Joseph, ``Flow induced by acoustic streaming on surface-acoustic-wave devices and its application in biofouling removal: {A} computational study and comparisons to experiment,'' \emph{Phys. Rev. E—Stat., Nonlinear, and Soft Matter Phys.}, vol.~77, no.~6, Jun. 2008, {Art. no. 066308}.

\bibitem{damasceno2023recent}
B.~S. Damasceno \emph{et~al.}, ``Recent improvements on surface acoustic wave sensors based on graphenic nanomaterials,'' \emph{Mater. Sci. in Semicond. Process.}, vol. 167, Nov. 2023, {Art. no. 107811}.

\bibitem{mcgill1998performance}
R.~A. McGill \emph{et~al.}, ``Performance optimization of surface acoustic wave chemical sensors,'' \emph{IEEE Trans. Ultrason., Ferroelectr., Freq. Control.}, vol.~45, no.~5, pp. 1370--1380, Aug. 1998.

\bibitem{lange2006integration}
K.~L{\"a}nge, G.~Blaess, A.~Voigt, R.~G{\"o}tzen, and M.~Rapp, ``Integration of a surface acoustic wave biosensor in a microfluidic polymer chip,'' \emph{Biosens. and Bioelectron.}, vol.~22, no.~2, pp. 227--232, Aug. 2006.

\bibitem{griffiths2008brief}
J.~Griffiths \emph{et~al.}, ``A brief history of mass spectrometry,'' \emph{Anal. Chem}, vol.~80, no.~15, pp. 5678--5683, Jul. 2008.

\bibitem{mirsaleh2008electron}
N.~Mirsaleh-Kohan, W.~D. Robertson, and R.~N. Compton, ``Electron ionization time-of-flight mass spectrometry: {Historical} review and current applications,'' \emph{Mass spectrom. rev.}, vol.~27, no.~3, pp. 237--285, Mar. 2008.

\bibitem{mbughuni2016mass}
M.~M. Mbughuni, P.~J. Jannetto, and L.~J. Langman, ``Mass spectrometry applications for toxicology,'' \emph{Ejifcc}, vol.~27, no.~4, Dec. 2016, {Art. no. 272}.

\bibitem{de2007mass}
E.~De~Hoffmann and V.~Stroobant, \emph{Mass spectrometry: {Principles} and applications}.\hskip 1em plus 0.5em minus 0.4em\relax John Wiley \& Sons, 2007.

\bibitem{hiraoka2013fundamentals}
K.~Hiraoka, \emph{Fundamentals of mass spectrometry}.\hskip 1em plus 0.5em minus 0.4em\relax Springer, 2013, vol.~8.

\bibitem{elpa2020automation}
D.~P. Elpa, G.~R.~D. Prabhu, S.-P. Wu, K.~S. Tay, and P.~L. Urban, ``Automation of mass spectrometric detection of analytes and related workflows: {A} review,'' \emph{Talanta}, vol. 208, Feb. 2020, {Art. no. 120304}.

\bibitem{maciel2020miniaturized}
E.~V.~S. Maciel, A.~L. de~Toffoli, E.~Sobieski, C.~E.~D. Naz{\'a}rio, and F.~M. Lancas, ``Miniaturized liquid chromatography focusing on analytical columns and mass spectrometry: {A} review,'' \emph{Anal. chim. acta}, vol. 1103, pp. 11--31, Mar. 2020.

\bibitem{santos2014handbook}
J.~L. Santos and F.~Farahi, \emph{Handbook of optical sensors}.\hskip 1em plus 0.5em minus 0.4em\relax CRC Press, 2014.

\bibitem{khonina2023optical}
S.~N. Khonina, N.~L. Kazanskiy, and M.~A. Butt, ``Optical fibre-based sensors—{An} assessment of current innovations,'' \emph{Biosens.}, vol.~13, no.~9, Aug. 2023, {Art. no. 835}.

\bibitem{nair2010photonic}
R.~V. Nair and R.~Vijaya, ``Photonic crystal sensors: {An} overview,'' \emph{Prog. in Quantum Electron.}, vol.~34, no.~3, pp. 89--134, May 2010.

\bibitem{homola2008surface}
J.~Homola, ``Surface plasmon resonance sensors for detection of chemical and biological species,'' \emph{Chem. Rev.}, vol. 108, no.~2, pp. 462--493, Jan. 2008.

\bibitem{dey2018semiconductor}
A.~Dey, ``Semiconductor metal oxide gas sensors: {A} review,'' \emph{Mater. Sci. and Eng.: B}, vol. 229, pp. 206--217, Mar. 2018.

\bibitem{du2022advances}
J.~Du \emph{et~al.}, ``Advances in flexible optoelectronics based on chemical vapor deposition-grown graphene,'' \emph{Adv. Funct. Mater.}, vol.~32, no.~42, Jan. 2022, {Art. no. 2203115}.

\bibitem{gupta2017optical}
B.~D. Gupta, A.~M. Shrivastav, and S.~P. Usha, \emph{Optical Sensors for Biomedical Diagnostics and Environmental Monitoring}.\hskip 1em plus 0.5em minus 0.4em\relax CRC Press, 2017.

\bibitem{narsaiah2012optical}
K.~Narsaiah, S.~N. Jha, R.~Bhardwaj, R.~Sharma, and R.~Kumar, ``Optical biosensors for food quality and safety assurance—{A} review,'' \emph{J. of Food Sci. and Technol.}, vol.~49, pp. 383--406, Aug. 2012.

\bibitem{olle2020advancements}
E.~P. Oll{\'e}, J.~Farr{\'e}-Llad{\'o}s, and J.~Casals-Terr{\'e}, ``Advancements in microfabricated gas sensors and microanalytical tools for the sensitive and selective detection of odors,'' \emph{Sensors}, vol.~20, no.~19, Sep. 2020, {Art. no. 5478}.

\bibitem{akbar2015chip}
M.~Akbar, M.~Restaino, and M.~Agah, ``Chip-scale gas chromatography: {From} injection through detection,'' \emph{Microsystems \& Nanoeng.}, vol.~1, no.~1, pp. 1--8, Dec. 2015.

\bibitem{mcnair2019basic}
H.~M. McNair, J.~M. Miller, and N.~H. Snow, \emph{Basic gas chromatography}.\hskip 1em plus 0.5em minus 0.4em\relax John Wiley \& Sons, 2019.

\bibitem{koblentz2003pathogens}
G.~Koblentz, ``Pathogens as weapons: {The} international security implications of biological warfare,'' \emph{Int. Secur.}, pp. 84--122, May 2004.

\bibitem{rydosz2018sensors}
A.~Rydosz, ``Sensors for enhanced detection of acetone as a potential tool for noninvasive diabetes monitoring,'' \emph{Sensors}, vol.~18, no.~7, Jul. 2018, {Art. no. 2298}.

\bibitem{hubbard1920determination}
R.~S. Hubbard, ``Determination of acetone in expired air,'' \emph{J. of Biol. Chem.}, vol.~43, no.~1, p. 57–65, Aug. 1920.

\bibitem{wang2013breath}
Z.~Wang and C.~Wang, ``Is breath acetone a biomarker of diabetes? {A} historical review on breath acetone measurements,'' \emph{J. Breath Res.}, vol.~7, no.~3, Aug. 2013, {Art. no. 037109}.

\bibitem{ghimenti2013monitoring}
S.~Ghimenti \emph{et~al.}, ``Monitoring breath during oral glucose tolerance tests,'' \emph{J. Breath Res.}, vol.~7, no.~1, Feb. 2013, {Art. no. 017115}.

\bibitem{sun2015determination}
M.~Sun \emph{et~al.}, ``Determination of breath acetone in 149 {Type} 2 diabetic patients using a ringdown breath-acetone analyzer,'' \emph{Anal. and Bioanal. Chem.}, vol. 407, no.~6, p. 1641–1650, Jan. 2015.

\bibitem{harkins2004exhaled}
M.~S. Harkins, K.~Fiato, and G.~K. Iwamoto, ``Exhaled nitric oxide predicts asthma exacerbation,'' \emph{J. of Asthma}, vol.~41, no.~4, p. 471–476, Jan. 2004.

\bibitem{gilchrist2012investigation}
F.~J. Gilchrist \emph{et~al.}, ``An investigation of suitable bag materials for the collection and storage of breath samples containing hydrogen cyanide,'' \emph{J. Breath Res.}, vol.~6, no.~3, Jul. 2012, {Art. no. 036004}.

\bibitem{smith2013hydrogen}
D.~Smith, P.~Španěl, F.~J. Gilchrist, and W.~Lenney, ``Hydrogen cyanide, a volatile biomarker of {Pseudomonas aeruginosa} infection,'' \emph{J. Breath Res.}, vol.~7, no.~4, Nov. 2013, {Art. no. 044001}.

\bibitem{dent2013exhaled}
A.~G. Dent, T.~G. Sutedja, and P.~V. Zimmerman, ``Exhaled breath analysis for lung cancer,'' \emph{J. of Thorac. Dis.}, vol.~5, no.~5, Aug. 2013.

\bibitem{sakumura2017diagnosis}
Y.~Sakumura \emph{et~al.}, ``Diagnosis by volatile organic compounds in exhaled breath from lung cancer patients using support vector machine algorithm,'' \emph{Sensors}, vol.~17, no.~2, Feb. 2017, {Art. no. 287}.

\bibitem{buszewski2007human}
B.~Buszewski, M.~Kęsy, T.~Ligor, and A.~Amann, ``Human exhaled air analytics: {Biomarkers} of diseases,'' \emph{Biomed. Chromatogr.}, vol.~21, no.~6, p. 553–566, Apr. 2007.

\bibitem{carpagnano2002interleukin}
G.~Carpagnano, O.~Resta, M.~Foschino-Barbaro, E.~Gramiccioni, and F.~Carpagnano, ``Interleukin-6 is increased in breath condensate of patients with non-small cell lung cancer,'' \emph{The Int. J. of Biol. Markers}, vol.~17, no.~2, p. 141–145, Apr. 2002.

\bibitem{Haines1977breath}
A.~Haines, J.~Dilawari, G.~Metz, L.~Blendis, and H.~Wiggins, ``Breath-methane in patients with cancer of the large bowel,'' \emph{The Lancet}, vol. 310, no. 8036, p. 481–483, Sep. 1977.

\bibitem{Sivertsen1992breath}
S.~M. Sivertsen, A.~Bjørneklett, H.~P. Gullestad, and K.~Nygaard, ``Breath methane and colorectal cancer,'' \emph{Scand. J. of Gastroenterol.}, vol.~27, no.~1, p. 25–28, Jan. 1992.

\bibitem{Grabowska2013detection}
B.~Grabowska-Polanowska \emph{et~al.}, ``Detection of potential chronic kidney disease markers in breath using gas chromatography with mass-spectral detection coupled with thermal desorption method,'' \emph{J. of Chromatogr. A}, vol. 1301, p. 179–189, Aug. 2013.

\bibitem{Weitz1991high}
Z.~Weitz, A.~Birnbaum, J.~Skosey, P.~Sobotka, and E.~Zarling, ``High breath pentane concentrations during acute myocardial infarction,'' \emph{The Lancet}, vol. 337, no. 8747, p. 933–935, Apr. 1991.

\bibitem{Mendis1995expired}
S.~Mendis, P.~A. Sobotka, and D.~E. Euler, ``Expired hydrocarbons in patients with acute myocardial infarction,'' \emph{Free. Radic. Res.}, vol.~23, no.~2, p. 117–122, Jan. 1995.

\bibitem{Olopade1997exhalded}
C.~O. Olopade \emph{et~al.}, ``Exhaled pentane and nitric oxide levels in patients with obstructive sleep apnea,'' \emph{Chest}, vol. 111, no.~6, p. 1500–1504, Jun. 1997.

\bibitem{carpagnano2002increased}
G.~E. Carpagnano, S.~A. Kharitonov, O.~Resta, M.~P. Foschino-Barbaro, E.~Gramiccioni, and P.~J. Barnes, ``Increased 8-isoprostane and interleukin-6 in breath condensate of obstructive sleep apnea patients,'' \emph{Chest}, vol. 122, no.~4, p. 1162–1167, Oct. 2002.

\bibitem{Davies1997quantitative}
S.~Davies, P.~Spanel, and D.~Smith, ``Quantitative analysis of ammonia on the breath of patients in end-stage renal failure,'' \emph{Kidney Int.}, vol.~52, no.~1, p. 223–228, Jul. 1997.

\bibitem{Popa2011ethylene}
C.~Popa \emph{et~al.}, ``Ethylene and ammonia traces measurements from the patients’ breath with renal failure via {LPAS} method,'' \emph{Appl. Phys. B}, vol. 105, no.~3, p. 669–674, Sep. 2011.

\bibitem{kearney2002breath}
D.~J. Kearney, T.~Hubbard, and D.~Putnam, ``Breath ammonia measurement in {Helicobacter} pylori infection,'' \emph{Dig. Dis. and Sci.}, vol.~47, pp. 2523--2530, Nov. 2002.

\bibitem{kapishon2013atmospheric}
V.~Kapishon, G.~K. Koyanagi, V.~Blagojevic, and D.~K. Bohme, ``Atmospheric pressure chemical ionization mass spectrometry of pyridine and isoprene: {Potential breath exposure and disease biomarkers},'' \emph{J. Breath Res.}, vol.~7, no.~2, Apr. 2013, {Art. no. 026005}.

\bibitem{selvaraj2020advances}
R.~Selvaraj, N.~J. Vasa, S.~M.~S. Nagendra, and B.~Mizaikoff, ``Advances in mid-infrared spectroscopy-based sensing techniques for exhaled breath diagnostics,'' \emph{Mol.}, vol.~25, no.~9, May 2020, {Art. no. 2227}.

\bibitem{phillips1993increased}
M.~Phillips, M.~Sabas, and J.~Greenberg, ``Increased pentane and carbon disulfide in the breath of patients with schizophrenia,'' \emph{J. of Clin. Pathol.}, vol.~46, no.~9, p. 861–864, Sep. 1993.

\bibitem{sedghi1994elevated}
S.~Sedghi, A.~Keshavarzian, M.~Klamut, D.~A. Eiznhamer, and E.~J. Zarling, ``Elevated breath ethane levels in active ulcerative colitis: {Evidence} for excessive lipid peroxidation,'' \emph{The Am. J. of Gastroenterol.}, vol.~89, no.~12, Dec. 1994, {Art. no. 2217-21}.

\bibitem{vasthare2018carbon}
R.~Vasthare, S.~Kumar, and L.~R. Arron, ``Carbon monoxide breath analyzers and its role in tobacco cessation: {A} narrative review of literature,'' \emph{J. of Int. Oral Heal.}, vol.~10, no.~2, pp. 71--76, Mar. 2018.

\bibitem{ryter2013carbon}
S.~W. Ryter and A.~M.~K. Choi, ``Carbon monoxide in exhaled breath testing and therapeutics,'' \emph{J. Breath Res.}, vol.~7, no.~1, Feb. 2013, {Art. no. 017111}.

\bibitem{zheng2018bacterial}
Y.~Zheng, H.~Chen, M.~Yao, and X.~Li, ``Bacterial pathogens were detected from human exhaled breath using a novel protocol,'' \emph{J. Aerosol Sci.}, vol. 117, pp. 224--234, Mar. 2018.

\bibitem{cheng2025visual}
J.~Cheng, Y.~An, Q.~Wang, Z.~Chen, and Y.~Tong, ``Visual detection of {Mycobacterium} tuberculosis in exhaled breath using {N95} enrichment respirator, {RPA}, and lateral flow assay,'' \emph{Talanta}, vol. 286, May 2025, {Art. no. 127490}.

\bibitem{bellarmino2024covid}
N.~Bellarmino \emph{et~al.}, ``{COVID-19} detection from exhaled breath,'' \emph{Sci. Rep.}, vol.~14, no.~1, Oct. 2024, {Art. no. 23245}.

\bibitem{ryan2020use}
D.~J. Ryan \emph{et~al.}, ``Use of exhaled breath condensate {(EBC)} in the diagnosis of {SARS-COV-2 (COVID-19)},'' \emph{Thorax}, vol.~76, no.~1, p. 86–88, Oct. 2020.

\bibitem{fabian2008influenza}
P.~Fabian \emph{et~al.}, ``Influenza virus in human exhaled breath: {An} observational study,'' \emph{PLoS One}, vol.~3, no.~7, Jul. 2008, {Art. no. e2691}.

\bibitem{carpagnano2011hpv}
G.~E. Carpagnano \emph{et~al.}, ``{HPV} in exhaled breath condensate of lung cancer patients,'' \emph{Br. J. Cancer}, vol. 105, no.~8, pp. 1183--1190, Sep. 2011.

\bibitem{diefenderfer2024new}
J.~Diefenderfer, H.~D. Bean, and E.~A. Higgins~Keppler, ``New breath diagnostics for fungal disease,'' \emph{Curr. Clin. Microbiol. Rep.}, vol.~11, no.~2, p. 51–61, Feb. 2024.

\bibitem{carpagnano2016analysis}
G.~E. Carpagnano \emph{et~al.}, ``Analysis of the fungal microbiome in exhaled breath condensate of patients with asthma,'' \emph{Allergy Asthma Proc.}, vol.~37, no.~3, p. 41–46, May 2016.

\bibitem{bhimji2018asper}
A.~Bhimji \emph{et~al.}, ``Aspergillus galactomannan detection in exhaled breath condensate compared to bronchoalveolar lavage fluid for the diagnosis of invasive aspergillosis in immunocompromised patients,'' \emph{Clin. Microbiol. Infect.}, vol.~24, no.~6, p. 640–645, Jun. 2018.

\bibitem{bitterman2024exhaled}
R.~Bitterman \emph{et~al.}, ``Exhaled breath condensate surveillance for {Aspergillus} in acute leukemia—{A} pilot trial,'' \emph{Open Forum Infect. Dis.}, vol.~11, no.~10, Sep. 2024.

\bibitem{breathbase_data}
\BIBentryALTinterwordspacing
{Breathomix}, ``{BreathBase® Data},'' 2024, {Accessed: 2024-11-13}. [Online]. Available: \url{https://www.breathomix.com/breathbase-data/}
\BIBentrySTDinterwordspacing

\bibitem{hbdb_data}
\BIBentryALTinterwordspacing
{Computational Molecular Design \& Metabolomics Laboratory, National Taiwan University}, ``{Human Breathomics Database},'' 2024, accessed: 2024-11-13. [Online]. Available: \url{https://hbdb.cmdm.tw/}
\BIBentrySTDinterwordspacing

\bibitem{melo2010exhaled}
R.~E. Melo, T.~A. Popov, and D.~Solé, ``Exhaled breath temperature, a new biomarker in asthma control: {A} pilot study,'' \emph{J. Bras. de Pneumol.}, vol.~36, no.~6, p. 693–699, Dec. 2010.

\bibitem{garcia2013increased}
G.~García, M.~Bergna, E.~Uribe, A.~Yañez, and J.~B. Soriano, ``Increased exhaled breath temperature in subjects with uncontrolled asthma,'' \emph{The Int. J. of Tuberc. and Lung Dis.}, vol.~17, no.~7, p. 969–972, Jul. 2013.

\bibitem{wojsyk2017evaluation}
I.~Wojsyk-Banaszak \emph{et~al.}, ``Evaluation of exhaled breath temperature {(EBT)} as a marker and predictor of asthma exacerbation in children and adolescents,'' \emph{J. of Asthma}, vol.~54, no.~7, pp. 699--705, Mar. 2017.

\bibitem{paredi2003exhaled}
P.~Paredi, S.~Kharitonov, and P.~Barnes, ``Exhaled breath temperature in asthma,'' \emph{Eur. Respir. J.}, vol.~21, no.~1, pp. 195--195, Jan. 2003.

\bibitem{klokstad2010effect}
S.~Kl{\o}kstad, A.~Bikov, Z.~Lazar, G.~Galffy, G.~Losonczy, and I.~Horvath, ``The effect of smoking and {COPD} on exhaled breath temperature,'' \emph{Eur. Respir J.}, vol.~36, Sep. 2010, {Art. no. 832s}.

\bibitem{lazar2014exhaled}
Z.~L{\'a}z{\'a}r, A.~Bikov, F.~Martinovszky, G.~G{\'a}lffy, G.~Losonczy, and I.~Horv{\'a}th, ``Exhaled breath temperature in patients with stable and exacerbated {COPD},'' \emph{J. Breath Res.}, vol.~8, no.~4, p. {Art no. 046002}, Sep. 2014.

\bibitem{carpagnano2014exhaled}
G.~E. Carpagnano \emph{et~al.}, ``Exhaled breath temperature in {NSCLC: C}ould be a new non-invasive marker?'' \emph{Med. Oncol.}, vol.~31, pp. 1--6, May 2014.

\bibitem{carpagnano2015exhaled}
------, ``Is the exhaled breath temperature in lung cancer influenced by airways neoangiogenesis or by inflammation?'' \emph{Med. Oncol.}, vol.~32, pp. 1--5, Oct. 2015.

\bibitem{popov2016relationship}
T.~A. Popov, T.~Kralimarkova, and L.~M. DuBuske, ``Relationship between exhaled breath temperature and ear temperature in otherwise healthy persons during febrile infectious illness,'' \emph{J. of Allergy and Clin. Immunol.}, vol. 137, no.~2, Feb. 2016, {Art. no. AB202}.

\bibitem{dixit2021exhaled}
K.~Dixit, S.~Fardindoost, A.~Ravishankara, N.~Tasnim, and M.~Hoorfar, ``Exhaled breath analysis for diabetes diagnosis and monitoring: {Relevance}, challenges and possibilities,'' \emph{Biosens.}, vol.~11, no.~12, Nov. 2021, {Art. no. 476}.

\bibitem{lagopati2023role}
N.~Lagopati \emph{et~al.}, ``The role of nano-sensors in breath analysis for early and non-invasive disease diagnosis,'' \emph{Chemosens.}, vol.~11, no.~6, May 2023, {Art. no. 317}.

\bibitem{sukul2017oral}
P.~Sukul, P.~Oertel, S.~Kamysek, and P.~Trefz, ``Oral or nasal breathing? {Real-time} effects of switching sampling route onto exhaled {VOC} concentrations,'' \emph{J. Breath Res.}, vol.~11, no.~2, Mar. 2017, {Art. no. 027101}.

\bibitem{huang2021nose}
Z.~Huang, W.~Wang, and G.~d. Haan, ``Nose breathing or mouth breathing - {A} thermography-based new measurement for sleep monitoring,'' in \emph{Proc. of the IEEE/CVF Conf. on Comput. Vis. and Pattern Recognit. Work. (CVPRW)}, Jun. 2021, p. 3877–3883.

\bibitem{curran2012using}
K.~Curran, P.~Yuan, and D.~Coyle, ``Using acoustic sensors to discriminate between nasal and mouth breathing,'' \emph{Int. J. of Bioinform. Res. and Appl.}, vol.~8, no. 5/6, Oct. 2012, {Art. no. 382}.

\bibitem{yu2017wrinkled}
Y.~Yu, L.~Ye, Y.~Song, Y.~Guan, and J.~Zang, ``Wrinkled nitrile rubber films for stretchable and ultra-sensitive respiration sensors,'' \emph{Extrem. Mech. Lett.}, vol.~11, p. 128–136, Feb. 2017.

\bibitem{whited2023abnormal}
\BIBentryALTinterwordspacing
L.~Whited, M.~F. Hashmi, and D.~D. Graham, ``Abnormal respirations,'' 2023, accessed: 2025-03-17. [Online]. Available: \url{https://www.ncbi.nlm.nih.gov/books/NBK470309/}
\BIBentrySTDinterwordspacing

\bibitem{buaruk2023respiratory}
S.~Buaruk, C.~Srisumarnk, S.~Seinglek, W.~Thaweekul, and S.~Deepaisarn, \emph{Respiratory Disease Classification Using Chest Movement Patterns Measured by Non-contact Sensor}.\hskip 1em plus 0.5em minus 0.4em\relax Springer Nature Switzerland, Jul. 2023, p. 397–409.

\bibitem{yang2022artificial}
Y.~Yang \emph{et~al.}, ``Artificial intelligence-enabled detection and assessment of {Parkinson’s} disease using nocturnal breathing signals,'' \emph{Nat. Med.}, vol.~28, no.~10, p. 2207–2215, Aug. 2022.

\bibitem{bjerkan2024neurovascular}
J.~Bjerkan \emph{et~al.}, ``Neurovascular phase coherence is altered in {Alzheimer’s} disease,'' \emph{Brain Commun.}, vol.~7, no.~1, Dec. 2024.

\bibitem{purnamadjaja2007guiding}
A.~H. Purnamadjaja and R.~A. Russell, ``Guiding robots’ behaviors using pheromone communication,'' \emph{Auton. Robot.}, vol.~23, pp. 113--130, May 2007.

\bibitem{na2021overview}
S.~Na~Songkhla and T.~Nakamoto, ``Overview of quartz crystal microbalance behavior analysis and measurement,'' \emph{Chemosens.}, vol.~9, no.~12, Dec. 2021, {Art. no. 350}.

\bibitem{swinehart1962beer}
D.~F. Swinehart, ``The {B}eer-{L}ambert {L}aw,'' \emph{Journal of chemical education}, vol.~39, no.~7, p. 333, 1962.

\bibitem{koo2016molecular}
B.-H. Koo, C.~Lee, H.~B. Yilmaz, N.~Farsad, A.~Eckford, and C.-B. Chae, ``Molecular {MIMO: From} theory to prototype,'' \emph{IEEE J. Sel. Areas Commun.}, vol.~34, no.~3, pp. 600--614, Feb. 2016.

\bibitem{qiu2016mobile}
S.~Qiu, T.~Asyhari, and W.~Guo, ``Mobile molecular communications: {Positional-distance} codes,'' in \emph{Proc. IEEE Int. Work. on Signal Process. Adv. in Wirel. Commun. (SPAWC)}, Jul. 2016, pp. 1--5.

\bibitem{shakya2018correlated}
P.~Shakya, E.~Kennedy, C.~Rose, and J.~K. Rosenstein, ``Correlated transmission and detection of concentration-modulated chemical vapor plumes,'' \emph{IEEE Sensors J.}, vol.~18, no.~16, pp. 6504--6509, Jun. 2018.

\bibitem{ozmen2018high}
M.~Ozmen, E.~Kennedy, J.~Rose, P.~Shakya, J.~K. Rosenstein, and C.~Rose, ``High speed chemical vapor communication using photoionization detectors in turbulent flow,'' \emph{IEEE Trans. Mol. Biol. Multi-Scale Commun.}, vol.~4, no.~3, pp. 160--170, Jul. 2019.

\bibitem{cole2009biomimetic}
M.~Cole \emph{et~al.}, ``Biomimetic insect infochemical communication system,'' in \emph{Proc. IEEE Sensors}, Oct. 2009, pp. 1358--1361.

\bibitem{giannoukos2017molecular}
S.~Giannoukos, A.~Marshall, S.~Taylor, and J.~Smith, ``Molecular communication over gas stream channels using portable mass spectrometry,'' \emph{J. Am. Soc. Mass Spectrom.}, vol.~28, no.~11, pp. 2371--2383, Jul. 2017.

\bibitem{giannoukos2018chemical}
S.~Giannoukos, D.~T. McGuiness, A.~Marshall, J.~Smith, and S.~Taylor, ``A chemical alphabet for macromolecular communications,'' \emph{Anal. Chem.}, vol.~90, no.~12, pp. 7739--7746, May 2018.

\bibitem{zhou2024pathogen}
Y.~Zhou \emph{et~al.}, ``Pathogen detection via inductively coupled plasma mass spectrometry analysis with nanoparticles,'' \emph{Talanta}, Sep. 2024, {Art. no. 126325}.

\bibitem{bhattacharjee2020testbed}
S.~Bhattacharjee \emph{et~al.}, ``A testbed and simulation framework for air-based molecular communication using fluorescein,'' in \emph{Proc. ACM Int. Conf. on Nanoscale Comput. and Commun. (NanoCom)}, Sep. 2020, pp. 1--6.

\bibitem{drees2020efficient}
J.~P. Drees \emph{et~al.}, ``Efficient simulation of macroscopic molecular communication: {The} pogona simulator,'' in \emph{Proc. ACM Int. Conf. on Nanoscale Comput. and Commun. (NanoCom)}, Sep. 2020, pp. 1--6.

\bibitem{schurwanz2021infectious}
M.~Schurwanz, P.~A. Hoeher, S.~Bhattacharjee, M.~Damrath, L.~Stratmann, and F.~Dressler, ``Infectious disease transmission via aerosol propagation from a molecular communication perspective: {Shannon} meets coronavirus,'' \emph{IEEE Commun. Mag.}, vol.~59, no.~5, pp. 40--46, Jun. 2021.

\bibitem{xie2009exhaled}
X.~Xie, Y.~Li, H.~Sun, and L.~Liu, ``Exhaled droplets due to talking and coughing,'' \emph{J. of the R. Soc. Interface}, vol.~6, no.~6, pp. S703--S714, Dec. 2009.

\bibitem{zhu2006study}
S.~Zhu, S.~Kato, and J.-H. Yang, ``Study on transport characteristics of saliva droplets produced by coughing in a calm indoor environment,'' \emph{Build. and Environ.}, vol.~41, no.~12, pp. 1691--1702, Dec. 2006.

\bibitem{chao2009characterization}
C.~Y.~H. Chao \emph{et~al.}, ``Characterization of expiration air jets and droplet size distributions immediately at the mouth opening,'' \emph{J. of Aerosol Sci.}, vol.~40, no.~2, pp. 122--133, Feb. 2009.

\bibitem{tang2011qualitative}
J.~W. Tang \emph{et~al.}, ``Qualitative real-time schlieren and shadowgraph imaging of human exhaled airflows: {An} aid to aerosol infection control,'' \emph{PLoS One}, vol.~6, no.~6, Jun. 2011, {Art. no. e21392}.

\bibitem{zhang2019distribution}
Y.~Zhang, G.~Feng, Y.~Bi, Y.~Cai, Z.~Zhang, and G.~Cao, ``Distribution of droplet aerosols generated by mouth coughing and nose breathing in an air-conditioned room,'' \emph{Sustain. Cities and Soc.}, vol.~51, Nov. 2019, {Art. no. 101721}.

\bibitem{perello2021cmos}
R.~Perell{\'o}-Roig, J.~Verd, S.~Bota, B.~Soberats, A.~Costa, and J.~Segura, ``{CMOS--MEMS VOC} sensors functionalized via inkjet polymer deposition for high-sensitivity acetone detection,'' \emph{Lab on a Chip}, vol.~21, no.~17, pp. 3307--3315, Jul. 2021.

\bibitem{jaeschke2018novel}
C.~Jaeschke, O.~Gonzalez, J.~J. Gl{\"o}ckler, L.~T. Hagemann, K.~E. Richardson, F.~Adrover, M.~Padilla, J.~Mitrovics, and B.~Mizaikoff, ``A novel modular {eNose} system based on commercial {MOX} sensors to detect low concentrations of {VOCs} for breath gas analysis,'' in \emph{Proc. Eurosensors}, vol.~2, no.~13, Sep. 2018, {Art. no. 993}.

\bibitem{shang2023chemiresistive}
G.~Shang \emph{et~al.}, ``Chemiresistive sensor array with nanostructured interfaces for detection of human breaths with simulated lung cancer breath {VOCs},'' \emph{ACS Sensors}, vol.~8, no.~3, pp. 1328--1338, Mar. 2023.

\bibitem{lommel2021novel}
M.~Lommel \emph{et~al.}, ``Novel measurement system for respiratory aerosols and droplets in indoor environments,'' \emph{Indoor Air}, vol.~31, no.~6, pp. 1860--1873, Jun. 2021.

\bibitem{schulz2024}
I.~Schulz \emph{et~al.}, ``Experimental device to evaluate aerosol dispersion in venues,'' \emph{Appl. Sci.}, vol.~14, no.~13, Jun. 2024, {Art. no. 5601}.

\bibitem{noti2013high}
J.~D. Noti \emph{et~al.}, ``High humidity leads to loss of infectious influenza virus from simulated coughs,'' \emph{PloS One}, vol.~8, no.~2, Feb. 2013, {Art. no. e57485}.

\bibitem{zhang2024experimental}
H.~Zhang, Y.~Pan, C.~Deng, Z.~Niu, R.~You, and C.~Chen, ``Experimental investigation of interpersonal particle transport in an aircraft cabin mockup with nanofiber air filters,'' \emph{Sci. of the Total. Environ.}, vol. 954, Dec. 2024, {Art. no. 176059}.

\bibitem{adrian2011particle}
R.~J. Adrian and J.~Westerweel, \emph{Particle image velocimetry}.\hskip 1em plus 0.5em minus 0.4em\relax Cambridge University Press, 2011, no.~30.

\bibitem{fens2013exhaled}
N.~Fens, M.~Van~der Schee, P.~Brinkman, and P.~Sterk, ``Exhaled breath analysis by electronic nose in airways disease. {Established} issues and key questions,'' \emph{Clin. \& Exp. Allergy}, vol.~43, no.~7, pp. 705--715, Jul. 2013.

\bibitem{das2020non}
S.~Das and M.~Pal, ``Non-invasive monitoring of human health by exhaled breath analysis: {A} comprehensive review,'' \emph{J. of the Electrochem. Soc.}, vol. 167, no.~3, Feb. 2020, {Art. no. 037562}.

\bibitem{queiroz2021time}
J.~d.~N. Queiroz, R.~C.~O. Macedo, G.~M. Tinsley, and A.~Reischak-Oliveira, ``Time-restricted eating and circadian rhythms: {The} biological clock is ticking,'' \emph{Crit. Rev. in Food Sci. and Nutr.}, vol.~61, no.~17, pp. 2863--2875, Jul. 2021.

\bibitem{samudrala2014breath}
D.~Samudrala \emph{et~al.}, ``Breath acetone to monitor life style interventions in field conditions: {An} exploratory study,'' \emph{Obes.}, vol.~22, no.~4, pp. 980--983, Apr. 2014.

\bibitem{zhang2022advanced}
Z.~Zhang \emph{et~al.}, ``Advanced point-of-care testing technologies for human acute respiratory virus detection,'' \emph{Adv. Mater.}, vol.~34, no.~1, Oct. 2021, {Art. no. 2103646}.

\bibitem{ates2021integrated}
H.~C. Ates, A.~Brunauer, F.~von Stetten, G.~A. Urban, F.~G{\"u}der, A.~Merko{\c{c}}i, S.~M. Fr{\"u}h, and C.~Dincer, ``Integrated devices for non-invasive diagnostics,'' \emph{Adv. Funct. Mater.}, vol.~31, no.~15, Jan. 2021, {Art. no. 2010388}.

\bibitem{mills1999targeted}
J.~K. Mills and D.~Needham, ``Targeted drug delivery,'' \emph{Expert. Opin. on Ther. Patents}, vol.~9, no.~11, pp. 1499--1513, Nov. 1999.

\bibitem{skarysz2022fast}
A.~Skarysz \emph{et~al.}, ``Fast and automated biomarker detection in breath samples with machine learning,'' \emph{PloS One}, vol.~17, no.~4, Apr. 2022, {Art. no. e0265399}.

\bibitem{intharawijitr2017simulation}
K.~Intharawijitr, K.~Iida, and H.~Koga, ``Simulation study of low latency network architecture using mobile edge computing,'' \emph{IEICE Trans. Inf. Syst.}, vol. 100, no.~5, pp. 963--972, May 2017.

\bibitem{schulzNegativeLatencyTactile2024}
J.~Schulz \emph{et~al.}, ``Negative {{latency}} in the {{Tactile Internet}} as {{enabler}} for {{global metaverse immersion}},'' \emph{IEEE Network}, vol.~38, no.~5, pp. 167--173, Sep. 2024.

\end{thebibliography}
